\documentclass[journal=jpclcd,manuscript=article]{achemso}
\usepackage[utf8]{inputenc}

\usepackage[labelsep=period]{caption}
\usepackage{pdflscape}
\usepackage{textcomp}
\usepackage{gensymb}
\usepackage{geometry}

\usepackage{subcaption} 

\usepackage{pifont}

\usepackage{amsmath}
\usepackage{amssymb}
\usepackage{multirow}
 
\usepackage{siunitx} 

\usepackage{graphicx}
\graphicspath{ {./images/} }

\usepackage{todonotes}

\usepackage[normalem]{ulem}

\author{Peter M. Attia}
\email{peter.m.attia@gmail.com}
\affiliation{\scriptsize{Department of Materials Science and Engineering, Stanford University, Stanford, CA, USA}}
\author{Alexander Bills} 
\affiliation{Department of Mechanical Engineering, Carnegie Mellon University, Pittsburgh, PA, USA}
\author{Ferran Brosa Planella} 
\affiliation{WMG, University of Warwick, Coventry, UK, and Faraday Institution, Harwell, UK}
\author{Philipp Dechent} 
\affiliation{Institute for Power Electronics and Electrical Drives (ISEA), RWTH Aachen University, Aachen, Germany}
\author{Gon\c{c}alo dos Reis} 
\affiliation{School of Mathematics, University of Edinburgh, Edinburgh, UK and Centro de Matem\'atica e Aplica\c c$\tilde{\text{o}}$es (CMA), FCT, UNL, Caparica, Portugal}
\author{Matthieu Dubarry}
\affiliation{Hawaii Natural Energy Institute, University of Hawaii at Manoa, Honolulu, HI, USA}
\author{Paul Gasper} 
\affiliation{National Renewable Energy Laboratory, Golden, CO, USA}
\author{Richard Gilchrist} 
\affiliation{School of Mathematics, University of Edinburgh, Edinburgh, UK}
\author{Samuel Greenbank} 
\affiliation{Department of Engineering Science, University of Oxford, Oxford, UK}
\author{David Howey} 
\affiliation{Department of Engineering Science, University of Oxford,  Oxford, UK, and Faraday Institution, Harwell, UK}
\author{Ouyang Liu} 
\affiliation{Institute for Infocomm Research, Agency for Science, Technology, and Research (A*STAR), Connexis, Singapore}
\author{Edwin Khoo}  
\affiliation{Institute for Infocomm Research, Agency for Science, Technology, and Research (A*STAR), Connexis, Singapore}
\author{Yuliya Preger}  
\affiliation{Sandia National Laboratories, Albuquerque, NM, USA}
\author{Abhishek Soni}
\affiliation{Department of Mechanical Engineering, University of Cincinnati, Cincinnati, OH, USA}
\author{Shashank Sripad} 
\affiliation{Department of Mechanical Engineering, Carnegie Mellon University, Pittsburgh, PA, USA}
\author{Anna G. Stefanopoulou}  
\affiliation{Department of Mechanical Engineering, University of Michigan, Ann Arbor, MI, USA}
\author{Valentin Sulzer}
\affiliation{Department of Mechanical Engineering, University of Michigan, Ann Arbor, MI, USA}

\title{``Knees'' in lithium-ion battery aging trajectories}

\date{}

\begin{document}

\maketitle


\section{Abstract} 

Lithium-ion batteries can last many years but sometimes exhibit rapid, nonlinear degradation that severely limits battery lifetime.
In this work, we review prior work on ``knees'' in lithium-ion battery aging trajectories.
We first review definitions for knees and three classes of ``internal state trajectories'' (termed snowball, hidden, and threshold trajectories) that can cause a knee.
We then discuss six knee ``pathways'', including lithium plating, electrode saturation, resistance growth, electrolyte and additive depletion, percolation-limited connectivity, and mechanical deformation---some of which have internal state trajectories with signals that are electrochemically undetectable.
We also identify key design and usage sensitivities for knees.
Finally, we discuss challenges and opportunities for knee modeling and prediction.
Our findings illustrate the complexity and subtlety of lithium-ion battery degradation and can aid both academic and industrial efforts to improve battery lifetime.



\newpage

Lithium-ion batteries are expected to play a critical role in decarbonization via their use in electric vehicles and stationary energy storage. One of the most challenging requirements for these demanding applications is long lifetime, with typical warranties of eight years for electric vehicles and ten years for grid storage.\cite{hesse_lithium-ion_2017, bocca_optimal_2020, beltran_lifetime_2020} Battery lifetime requirements will become increasingly challenging as ``million-mile batteries''\cite{harlow_wide_2019} become the expectation for next-generation electric vehicles. Furthermore, as concerns around battery materials mining, manufacturing, and disposal increase\cite{harper_recycling_2019}, improving battery lifetime is a straightforward way to decrease the environmental impact of the lithium-ion battery lifecycle. Thus, understanding and improving the lifetime of lithium-ion batteries is a critical research direction.

\begin{figure}[t]
\centering
\includegraphics[scale=1]{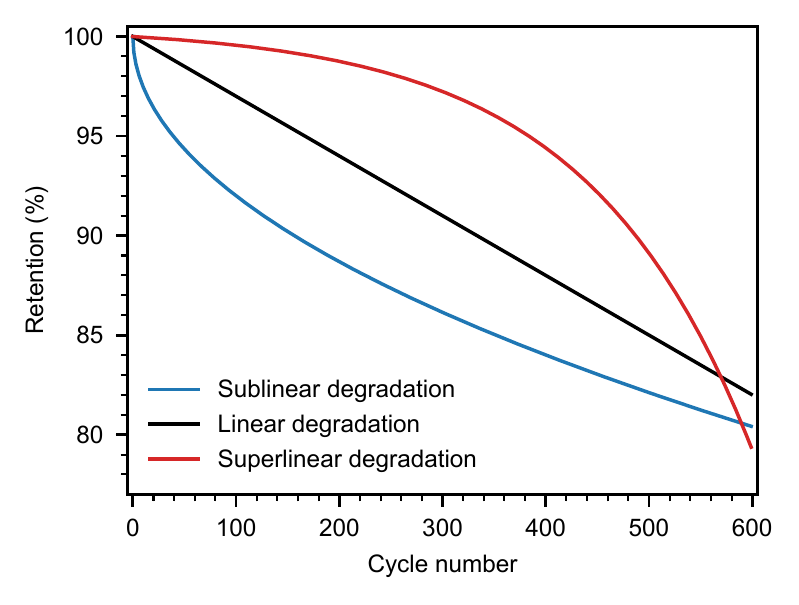}
\caption{Schematic of the three lithium-ion battery aging trajectories: sublinear, linear, and superlinear degradation (``knees''). Here, the $x$ axis is labeled ``cycle number'', although it could also represent equivalent full cycles, capacity or energy throughput, time, or similar. Similarly, the $y$ axis is labeled ``retention'', which could represent capacity, energy, or power retention; furthermore, these values can be either at moderate-high rates from a cycling experiment or from low-rate periodic diagnostic tests. We use this convention (``retention vs. cycle number'') in conceptual figures throughout this work.}
\label{fig:degradation_shapes}
\end{figure}

By definition, lithium-ion batteries can exhibit either linear, sublinear, or superlinear aging trajectories (Figure \ref{fig:degradation_shapes}). In laboratory settings (i.e., single-cell testing using battery cyclers), these aging trajectories are often presented as capacity vs. cycle number or similar. Battery aging trajectories are often linear\cite{ma_editors_2019, keil_electrochemical_2020,  preger_degradation_2020} or sublinear\cite{bloom_accelerated_2001, broussely_aging_2001, wright_calendar-_2002, smith_high_2011, attia_revisiting_2020}{}. Sublinear degradation is often attributed to side reactions such as solid-electrolyte interphase (SEI) growth, which grows approximately with the square root of time or cycle number due to its self-passivating nature.\cite{bloom_accelerated_2001, broussely_aging_2001, broussely_main_2005, wright_calendar-_2002, smith_high_2011, attia_revisiting_2020} While this type of degradation is largely unavoidable, the decelerating degradation rate is a fortunate property for long-lifetime applications. However, superlinear battery degradation is also commonly observed. This type of degradation goes by many names in the battery literature, including ``knee''\cite{diao_algorithm_2019, fermin-cueto_identification_2020}{}, ``rollover failure''\cite{ma_editors_2019}{}, ``nonlinear aging''\cite{schuster_nonlinear_2015, bach_nonlinear_2016, yang_modeling_2017, mandli_analysis_2019, keil_linear_2019, keil_electrochemical_2020, atalay_theory_2020}{}, ``sudden death''\cite{muller_model-based_2019, willenberg_development_2020, kupper_end--life_2018}{}, ``saturation''\cite{lin_comprehensive_2013}{}, ``second-stage degradation''\cite{dubarry_perspective_2020} or ``two-phase degradation''\cite{pugalenthi_piecewise_2020}{}, ``capacity plunge''\cite{fang_capacity_2021}{}, etc.; we use the term ``knee'' in the remainder of this work.
Avoiding or delaying knees is critical to ensure long battery lifetimes; furthermore, knees pose challenges for accurate onboard state-of-health estimation, as batteries with identical states of health (i.e., estimated capacity or energy retention) may have entirely different remaining useful lives\cite{dubarry_perspective_2020, braco_experimental_2020}. However, despite many reports on this topic, a comprehensive understanding of knees is lacking, likely due to the variety and complexity of the observed degradation mechanisms.

In this review, we survey the literature and critically examine both experimental and modeling work on the subject of knees in lithium-ion battery aging. We first review methods to identify the knee point from an aging trajectory. We then identify six knee ``pathways'' from the literature, including lithium plating, electrode saturation, resistance growth, electrolyte and additive depletion, percolation-limited connectivity, and mechanical deformation; each of these knee pathways can be categorized into one or more of three classes of ``internal state trajectories'' (``snowball'', ``hidden'', or ``threshold'') that reflect the measurement requirements for modeling and prediction. We also classify differences in experimentally observed knee behavior as either differences in design, differences in usage conditions, or cell-to-cell/testing variation. Finally, we discuss the implications of our findings to modeling, predicting, and avoiding knees; as a whole, knee prediction is challenging, but a better understanding of the underlying physics will help. This review can serve both academic and industrial efforts to understand and improve lithium-ion battery lifetime.

\section{Defining the knee point}
\label{sec:defining-knee-points}

Knees are often straightforward to identify by eye, especially in single, smooth and ideal aging trajectories (e.g.,~the superlinear aging trend in Figure \ref{fig:degradation_shapes}).
While identifying the \textit{presence} of a knee may be sufficient for some analyses, we are often interested in the \textit{location} of the knee, known as the ``knee point''.
The battery community has not aligned on a standardized definition of the knee point; for instance, while
IEEE Standard 485\texttrademark-2020 defines a capacity knee as when ``the capacity slowly declines throughout
most of the battery’s life, but begins to decrease rapidly in the latter stages''\cite{ieee_power_and_energy_society_ieee_2020}{}, this definition is qualitative and thus unusable for quantitative analysis.
Here, we discuss approaches for quantitative knee point estimation.
This problem can be considered in both the ``offline'' (i.e., methods that identify knee points given the complete aging trajectory) and ``online'' (i.e., methods that identify knee points during use) settings.

\begin{figure}[!ht]
\centering
\includegraphics[scale=1]{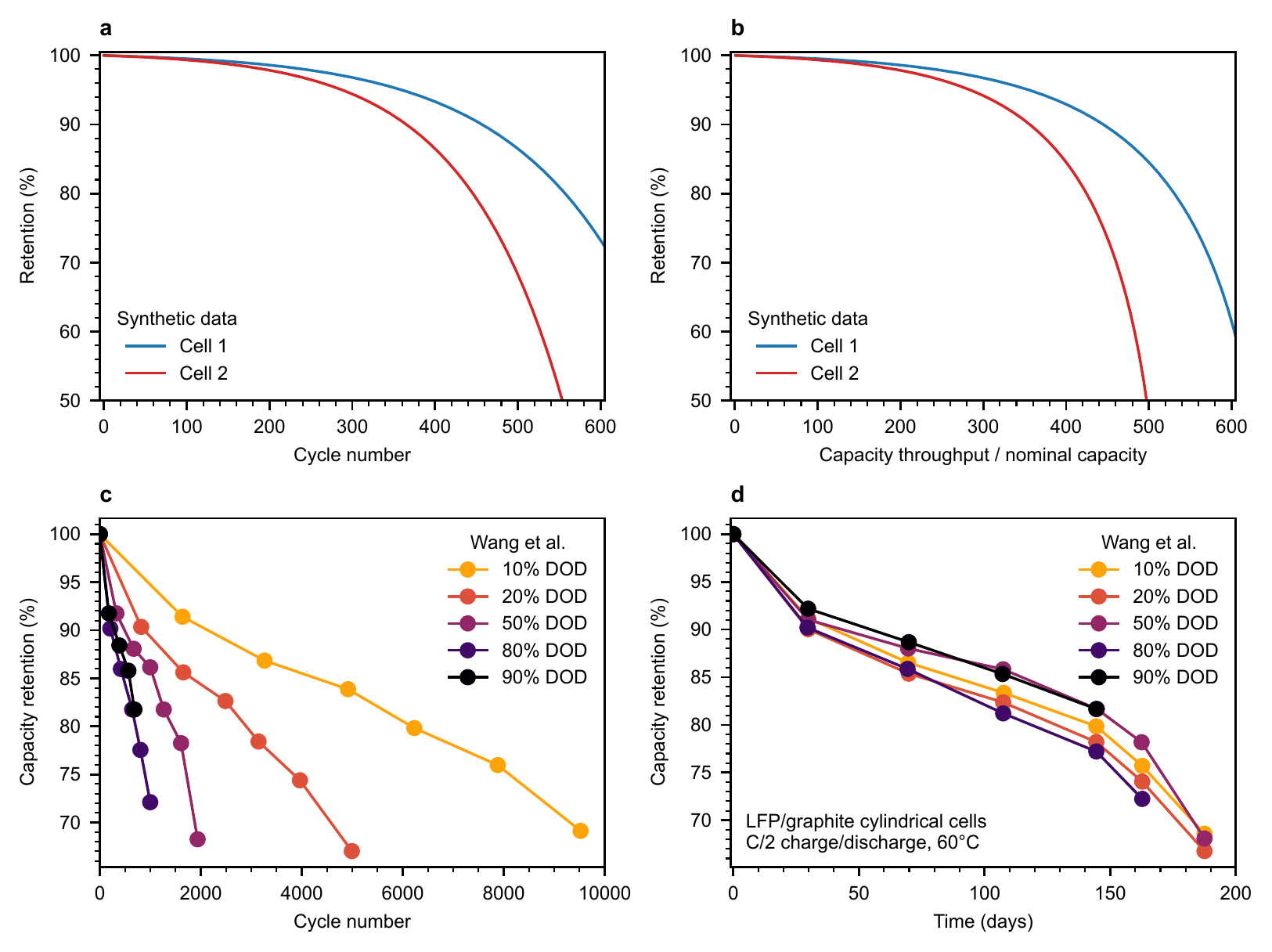}
\caption{Sensitivity of knees to data visualization choices.
(a) Retention vs.~cycle number. The data is artificially generated from an exponential function. 
(b) Retention vs.~capacity throughput, normalized by the nominal capacity. The throughput is calculated by taking the cumulative sum of the retention values.
Note how the knee appears earlier and sharper when viewed with throughput on the $x$ axis.
(c) Capacity retention vs.~cycle number for lithium iron phosphate/graphite cells cycled at varying depth of discharge (DOD). Adopted from Figure 3 of Wang et al.\cite{wang_cycle-life_2011}
(d) Capacity retention vs.~time for lithium iron phosphate/graphite cells cycled at varying DOD.
Note how the curves collapse when plotted as a function of time.
Adopted from Figure 4 of Wang et al.\cite{wang_cycle-life_2011}
}
\label{fig:x_axis}
\end{figure}

First, we note that the convention used for analyzing and visualizing lifetime data impacts the definition and location of the knee. The battery community has many such conventions. For instance, time, cycle number, equivalent full cycles, or capacity/energy throughput can be used to represent the $x$ axis of a lifetime plot. Similarly, the capacity, energy, or power can be used on the $y$ axis (power, or energy divided by time, is less commonly reported but useful in some contexts, such as aviation).
These values can be reported as either absolute or normalized to the initial or nominal value, and for either charge or discharge. Furthermore, these $y$ values can be either at moderate or high rates from a cycling experiment or from low-rate periodic diagnostic tests. Figure \ref{fig:x_axis} illustrates how the same data plotted as a function of either cycle number or capacity throughput (\ref{fig:x_axis}a--\ref{fig:x_axis}b), and cycle number or time (\ref{fig:x_axis}c--\ref{fig:x_axis}d), can change the apparent severity of the knee.
Finally, we mention that resistance can also be used on the $y$ axis, but these curves have been referred to as ``resistance elbows'' instead of ``knees'' since the resistance \textit{increases} superlinearly.\cite{strange_elbows_2021}

While the knee point is mathematically well defined in the offline setting, the mathematical definition is difficult to apply in practice. For a continuous function, the knee point is mathematically defined as the maximum of its curvature, i.e., when the function deviates most from a straight line.
However, the curvature calculation requires an estimate of the second derivative.
Real-world battery aging datasets are discrete (e.g., capacity vs. cycle number is only measured at cardinal number values of cycle number), noisy (e.g., due to environmental temperature fluctuations in lab data or due to varying duty cycles and temperature fluctuations in field data), and sometimes infrequently sampled (e.g., datasets where capacity estimates come only from periodic diagnostic cycles). 
Numerical differentiation is challenging under these conditions due to noise amplification, and numerical second differentiation is even more challenging as the noise amplification challenge becomes overwhelming.
While many methods have been proposed to obtain less noisy numerical derivatives\cite{ahnert_numerical_2007, van_breugel_numerical_2020}{}, often involving smoothing or curve fitting, the numerical second derivative is highly sensitive to the method used for the numerical first derivative.
In summary, identifying a knee point by calculating the maximum curvature---the mathematically correct definition of the knee point---is difficult for real-world battery aging trajectories.

\begin{figure}[h!tb]
\centering
\includegraphics[scale=1]{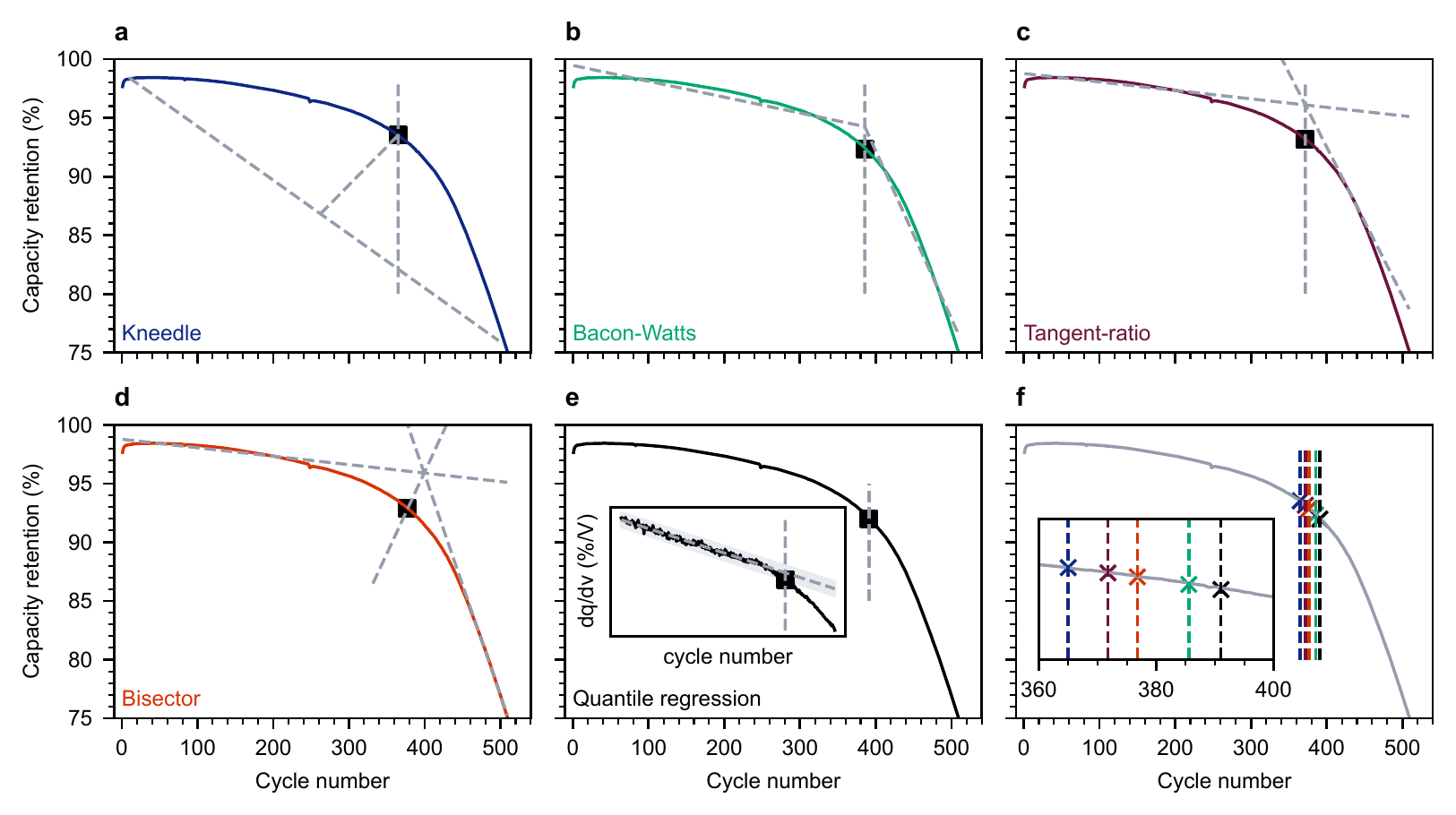}
\caption{Results of various knee identification methods proposed in the literature illustrated on the batch 2, channel 12 cell from the Severson et al.\cite{severson_data-driven_2019} dataset (arbitrarily selected). The capacity is normalized by the nominal capacity of the cell.
(a) Kneedle method, proposed by Satop{\"a}{\"a} et al.\cite{satopaa_finding_2011}
(b) Bacon-Watts method, proposed by Fermín-Cueto et al. \cite{fermin-cueto_identification_2020} for use in battery aging trajectories.
(c) Tangent-ratio method, proposed by Diao et al.\cite{diao_algorithm_2019}
(d) Bisector method, proposed by Greenbank and Howey.\cite{greenbank_automated_2021}
(e) Quantile regression method, proposed by Zhang et al.\cite{zhang_accelerated_2019}.
The inset illustrates how voltage data is used to determine the initial aging trajectory, and knees occur when this feature from voltage data falls below this initial trajectory.
(f) Comparison of knee points as identified via these five methods.}
\label{fig:knee_identification_methods}
\end{figure}

A few methods have been proposed to deal with the specific problem of offline knee point detection without requiring numerical differentiation. Some of these methods apply to knee point detection in any domain, while others are specific to knee point detection in battery aging trajectories. 
In Figure \ref{fig:knee_identification_methods}, we implemented and applied these methods to the capacity curve from a single cell in the Severson et al.\cite{severson_data-driven_2019} dataset.
The well-accepted ``Kneedle'' method (Figure \ref{fig:knee_identification_methods}a), proposed by Satop{\"a}{\"a} et al.\cite{satopaa_finding_2011}{}, calculates the knee as the maximum distance of the aging trajectory from a line drawn from beginning to end of life.
The most common approaches use the intersection of two lines fit to the beginning and end of the aging trajectory.
The ``Bacon-Watts'' method (Figure \ref{fig:knee_identification_methods}b), proposed by Fermín-Cueto et al. \cite{fermin-cueto_identification_2020} for use in battery aging trajectories, estimates the transition between two intersecting lines fitted to an aging trajectory; this method also provides an estimate of the ``knee-onset'', or the point where the aging trajectory is no longer linear.
The ``tangent-ratio'' method (Figure \ref{fig:knee_identification_methods}c), proposed by Diao et al.\cite{diao_algorithm_2019}{}, defines the knee based on a tangent ratio at the inflection and maximum slope points of the aging trajectory.
Similarly, the bisector method (Figure \ref{fig:knee_identification_methods}d), proposed by Greenbank and Howey\cite{greenbank_automated_2021}{}, combines linear extrapolations of early and late life with an angle bisector to identify the knee.
A final method was designed specifically for the battery use case: Zhang et al.\cite{zhang_accelerated_2019} proposed the quantile regression method (Figure \ref{fig:knee_identification_methods}e), which approximates early life with a linear regression and then defines the knee as when the aging trajectory falls below a defined band below that regression line.
Unlike other methods that use only the aging trajectory, this method requires the use of voltage data.
For each of these methods, smoothing the aging trajectories prior to knee identification may result in more accurate or consistent knee point detection\cite{strange_elbows_2021}{}, but again,
the results may be sensitive to the smoothing choices made.
We mention that Satop{\"a}{\"a} et al.\cite{satopaa_finding_2011} review additional knee point identification algorithms that were not considered here.

We compare the knee points as estimated by these five offline methods in Figure \ref{fig:knee_identification_methods}f.
All methods estimate the knee point at cycle numbers within a 26 cycle range (365--391 cycles). 
In general, the knee points estimated by these methods were highly correlated across most cells in the Severson et al.\cite{severson_data-driven_2019} dataset (Figure S1; quantile regression method not included).
The Kneedle and Bacon-Watts knee points were highly correlated with each other ($R^2\approx 1.00$); the bisector method also correlated well to these methods ($R^2\approx 0.96$), but the tangent-ratio method correlated fairly poorly to the others ($R^2\approx 0.86$).
These results suggest that the Kneedle, Bacon-Watts, and bisector methods are generally comparable for offline knee point estimation.
Of these five methods, the Kneedle, Bacon-Watts, and bisector methods are arguably also the easiest to implement, since all avoid the use of derivatives and voltage data.

Finding the knee ``online'' is difficult because the end-of-life capacity profile is not known and because the discharging conditions are often inconsistent.
Transforming the offline methods into online methods is challenging since many of these methods require the entire aging trajectory for knee point estimation (e.g., many require data after the knee to fit an intersecting line). However, the quantile regression method proposed by Zhang et al.\cite{zhang_accelerated_2019} can accommodate online knee point estimation since only the initial aging trajectory is required.
Of course, the challenge of uncontrolled usage conditions and thus higher noise observations\cite{aitio_predicting_2021} is inherent to online state estimation; varying duty cycles and varying environmental temperature in deployed systems could mask the knee.
In principle, knowledge of how similar cells knee under similar usage conditions may enable more accurate online knee point estimation.
These issues remain opportunities for future work.

\section{Pathways and internal state trajectories for knee points}

\begin{figure}[h!tb]
\centering
\includegraphics[scale=0.9]{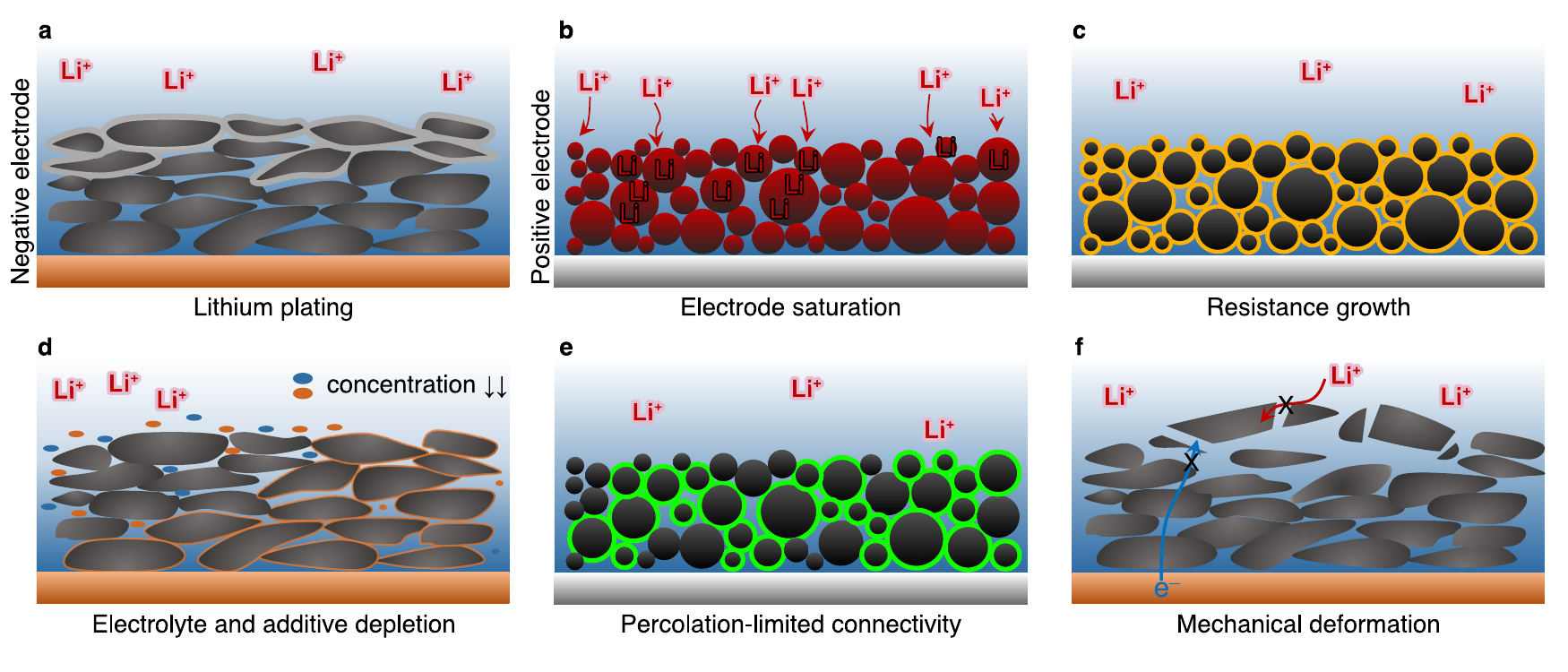}
\caption{Schematics of the six knee ``pathways'' identified in the literature. Each of these pathways may have multiple degradation modes (e.g., loss of active material), and each of these modes may have multiple degradation mechanisms (e.g., electrode delamination). This figure emphasizes particle- and electrode-level effects, although many of these mechanisms occur on the nano- and macroscales as well.
(a) Lithium plating, in which metallic lithium deposits on the surface of the negative electrode particles.
(b) Electrode saturation, in which the number of active sites in the electrode constrains the lithium inventory.
(c) Resistance growth, in which high overpotentials lead to a rapid drop in available capacity.
(d) Electrolyte depletion, in which the local depletion of electrolyte leads to loss of active material, and additive depletion, in which the depletion of a critical electrolyte additive triggers a knee.
(e) Percolation-limited connectivity, in which a small change in ionic or electronic electrode connectivity leads to a large change in electrode active material.
(f) Mechanical deformation, in which microscale, mesoscale, or macroscale mechanical effects trigger an increasing rate of active material loss.}
\label{fig:knee_pathways}
\end{figure}

In our review of the literature, we classified each proposed knee observation/hypothesis into both ``pathway'' and ``internal state trajectory'' categories.
First, we identified six knee pathways, or failure modes leading to knees grouped by the fundamentals of their degradation. These pathways are schematically illustrated in Figure \ref{fig:knee_pathways}. Some of these pathways (e.g., lithium plating) have been extensively characterized and modeled, while others (e.g., percolation-limited knees) are primarily hypotheses at this stage. Here, we critically examine the evidence for each pathway. For more extensively studied pathways such as lithium plating, we consider both \textit{modes}, defined as high-level, mechanism-agnostic changes in cell state, and \textit{mechanisms}, defined as the specific failure that leads to a change in cell state. For instance, loss of active material (LAM) is a degradation mode that can be caused by electrode delamination, one of several possible degradation mechanisms for this degradation mode. Degradation mechanisms are often challenging to pinpoint exactly or experimentally isolate, but degradation modes are usually identifiable through common electrochemical measurements or characterization methods and can help conceptually validate a proposed pathway.

\begin{figure}[htp]
    \centering
    \includegraphics[scale=1]{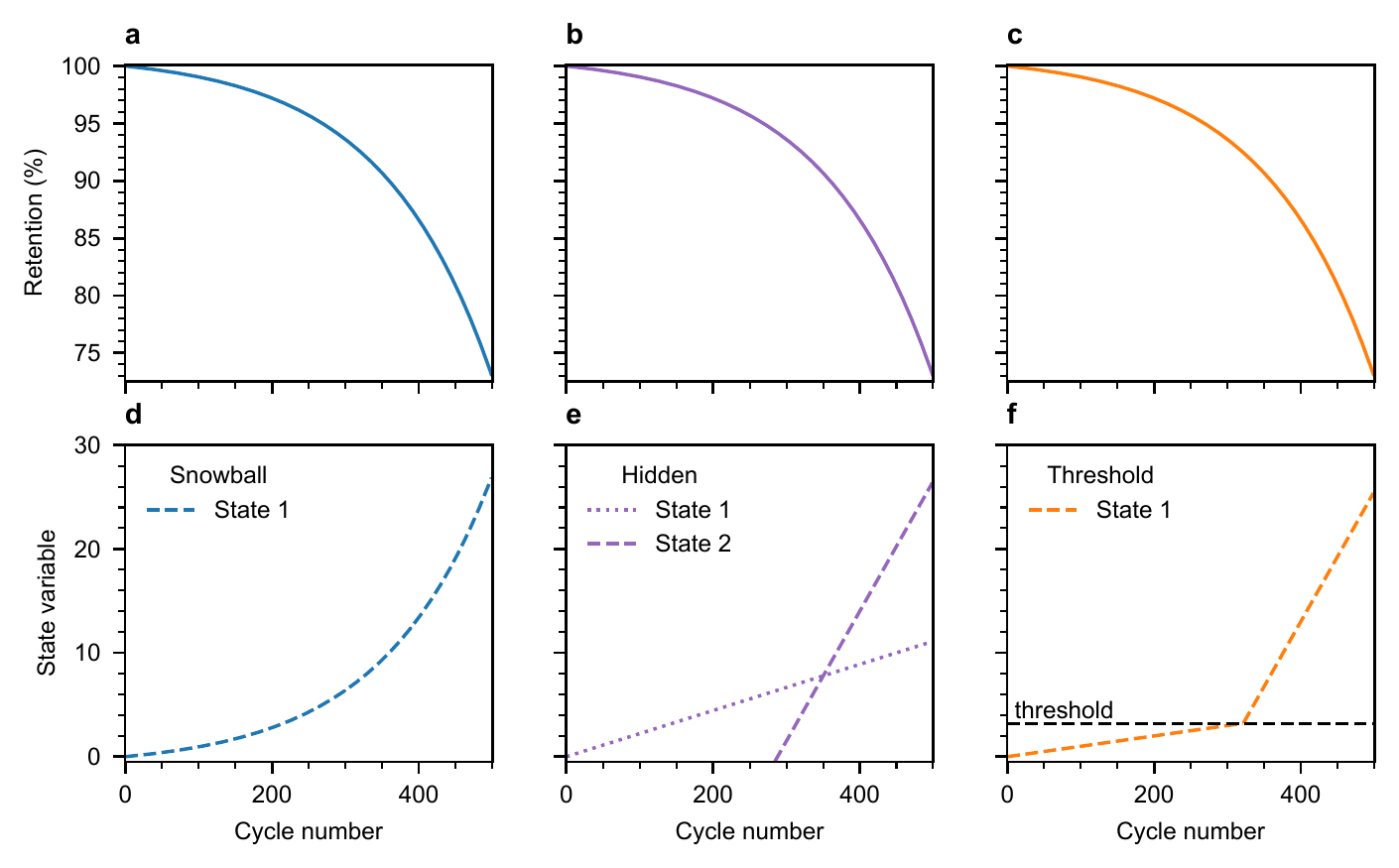}
    \caption{Schematic of the three types of ``internal state trajectories'' leading to a knee. In each case, the retention curve looks the same (a--c), but the underlying internal trajectories lead to knees via different mechanisms. (d) ``Snowball'' trajectory, in which the state variable is exponentially increasing. (e) ``Hidden'' trajectory, in which a slowly-increasing state variable (state 1) is dominated by a rapidly-increasing state variable (state 2). (f) ``Threshold'' trajectory, in which a dramatic change in observable state is triggered by a state variable reaching a threshold.
    The functional forms for the internal state variables for the hidden and threshold trajectories may be linear, sublinear, or superlinear.
    Note that the curves in panels a--c are illustrative and not explicitly derived from panels d--f.
    }
    \label{fig:snowball_vs_hidden_vs_threshold}
\end{figure}

Second, we considered the relationship between the directly observable state variables (i.e., capacity, energy, power, or resistance) and the trajectory of the internal states underlying their knees.
Figure \ref{fig:snowball_vs_hidden_vs_threshold} illustrates these three underlying ``internal state trajectories'' that can lead to a knee.
These internal states can be any internal variable---e.g., remaining active material in the negative electrode, charge-transfer kinetic parameters at the positive electrode, concentration of a specific additive remaining in the electrolyte---that dominates the observed variable.
\textit{Snowball} trajectories (Figure \ref{fig:snowball_vs_hidden_vs_threshold}a and \ref{fig:snowball_vs_hidden_vs_threshold}d) occur when the underlying state variable has a direct relationship with the observable state variable, but the underlying state variable is exponentially increasing.
Positive feedback between two degradation mechanisms is a special case of a snowball trajectory, a point we return to in our discussion of interactions and heterogeneity.
\textit{Hidden} trajectories (Figure \ref{fig:snowball_vs_hidden_vs_threshold}b and \ref{fig:snowball_vs_hidden_vs_threshold}e) occur when the observable state variable, originally controlled by a slowly-increasing state variable, becomes dominated by a second rapidly-increasing state variable.
Finally, \textit{threshold} trajectories (Figure \ref{fig:snowball_vs_hidden_vs_threshold}c and \ref{fig:snowball_vs_hidden_vs_threshold}f) occur when the observable state variable changes when the underlying state variable reaches a threshold. Each of these underlying internal state trajectories has unique implications for detectability and prediction, a point we return to throughout this work. Note that these classes of internal state trajectories cannot always be precisely distinguished; for instance, a hidden trajectory can sometimes be considered a threshold trajectory, i.e., the threshold can be considered the crossing point between two internal states.

Our ultimate goal is to provide a framework for modeling and predicting knees in a new cell chemistry/form factor.
Since we primarily take a physics-driven approach to understanding knees, we primarily focus on physics-driven and semi-empirical prediction approaches here.
Thus, measuring and extrapolating internal state trajectories for all relevant knee pathways is often the most straightforward way to predict knee onset for a given cell design and usage condition---especially in the absence of a suitable training set for data-driven approaches.
However, each class of internal state trajectories holds unique challenges for this prediction approach.
For instance, snowball trajectories require extrapolation of an exponential function, which is often an error-prone exercise and is exacerbated by noisy measurements.
Hidden trajectories require knowledge of the functional forms and simultaneous measurements for two internal state variables.
Lastly, threshold trajectories require knowledge of the functional form, the threshold, and measurements for one internal state variable.
In short, the requirements for each of these internal state trajectories are nontrivial.
Additionally, the difficulty of obtaining some of these required components can vary substantially, as we discuss throughout this work.

\begin{figure}[p]
    \centering
    \includegraphics[scale=0.85]{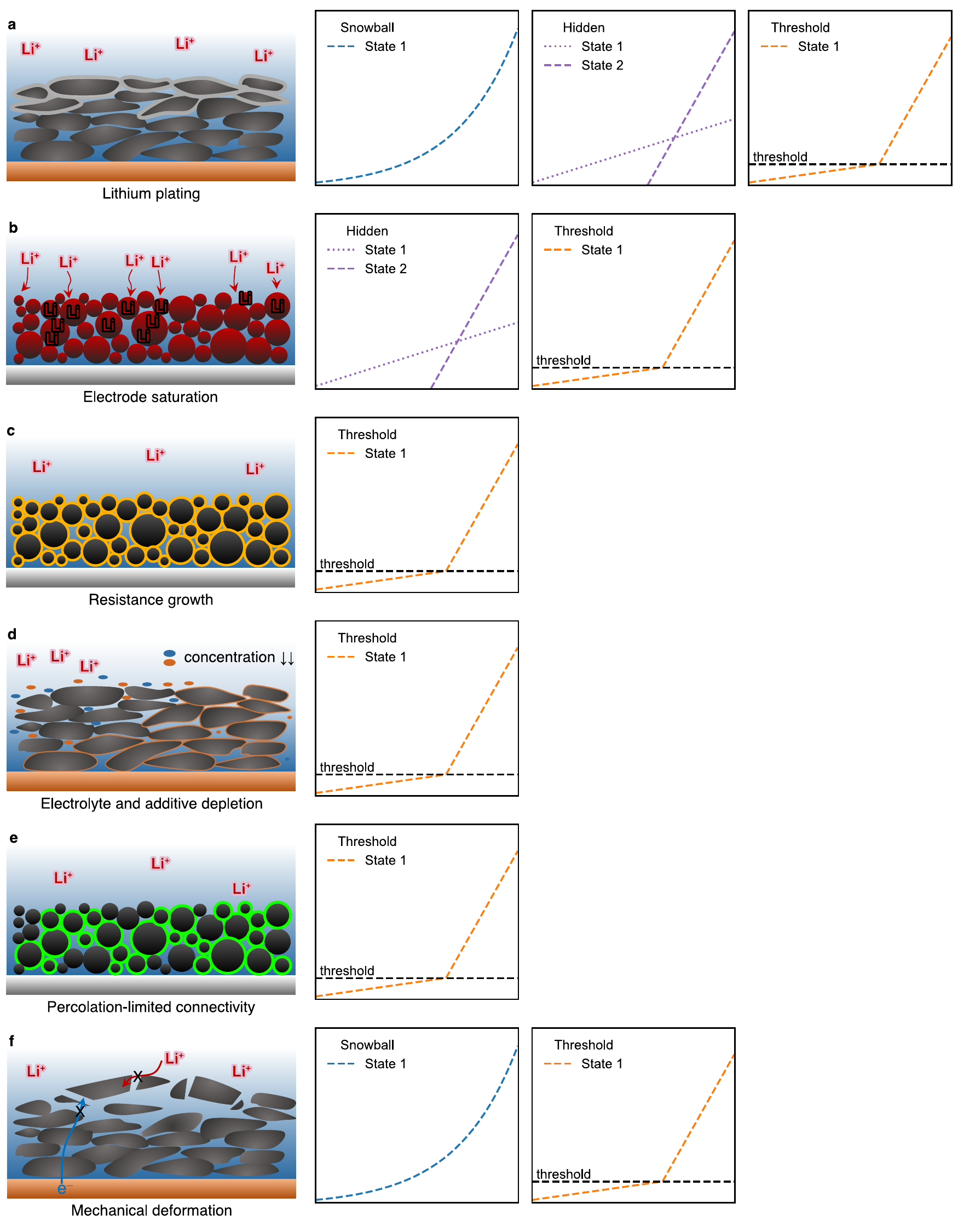}
    \caption{
    Connection between the six knee pathways (Figure \ref{fig:knee_pathways}) and the three internal state trajectories (Figure \ref{fig:snowball_vs_hidden_vs_threshold}).
    A knee pathway may have multiple internal state trajectories if multiple degradation mechanisms can occur for the same pathway.
    }
    \label{fig:summary}
\end{figure}

Figure \ref{fig:summary} displays the connection between the six knee pathways and the three internal state trajectories. Some knee pathways correspond to multiple internal state trajectories if multiple degradation mechanisms can occur for a pathway (e.g., lithium plating). In theory, similar modeling and prediction approaches can apply to pathways with the same internal state trajectories. In this section, we discuss the challenges and opportunities for modeling and prediction for each pathway and internal state trajectory.

\subsection{Lithium plating knees}

Lithium plating occurs when lithium ions form metallic lithium on the surface of the electrode rather than intercalating into it. The plating reaction is favorable when the reaction potential of Li/Li$^+$ is greater than the equilibrium potential for other alternative reaction pathways for Li$^+$ (i.e., graphite intercalation).\cite{gao_interplay_2021} Plating can be either ``rate-independent'', i.e.\ plating that occurs independent of the applied current (``overcharging''), or ``rate-dependent'', i.e.\ plating that only occurs if the applied current exceeds some value (``fast charging'').
Lithium plating can also occur in either fresh cells or aged cells.
Furthermore, lithium plating can occur either reversibly or irreversibly, depending on the ratio of lithium plated during charge that is recovered in the subsequent discharge.\cite{baure_synthetic_2019, dubarry_big_2020} In contrast to irreversible plating, reversible plating does not contribute to long-term degradation. Thus, we use ``plating'' to refer to irreversible plating throughout the remainder of this discussion.

Generally, lithium plating on graphite follows heterogeneous nucleation and growth kinetics, in which rapid growth proceeds quickly after an initial nucleation phase.\cite{ely_heterogeneous_2013, pei_nanoscale_2017, gao_interplay_2021}
Thus, lithium plating can often be considered a ``snowballing'' knee (Figure \ref{fig:snowball_vs_hidden_vs_threshold}a and \ref{fig:snowball_vs_hidden_vs_threshold}d). However, some lithium plating pathways (e.g., lithium plating driven by active material loss from the negative electrode, a hidden trajectory) leads to knees independent of the nucleation and growth of plated lithium.
In these cases, the nucleation and growth kinetics of lithium plating will only exacerbate these degradation mechanisms, and the degradation will be a combination of the hidden and snowball trajectories.

\begin{figure}[tbph]
    \centering
    \includegraphics[scale=0.9]{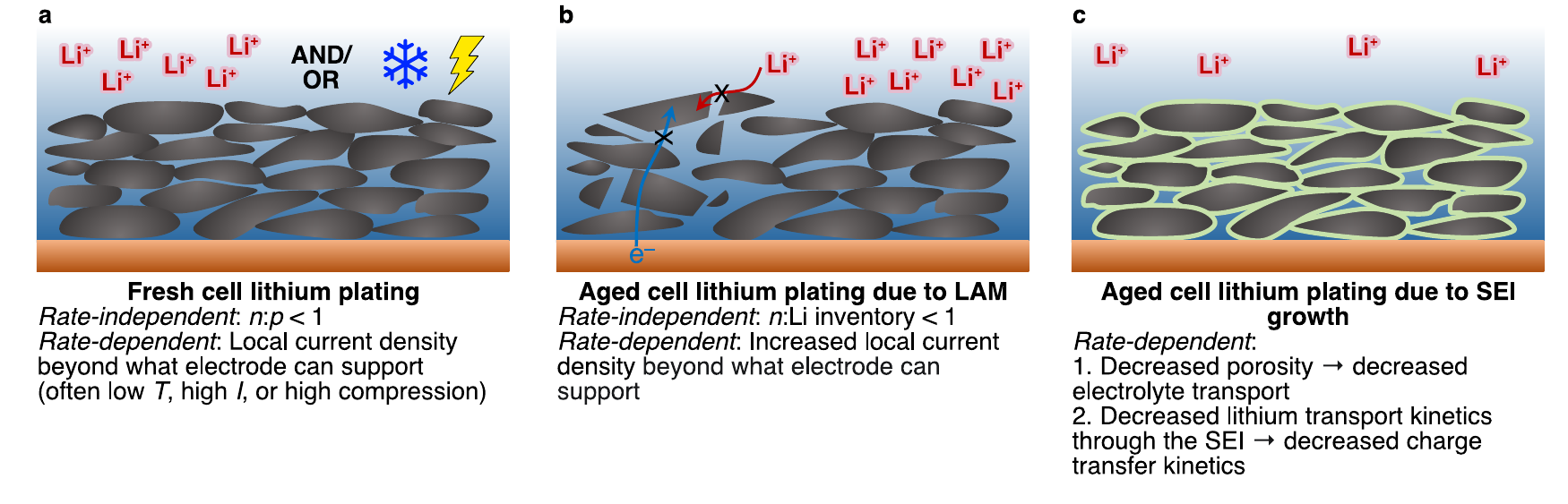}
    \caption{Sub-pathways for lithium plating knees.
    (a) Fresh cell lithium plating. Rate-independent lithium plating (``overcharging'') in fresh cells will occur if the cell has more positive electrode capacity than negative electrode capacity.
    Rate-dependent lithium plating (``fast charging'') in fresh cells will occur if the use case is too aggressive for what the cell can support; often, rate-dependent plating in fresh cells occurs due to low temperatures, high currents, or high compression.
    (b) Aged cell lithium plating due to loss of active material (LAM). LAM can lead to rate-independent lithium plating if the remaining negative electrode capacity falls below the remaining lithium inventory. LAM can lead to rate-dependent lithium plating if the local current density exceeds what the active negative electrode material is able to support.
    (c) Aged cell lithium plating due to SEI growth. SEI growth can lead to rate-dependent lithium plating by decreasing the porosity of the electrode, which will decrease electrolyte transport kinetics, or by decreasing the lithium transport kinetics through the SEI, which will decrease the charge transfer kinetics.
    Note that the nucleation and growth kinetics of lithium plating adds an additional snowball trajectory on top of the other trajectories associated with these sub-pathways.}
    \label{fig:plating_pathways}
\end{figure}

Historically, lithium plating has been considered to be a primary driver for capacity knees. Here, we discuss the mechanisms and sub-pathways by which plating can lead to a knee (Figure \ref{fig:plating_pathways}). We suggest Waldmann et al.\cite{waldmann_li_2018} and Gao et al.\cite{gao_interplay_2021} for comprehensive general overviews of lithium plating.

\subsubsection{Rate-independent lithium plating}


Rate-independent lithium plating occurs whenever the lithium capacity during charging exceeds the negative electrode capacity, i.e., the negative electrode is unable to host all lithium coming from the positive electrode. Generally, the latter can be avoided in fresh cells by simply using a negative electrode to positive electrode capacity ratio ($n$:$p$ ratio) greater than 1 (Figure \ref{fig:plating_pathways}a). However, if active material from the negative electrode is lost during aging, rate-independent lithium plating will occur even in cells with excess negative electrode capacity (Figure \ref{fig:plating_pathways}b).

\paragraph{Rate-independent lithium plating in fresh cells}

While rate-independent lithium plating in fresh cells can be easily avoided by proper cell design (Figure \ref{fig:plating_pathways}a), this degradation mechanism is often exploited for scientific studies of lithium plating.
For instance, Deichmann et al.\cite{deichmann_investigating_2020} created cells with $n$:$p$ ratios of 0.75 and 0.5 to intentionally deposit lithium metal on graphite electrodes. The authors identified a relationship between decreased $n$:$p$ and capacity fade, which they attributed to high loss of lithium inventory using differential capacity analysis and scanning electron microscopy. In a creative study, Martin et al.\cite{martin_cycling_2020} used deposited lithium metal as a mechanism to store extra capacity, enabling the cell to occasionally discharge extra energy (i.e., when extra range is needed) without requiring a substantially larger negative electrode. This cell design used an $n$:$p$ ratio of 0.6, where $n$:$p$ is calculated using the lithium capacity of the conventional graphite. A high upper cutoff voltage during charging was used to intentionally plate lithium onto graphite; unsurprisingly, irreversible lithium plating was found to be the primary failure mechanism, with over 50\% capacity loss in two of the three electrolytes tested (although the cells did not exhibit knees). Rate-independent lithium plating in fresh cells is trivial to model and predict if the cell design is known; if rate-independent plating is expected in fresh cells, a snowballing lithium plating knee is likely to occur early in life.

\paragraph{Rate-independent lithium plating due to loss of active material}

\begin{figure}[p]
    \centering
    \includegraphics[scale=1]{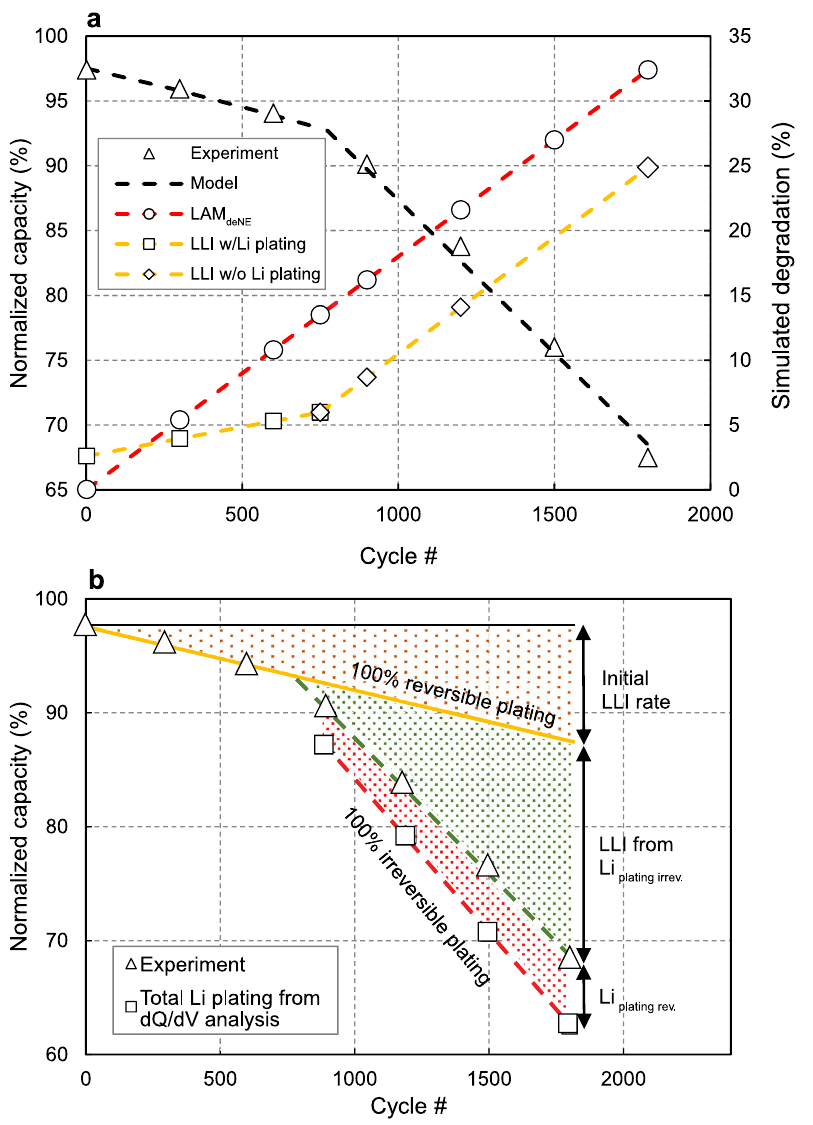}
    \caption{Rate-independent lithium plating driven by loss of active negative electrode material. (a) Evolution of aging parameters with cycling of cell degradation. The left axis shows the experimental normalized cell capacity at C/25 (triangle markers) from reference performance tests occurring throughout cycling, and the dashed black line shows the results of cell capacity simulations with the calculated aging modes at C/25. The right axis shows the evolution of the degradation induced by the calculated aging modes (markers and dashed lines) with cycling. Note that $\mathrm{LAM_{deNE}}$ increases linearly, and at a rate higher than LLI. At cycle 750, the negative electrode becomes the capacity-limiting electrode during charge, at which point plating begins.
    (b) Capacity vs. cycle number at C/25, depicting the contributions to the total capacity fade as a function of cycle number. The yellow region represents LLI from non-plating sources (i.e., SEI growth), the green region represents LLI from irreversible lithium plating, and the red region represents reversible plating estimated from incremental capacity analysis (the reversible plating does not contribute to the capacity fade). The total capacity fade, represented by triangles (same as above panel), comes from the sum of LLI from SEI growth and LLI from irreversible plating.
    Reproduced with permission from Figures 7 and 8 of Anse\'an et al.\cite{ansean_operando_2017} Copyright 2017, Elsevier.} 
    \label{fig:thermo_plating}
\end{figure}

Loss of active material --- specifically, loss of active material from the delithiated negative electrode ($\mathrm{LAM_{deNE}}$) --- during aging may result in rate-independent lithium plating if the lithium capacity of the negative electrode becomes limiting during charging (Figure \ref{fig:plating_pathways}b). For instance, if the rate of $\mathrm{LAM_{deNE}}$ exceeds that of the loss of lithium inventory (LLI), the negative electrode will eventually be unable to accommodate all lithium during charging, which will lead to rate-independent lithium plating and thus a knee. Dubarry and colleagues\cite{ansean_operando_2017, dubarry_durability_2018, baure_synthetic_2019, dubarry_big_2020} have extensively explored this scenario by considering both different ratios of $\mathrm{LAM_{deNE}}$ to LLI and different extents of reversible and irreversible plating (Figure \ref{fig:thermo_plating}). This scenario is a prototypical case of a hidden state (i.e., loss of active negative electrode material) causing a knee: because the negative electrode is typically oversized relative to the positive electrode, active material loss from the negative electrode is hidden from the measured capacity until the negative electrode capacity falls below the positive electrode capacity. We discuss this effect more generally in our discussion of the electrode saturation pathway. Fortunately, the onset of rate-independent lithium plating due to $\mathrm{LAM_{deNE}}$ can often be modeled and predicted via differential capacity analysis\cite{ansean_operando_2017, dubarry_durability_2018, baure_synthetic_2019, dubarry_big_2020} to identify the rates of LLI and $\mathrm{LAM_{deNE}}$. Note that differential capacity analysis generally requires periodic low-rate cycling interspersed within the cycling test.
Lastly, we note that the high predictive performance of features sourced from voltage curves over the discharge capacity curves in Severson et al.\cite{severson_data-driven_2019} was largely attributed to this sub-pathway.

Various degradation mechanisms can lead to $\mathrm{LAM_{deNE}}$, which occurs when active sites lose either ionic or electronic connectivity with the electrode.
Often, several of these mechanisms can occur in parallel, leading to a snowball effect where loss of active material due to one mechanism may result in further stress on the remaining active sites, accelerating degradation via rate-independent lithium plating.
Electrode sites can lose electronic connection via delamination\cite{liu_aging_2010, cannarella_stress_2014, somerville_effect_2016, willenberg_high-precision_2020}, particularly for cells with low external pressure\cite{cannarella_stress_2014}. Particle cracking is another mechanism for electronic disconnection of active sites, although graphite particles are not expected to crack appreciably\cite{takahashi_examination_2015}{}.
Electrode sites can lose ionic connection via electrolyte dry-out, which may be driven by gas generation during cycling\cite{mao_calendar_2017, kupper_end--life_2018}{}, or the growth of microns-thick ``covering layers'', an effect that we mention now but explore in depth in our discussion on mechanical deformation knees.
While the exact mechanisms leading to $\mathrm{LAM_{deNE}}$ are challenging to pinpoint exactly without extensive destructive analysis, LAM can often be identified via differential capacity analysis.\cite{ansean_operando_2017, dubarry_durability_2018, baure_synthetic_2019, dubarry_big_2020}

\subsubsection{Rate-dependent lithium plating}

``Rate-dependent'' lithium plating occurs when excessive transport or reaction overpotentials cause the local electrode potential to drop below that of Li/Li$\mathrm{^+}$.
In other words, rate-dependent lithium plating occurs at conditions when the plating could otherwise be mitigated by lithiating the graphite at a sufficiently low current.
While rate-dependent lithium plating has the same criterion as rate-independent lithium plating (i.e., the local potential falls below that of Li/Li$^+$), the dynamic nature of this process introduces additional avenues for lithium plating knees to occur.

As Gao et al.\cite{gao_interplay_2021} describe, salt depletion in the electrolyte, poor charge transfer kinetics, and surface crowding in the negative electrode particles at the graphite surface further favor lithium plating over intercalation.
These three effects mirror the transport of lithium from the electrolyte to the negative electrode (electrolyte transport, charge transfer from the electrolyte to the negative electrode particles, and solid-state transport within the negative electrode particles).
While solid-state transport within the negative electrode particles is generally not expected to degrade with aging,
both electrolyte transport and charge transfer from the electrolyte to the negative electrode particles can degrade with aging due to SEI growth (Figure \ref{fig:plating_pathways}c).

\paragraph{Rate-dependent lithium plating in fresh cells}
Rate-dependent lithium plating can be driven by a wide range of cell designs and usage conditions; the prototypical use case leading to lithium plating is high charging rates at low temperature \cite{waldmann_temperature_2014, petzl_lithium_2015} (Figure \ref{fig:plating_pathways}a). Waldmann et al.\cite{waldmann_temperature_2014} observed an increase in the rate of aging with a decrease in temperature below 25$^{\circ}$C, attributing the increased aging rate to lithium plating via dissections. Low temperatures increase both the transport overpotentials for lithium ions within the electrolyte and electrode and the reaction overpotential for lithium intercalation. Note that ``high'' charging rate or ``low'' temperature do not have consistent quantitative definitions, as plating will occur whenever the local potential exceeds the energy barrier for lithium nucleation. Thus, plating may be observed even at ``standard'' test conditions, such as 1C constant-current charging near room temperature \cite{waldmann_optimization_2015,burns_-situ_2015}. Increasing temperature and more rate-capable cell designs (i.e., thinner electrodes) may allow for more rapid charging before lithium plating and these knees are observed\cite{yang_understanding_2018, coron_impact_2020}; Lewerenz et al.\cite{lewerenz_systematic_2017} cycled cells at rates up to 8C, observing no knees at rates as high as 4C, though microstructural evidence of plating was found even at 1C. The onset of lithium plating is also sensitive to the charging protocol, with many studies demonstrating that informed design of charging protocols can substantially extend cell lifetime by preventing lithium plating.\cite{waldmann_optimization_2015,schindler_fast_2018, attia_closed-loop_2020}. Optimizing electrode architectures to improve electrolyte transport kinetics is a further path forward to increase charging rates without lithium plating.\cite{nemani_design_2015, usseglio-viretta_enabling_2020}
Careful electrochemical modeling can enable estimates of minimum negative electrode potential (and thus the plating risk) as a function of cell design and charging protocol\cite{yang_understanding_2018}{}; if accurate modeling suggests rate-dependent plating is expected in fresh cells, a snowballing lithium plating knee is likely to occur early in life.

Mechanical stress may also lead to rate-dependent lithium plating, as applied stress can compress the electrode or separator. This compression decreases the local porosity of the electrodes, which decreases the apparent diffusivity of electrolyte, increases the local polarization, and thus can cause lithium plating. Cannarella and Arnold\cite{cannarella_stress_2014} conducted a direct test of this mechanism, finding that high external pressures can induce lithium plating in pouch cells and lead to a knee. In a follow-up experiment, Liu and Arnold\cite{liu_effects_2020} demonstrated that localized lithium plating could be induced in densified regions of the separator. Bach et al.\cite{bach_nonlinear_2016} applied a hose clamp around the circumference of an 18650 cylindrical cell, and a post-test teardown clearly showed lithium plating localized to the regions of the electrodes that were under compressive stress. From this test, the authors concluded that internal gradients in pressure induced by the current-collecting tab can also cause lithium plating. Coin cells and pouch cells are also sensitive to localized external pressures.\cite{liu_size_2018, fuchs_post-mortem_2019, okasinski_situ_2020}
Rate-dependent lithium plating induced by mechanical gradients can be considered a threshold trajectory, as the lithium plating begins once the negative electrode porosity falls below a critical porosity (again, lithium plating itself can be considered a snowball trajectory).
This effect may be challenging to model and predict without a detailed understanding of the heterogeneous porosity distributions in the electrodes and separator, which is difficult to experimentally measure.

\paragraph{Rate-dependent lithium plating due to loss of active material}
As previously discussed, a hidden trajectory for rate-independent lithium plating is loss of delithiated negative electrode active material (Figure \ref{fig:plating_pathways}b).\cite{ansean_operando_2017, dubarry_durability_2018, baure_synthetic_2019, dubarry_big_2020} Mechanisms for loss of active material include delamination, particle cracking, electrolyte dry-out, and covering layer growth. However, $\mathrm{LAM_{deNE}}$ can also drive rate-dependent lithium plating, even if the negative electrode capacity does not limit the charging capacity. Active material loss without a corresponding loss in lithium flux will lead to an increased local current density on the negative electrode surface; these high local current densities can drive higher overpotentials and thus lithium plating.
Similarly to rate-independent lithium plating due to $\mathrm{LAM_{deNE}}$, rate-dependent lithium plating due to $\mathrm{LAM_{deNE}}$ can be modeled and predicted via differential capacity analysis;
however, a major complication is estimating if the increased local current density from rate-independent estimates of $\mathrm{LAM_{deNE}}$ is high enough to drive lithium plating.
Overall, rate-dependent lithium plating due to $\mathrm{LAM_{deNE}}$ can be considered a threshold trajectory, where the internal state is the minimum negative electrode potential and the threshold is the local plating potential.

Continuous active material loss will create increasingly larger local current densities, which will drive increasingly larger lithium plating potentials. Thus, in cell design/use case regimes where rate-dependent lithium plating is expected, linearly increasing active material loss can cause accelerating rates of lithium plating.
Furthermore, as previously discussed, the nucleation and growth kinetics of lithium plating adds an additional snowball trajectory, since additional growth of initially nucleated phases can occur rapidly. This ``double-snowball'' effect is especially pernicious and is expected to lead to sharp knees. To our knowledge, prior experimental or modeling work has not considered this effect. Overall, this effect highlights the high sensitivity of rate-dependent lithium plating to active material loss of the negative electrode.

\begin{figure}[hp]
    \centering
    \includegraphics[scale=1]{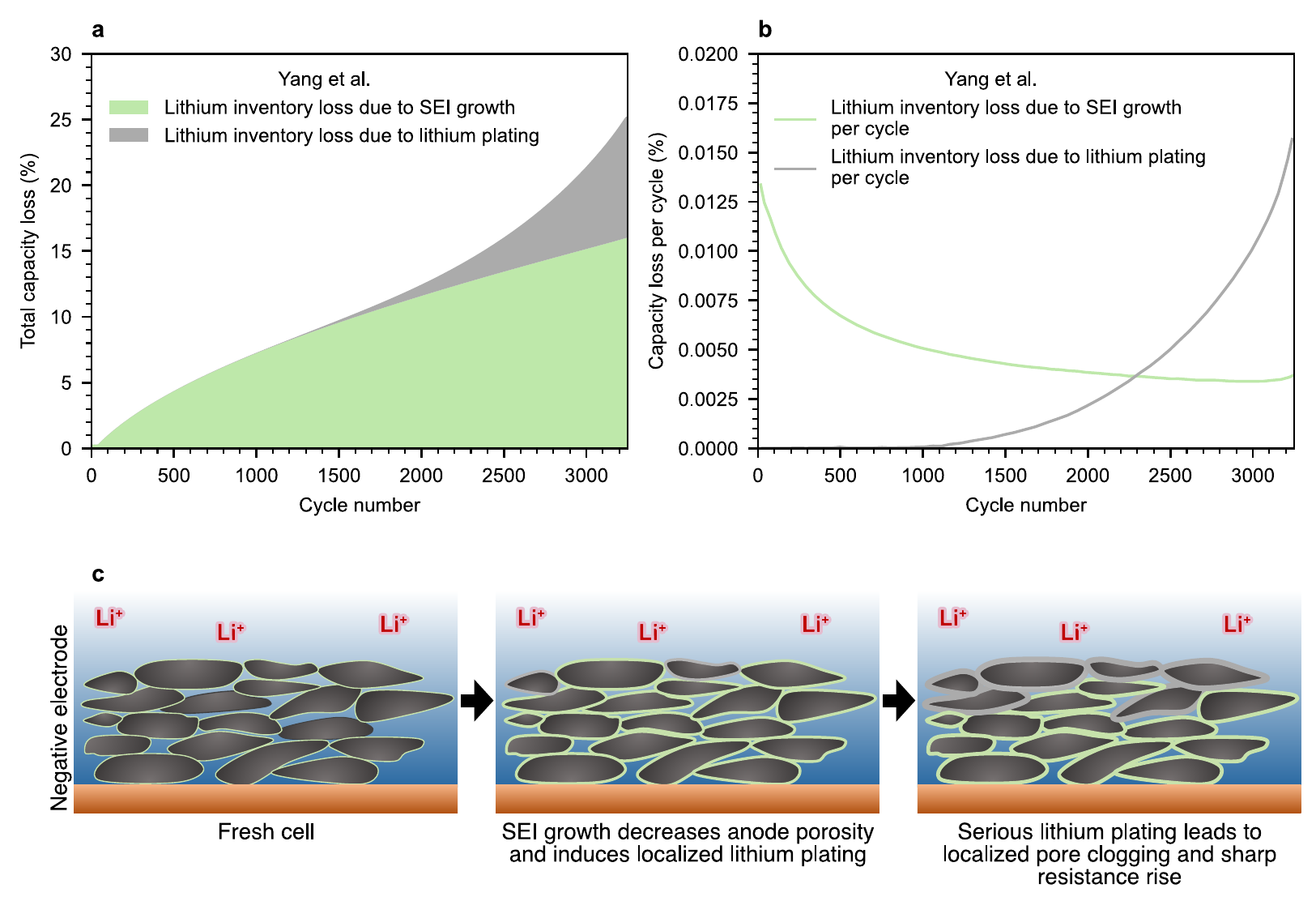}
    \caption{Rate-dependent lithium plating due to pore clogging, as modeled by Yang et al.\cite{yang_modeling_2017} (a) Lithium inventory loss contributed by SEI growth and lithium plating, respectively. (b) Lithium inventory loss per cycle contributed by SEI growth and lithium plating, respectively. The ``snowballing'' growth of lithium plating occurs due to the accelerating decrease in porosity, which creates high transport overpotentials in the electrolyte and drives additional lithium plating. (c) Schematic illustration of pore clogging driven initially by SEI growth and then by plating. Adapted from Figure 8 of Yang et al.\cite{yang_modeling_2017}}
    \label{fig:pore_clogging}
\end{figure}

\paragraph{Rate-dependent lithium plating due to pore clogging}
As SEI grows, it precipitates mainly in the pores of the negative electrode, decreasing the available volume fraction for electrolyte in the electrode \cite{sikha_effect_2004}
This decreased volume fraction increases the electrolyte transport overpotentials, which can ultimately lead to lithium plating.
The plated lithium, which has a much lower density than intercalated lithium, further decreases the porous volume fraction, creating a positive feedback loop for additional lithium plating\cite{yang_modeling_2017} (Figure \ref{fig:plating_pathways}c).
Thus, this effect can be considered a threshold trajectory, in which a knee is triggered when the porosity decreases below some critical porosity, after which plating begins.
A few works have modeled this phenomenon\cite{yang_modeling_2017, muller_model-based_2019, keil_electrochemical_2020, atalay_theory_2020}{}, namely Yang et al.\cite{yang_modeling_2017} (Figure \ref{fig:pore_clogging}).
While this mechanism has not been experimentally validated, decreasing negative electrode porosity with cycling has been observed via X-ray computed tomography\cite{frisco_understanding_2016, rahe_nanoscale_2019} and electrochemical impedance spectroscopy\cite{klett_uneven_2015, petzl_lithium_2015, sarasketa-zabala_understanding_2015, sarasketa-zabala_cycle_2015}{}.
Furthermore, the ``covering layer'' effect discussed at a later point may be related to this phenomenon, as models of lithium plating induced by pore clogging suggest that the pore clogging occurs primarily at the separator-electrode interface.\cite{yang_modeling_2017}
One proposed countermeasure for rate-dependent lithium plating due to pore clogging is to use a graded or stepped porosity profile through the thickness of the negative electrode; since most pore clogging occurs near the separator, having a higher porosity near the separator and a lower porosity near the current collector can slow the onset of the knee caused by pore clogging \cite{muller_model-based_2019}.
We note that measuring local porosity distributions in a commercially relevant form factor is challenging and generally requires extensive \textit{ex-situ} characterization. Furthermore, identifying the critical porosity at which plating starts is nontrivial and requires accurate electrochemical modeling of the porous electrode.

\paragraph{Rate-dependent lithium plating due to decreased charge-transfer kinetics}
SEI growth can also decrease the charge-transfer kinetics of the negative electrode particles, since the increased thickness of the SEI poses an additional barrier for lithium-ion intercalation.\cite{xu_differentiating_2010, pinson_theory_2013, li_li-desolvation_2017}
The chemistry and morphology of the additional SEI thickness likely influences its impact on the charge-transfer kinetics.\cite{lu_lithium_2011, xu_electrolytes_2014}
Similarly to the porosity mechanisms previously discussed, the charge-transfer kinetics could decrease to the point that the negative electrode can no longer accommodate the increased lithium flux.
A few authors have proposed this mechanism\cite{klett_non-uniform_2014, ecker_calendar_2014}{}; Schuster et al.\cite{schuster_nonlinear_2015} studied this mechanism in depth via electrochemical impedance spectroscopy and post-mortem analysis.
An open question is if (or under what conditions) SEI growth will cause lithium plating via decreased porosity or decreased charge-transfer resistance first.

Again, this effect can be considered a threshold trajectory, in which a knee is triggered when the charge-transfer kinetics decrease below some critical threshold, after which plating begins.
Tracking the charge-transfer kinetics over life can, in principle, be performed via electrochemical impedance spectroscopy of either the full cell or half cells harvested from the full cell\cite{klett_non-uniform_2014, ecker_calendar_2014, schuster_nonlinear_2015}{}, although interpreting impedance spectra can be challenging.
Furthermore, just as in the case of porosity decrease, identifying the critical charge-transfer kinetic parameters at which plating starts is nontrivial and requires accurate electrochemical modeling of the porous electrode. 

\subsection{Electrode saturation}

\begin{figure}[hp]
\centering
\includegraphics[scale = 1]{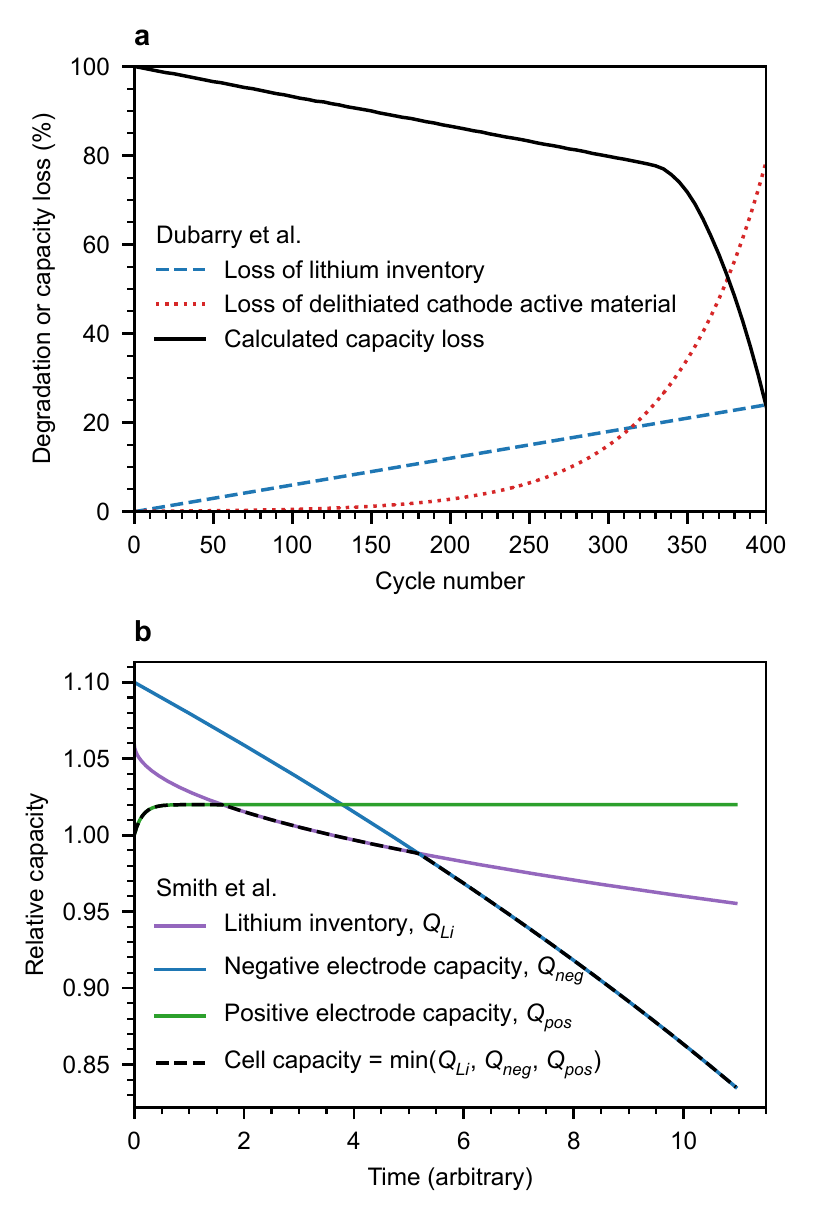}
\caption{Early models of ``hidden'' knee mechanisms due to electrode saturation. (a) The exponentially increasing positive electrode loss eventually limits the capacity and causes a knee. Adapted from Figure 17 of Dubarry et al.\cite{dubarry_synthesize_2012}
(b) The linearly decreasing negative electrode capacity eventually overtakes the sublinearly decreasing lithium inventory, causing a knee in the relative capacity. Adapted from Figure 1 of Smith et al.\cite{smith_life_2017}
}
\label{fig:electrode_sat_simple}
\end{figure}

As previously discussed, lithium plating can occur if the active sites in the negative electrode cannot accommodate the available lithium inventory, driving the local surface potential to potentials at which lithium metal deposition is favorable. More generally, if the rate of active material loss for one electrode outpaces the rate of lithium inventory loss, the electrode can ``saturate'' and reach the cutoff potential well before all lithium has transferred. If this electrode is not limiting, its loss of active material will be hidden from the overall capacity until this electrode becomes limiting; furthermore, if the rate of active material loss is higher than the initial rate of lithium inventory loss, a knee in capacity will manifest. Dubarry et al.\cite{dubarry_synthesize_2012} and Smith et al.\cite{smith_life_2017} captured this knee pathway by using a functional form for active material loss that increases more rapidly than that of lithium inventory loss (Figure \ref{fig:electrode_sat_simple}).
This pathway can apply for either electrode, but loss of active material from the negative electrode is more likely to be a hidden trajectory since this electrode is typically oversized relative to the positive electrode; the exception is cells with lithium titanate electrodes, in which the positive electrode is limiting and can cause a ``hidden'' knee\cite{baure_battery_2019, baure_battery_2020}.

\begin{figure}[p]
    \centering
    \includegraphics[width=\linewidth]{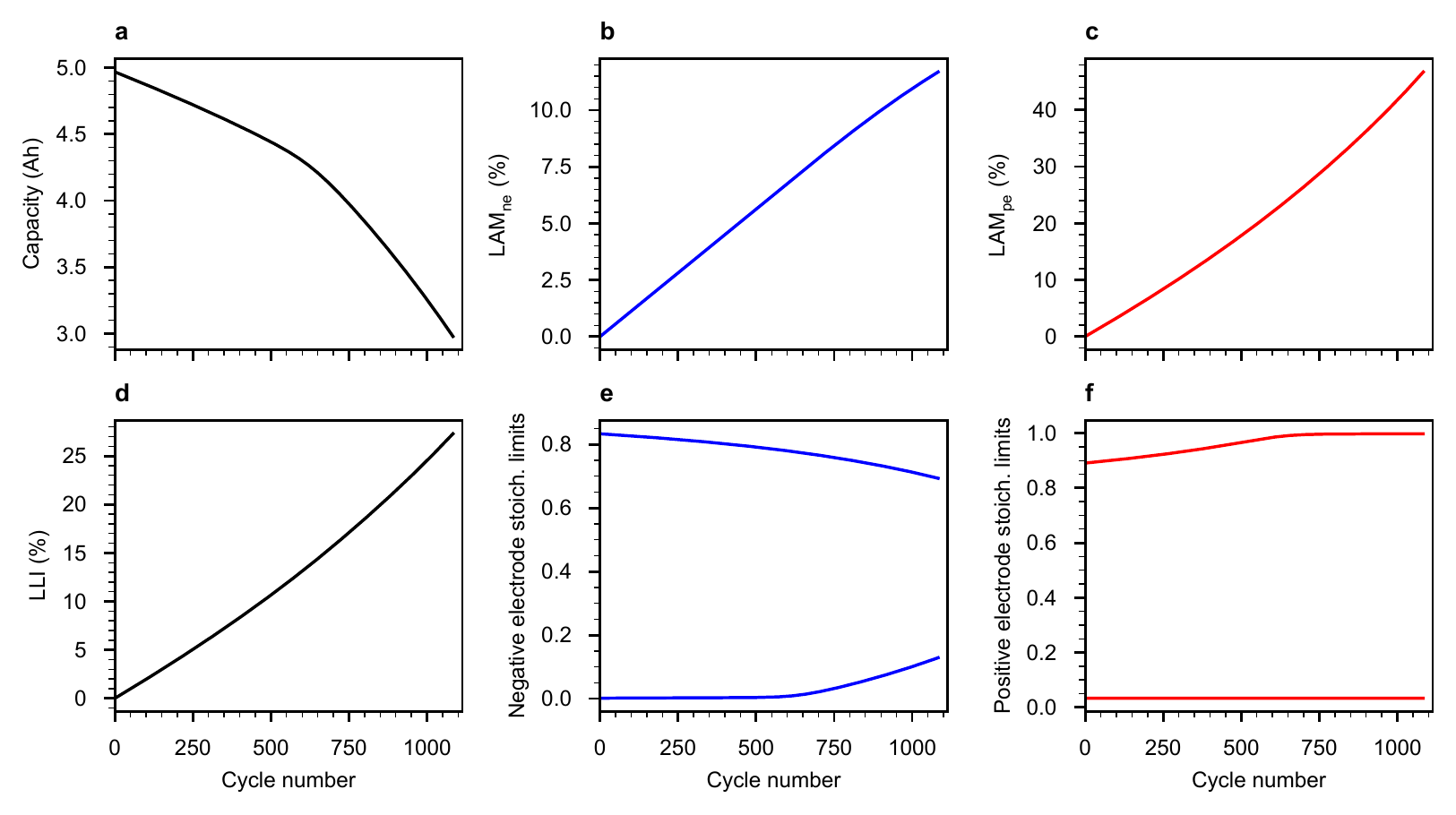}
    \caption{Simulation showing a knee point due to positive electrode saturation, from Sulzer et al.\cite{sulzer_accelerated_2021} Initially, all variables (LLI, $\mathrm{LAM_{ne}}$, and $\mathrm{LAM_{pe}}$) increase linearly with cycle number. Around cycle 600, the maximum stoichiometry in the positive electrode reaches unity, i.e., the electrode saturates. Electrode saturation causes a knee in the capacity curve despite nearly linear rates of LAM and LLI.
    Note that $\mathrm{LAM_{ne}}$ and $\mathrm{LAM_{pe}}$ do not directly correspond to LAM of lithiated or delithiated electrode sites; see Sulzer et al.\cite{sulzer_accelerated_2021} for more information.}
    \label{fig:electrode_sat_simulation}
\end{figure}

A richer picture emerges in models that capture the shifts in electrode stoichiometry with cycling. Lin et al.~\cite{lin_comprehensive_2013} and Kindermann et al.~\cite{kindermann_sei_2017} modeled loss of lithium inventory driven by SEI growth and loss of active material driven by mechanical effects in the positive electrode. Sulzer et al.\cite{sulzer_accelerated_2021} replicated a similar mechanism in Figure \ref{fig:electrode_sat_simulation} by simulating continuous constant-current discharge and constant-current, constant-voltage charge of a single particle model with SEI formation \cite{yang_modeling_2017} and loss of active material \cite{reniers_review_2019, laresgoiti_modeling_2015} due to particle swelling \cite{ai_electrochemical_2020, mohtat_differential_2020}.
When the aging parameters (SEI reaction rate and particle cracking rate) are chosen so that loss of active material in the positive electrode occurs at a faster rate than loss of lithium inventory, the stoichiometric window of the positive electrode widens (Figure \ref{fig:electrode_sat_simulation}f), which increases the cell voltage for a given amount of transferred lithium. This effect decreases the capacity between the voltage limits. The knee occurs when the positive electrode becomes fully saturated before the entire lithium inventory is transferred (around cycle 600 in Figure \ref{fig:electrode_sat_simulation}), despite the underlying rate of degradation remaining linear (Figure \ref{fig:electrode_sat_simulation}b-d).

Electrode saturation can also be rate dependent, sometimes in counterintuitive ways. Ma et al.\cite{ma_editors_2019} found that single-crystal nickel manganese cobalt oxide (NMC)/graphite cells exhibited no capacity fade in 1C diagnostic cycles but exhibited capacity fade in C/20 diagnostic cycles. The authors attributed this result to the poor rate capability of the single-crystal NMC particles. At low rates, the cells are ``negative electrode limited''; as lithium inventory loss shifts the negative electrode voltage curve, the available discharge capacity decreases and thus capacity loss is observed. At high rates, the cells are ``positive electrode limited'' because the positive electrode saturates before the negative electrode fully depletes; thus, the 1C capacities are unaffected. We refer the reader to Ma et al.\cite{ma_editors_2019} for further discussion of this phenomenon.

Overall, electrode saturation can be modeled and predicted using electrochemical modeling. This pathway can be considered either a threshold trajectory, where the knee is triggered by electrode saturation, or a hidden trajectory, where LAM of one electrode outpaces both LAM of the other electrode and LLI. While modeling of just the degradation modes (i.e., LLI, LAM, etc) can capture the key dynamics of this pathway, models that capture the shifts in stoichiometry as a function of cycling can capture more subtle effects. As Ma et al.\cite{ma_editors_2019} demonstrate, periodic diagnostic cycles at multiple rates can aid in identifying electrode saturation, especially if the saturation is rate-dependent.

\subsection{Resistance growth-induced knees}

Cell internal resistance often increases during aging, in part due to the growth of side reaction products on the surface of the electrode particles, particularly on the positive electrode\cite{ma_editors_2019}. Under constant current conditions, the additional overpotential from increased internal resistance will cause the cell to reach the cutoff voltage more quickly, decreasing the capacity, energy, and power per cycle. The magnitude of this overpotential growth rate is a product of both the resistance growth rate (i.e., electrolyte reaction rate) and the applied current.

Most modern lithium-ion batteries have voltage-capacity curves that are relatively flat at higher voltages/states of charge (SOCs) (i.e., small d$V$/d$Q$) and relatively steep at lower voltages/SOCs (i.e., large d$V$/d$Q$). Thus, when cycling at appreciable rates, the constant-current portion of charge capacity is highly sensitive to small increases in resistance growth. In contrast, the discharge capacity is less sensitive to resistance growth---until the overpotential is large enough such that the discharge ends in the flatter region of the voltage-capacity curve (i.e., a small increase in overpotential leads to a large decrease in capacity). When this flatter region is reached, the discharge capacity becomes more sensitive to small changes in overpotential (again, when cycling at appreciable rates), leading to a knee in capacity, energy, and/or power.

\begin{figure}[p]
\centering
\includegraphics[scale = 1]{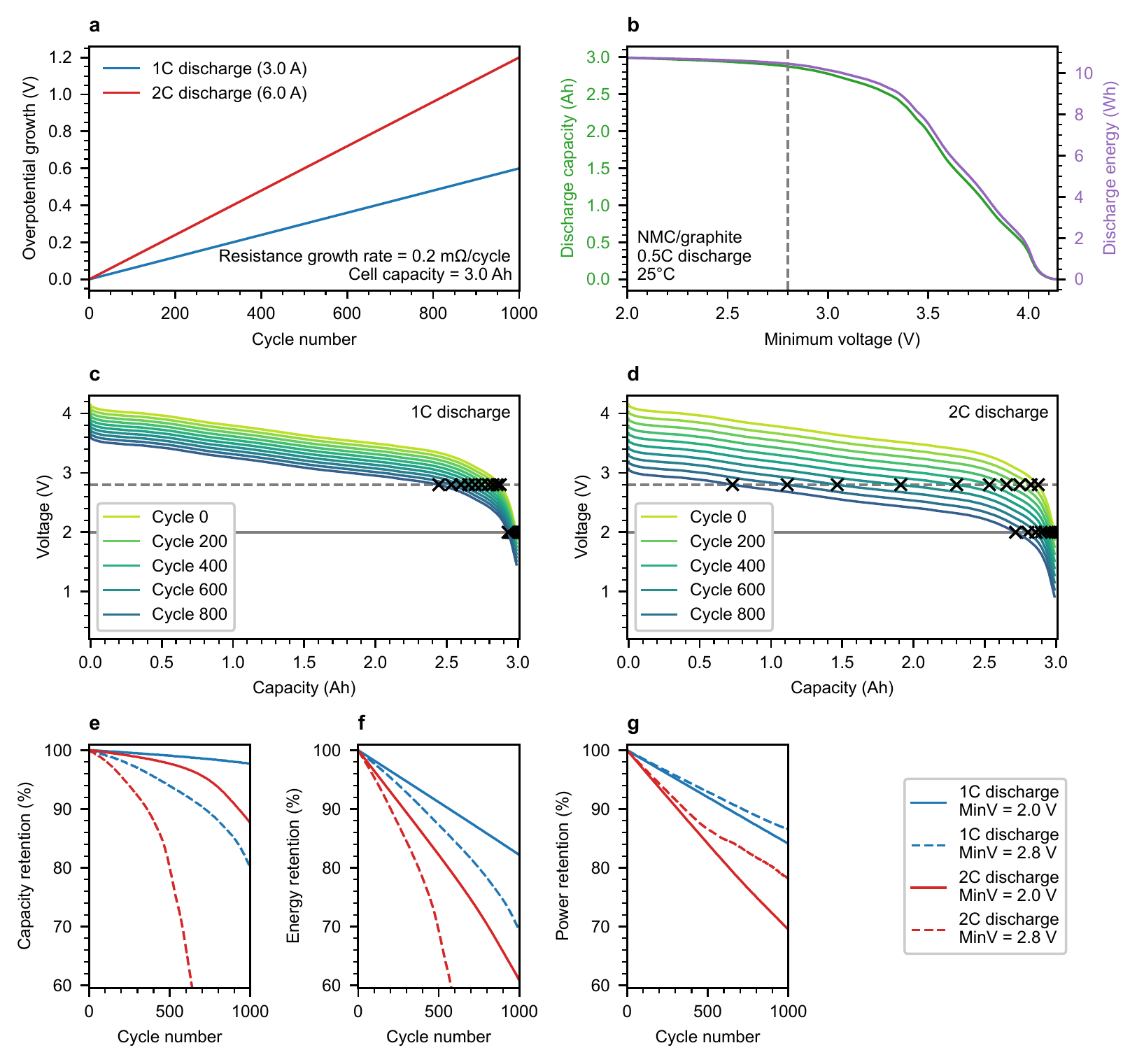}
\caption{Simple model illustrating ``pseudo-knees'' due to resistance growth; we use the term ``pseudo-knees'' here since the knee location is a function of rate and lower cutoff voltage. Inspired by Figure 16 of Ma et al.\cite{ma_editors_2019} and the work of Mandli et al.\cite{mandli_analysis_2019}. (a) Assumed overpotential growth vs. cycle number for a 1C and 2C discharge. The assumed resistance growth rate is 0.2 m$\Omega$/cycle. (b) Discharge capacity and energy vs. the minimum discharge voltage for an example NMC/graphite cell. Data obtained from Preger et al.\cite{preger_degradation_2020} (c, d) Voltage vs. capacity as a function of cycle number for the (c) 1C discharge and (d) 2C discharge cases. The final discharge capacity for each cycle is denoted by a marker. (e--g) (e) Discharge capacity, (f) discharge energy, and (g) discharge power (discharge energy/discharge time) retention vs.\ cycle number as a function of discharge current and the minimum voltage.
}
\label{fig:dcr_knee}
\end{figure}

Figure \ref{fig:dcr_knee} displays a simple model illustrating a knee due to ohmic resistance growth during cycling, inspired by the work of Ma et al.\cite{ma_editors_2019} and Mandli et al.\cite{mandli_analysis_2019}. The model arbitrarily assumes a constant resistance growth rate of 0.2 m$\Omega$ per cycle, occurring at all SOCs; the linearly increasing resistance with cycle number leads to linearly increasing ohmic overpotential (Figure \ref{fig:dcr_knee}a). The increased overpotential then shifts the voltage-capacity curve downwards.
To demonstrate the impact of increasing resistance due to this downshift in a ``real'' cell, we used voltage-capacity and voltage-energy relationships recorded from an NMC/graphite cell at beginning-of-life from Preger et al.\cite{preger_degradation_2020} (Figure \ref{fig:dcr_knee}b). Figures \ref{fig:dcr_knee}c and \ref{fig:dcr_knee}d show the impacts of this downshift on the voltage-capacity curve at lower voltage cutoffs of 2 V and 2.8 V and discharge currents of 1C (Figure \ref{fig:dcr_knee}c) and 2C (Figure \ref{fig:dcr_knee}d). In all cases, the discharge ends on the steep portion of the voltage-capacity curve at the beginning of life. However, as the cell ages and the resistance increases, the discharge ends on the flat region of the voltage-capacity curve, resulting in an increased rate of capacity loss (i.e., a knee) despite a linear increase in resistance. Thus, this knee pathway is a threshold trajectory, where the internal state variable is the overpotential and the threshold is the ``overpotential margin'' between the lower cutoff voltage and the flatter region of the voltage-capacity curve.

The impact of the resistance growth on the measured capacity (Figure \ref{fig:dcr_knee}e), energy (\ref{fig:dcr_knee}f), and power (\ref{fig:dcr_knee}g) during discharge is highly sensitive to the discharge rate and the lower cutoff voltage---usage parameters that are not often considered critical for their impact on knees. Decreasing the discharge rates, of course, decreases the overpotential and delays the onset of the knee.
Decreasing the lower cutoff voltage delays the knee by increasing the overpotential margin between the lower cutoff voltage and the flatter region of the voltage-capacity curve; however, note that low cutoff voltages can also induce additional degradation mechanisms such as copper dissolution\cite{fear_elucidating_2018, carter_x-ray_2018}. Naturally, energy and power knees (\ref{fig:dcr_knee}f, \ref{fig:dcr_knee}g) are more sensitive to rate than the capacity knees (\ref{fig:dcr_knee}e). Because these knees can ``disappear'' by cycling at lower rates or to lower cutoff voltages, we sometimes refer to these knees as ``pseudo-knees''; furthermore, this knee mechanism may not be observed in some practical settings (e.g., the slow weeks-long discharge of an electric vehicle battery pack under typical usage).
Finally, we note that ``resistance pseudo-knees'' may also occur due to a stoichiometric decrease of lithium available to cycle, as explored by Mandli et al \cite{mandli_analysis_2019}, or a stoichiometric shifting of lithium to one electrode preferentially during aging due to uneven loss of active material across both electrodes \cite{lin_comprehensive_2013}.

Ma et al.\cite{ma_editors_2019} extensively studied this knee pathway using lab-made 230 mAh NMC532/ graphite pouch cells, varying the upper cutoff potential, discharging rate, electrolyte composition, and positive electrode material coatings. Careful impedance measurements (on both full cells and symmetric coin cells of the positive and negative electrodes) were used to identify a dramatic increase of the positive electrode impedance during aging. This resistance growth was attributed to electrolyte oxidation at the positive electrode, which was accelerated by high upper cutoff voltages and the use of more reactive electrodes and electrolytes (i.e., uncoated positive electrode particles, lower salt concentrations, and the use of oxidation-prone additives such as methyl acetate). One practical consideration from the work\cite{ma_editors_2019} is that tests with high discharge rates exhibit resistance-growth-induced knees earlier than tests with low discharge rates; thus, tests with high discharge rates can be used as an early indicator of resistance-growth-induced knees at lower rates. Alternately, measurements of resistance throughout cycling (in tandem with the voltage curve, discharge rate, and lower cutoff voltage) can be used to estimate when the knee will occur.

This knee pathway is also sensitive to electrode chemistry, as each chemistry exhibits a unique voltage-capacity curve. For instance, lithium iron phosphate (LFP)/graphite cells experiencing high resistance growth would exhibit much sharper knees than NMC/graphite cells due to their flatter voltage-capacity curves. While LFP cells generally do not exhibit high resistance growth due to their lower operating voltages\cite{keil_calendar_2016, safari_aging_2011}, even moderate resistance growth coupled with high discharge rates and high lower cutoff voltages could lead to dramatic cell failure (although the magnitude of this effect depends on the cutoff voltages).

Lastly, we mention that capacity knees are often correlated with ``resistance elbows''---that is, a rapid rise in the internal resistance. While this correlation may be evidence for the prevalence of this knee pathway, we note that other knee pathways may also lead to a resistance increase at the knee (e.g., lithium plating due to loss of active material or porosity decrease). We return to this topic at a later point.

This ``threshold'' knee pathway is straightforward to model and predict using standard electrochemical models and measurements of resistance. However, convolutions with lithium inventory loss, active material loss, etc. require care. Overall, given the high sensitivity of this knee pathway to discharge rate and lower cutoff voltage---parameters that vary widely in real-world usage---care must be taken to transfer laboratory results to the field. 

\subsection{Electrolyte and additive depletion knees}

Both electrolyte and additive depletion have been linked to knees.
In principle, electrolyte depletion can lead to a knee by driving either loss of active material\cite{mao_calendar_2017, kupper_end--life_2018, fang_capacity_2021} or lithium plating\cite{sieg_fast-charging_2022}{}.
In turn, electrolyte depletion can be driven either by consumption via side reactions\cite{stevens_using_2014, park_semi-empirical_2017,sieg_fast-charging_2022} or via local gas generation leading to particles disconnecting from the electrolyte\cite{mao_calendar_2017, kupper_end--life_2018}{}.
Electrolyte depletion knees have been previously modeled, although often with limited experimental validation.
Park et al.\cite{park_semi-empirical_2017} provided the first empirical model of a capacity knee due to electrolyte depletion.
Fang et al.\cite{fang_capacity_2021} modeled electrolyte depletion knees occurring when the remaining electrolyte volume falls below the pore volume. The associated loss of active material increases local current density, which further increases the electrolyte dryout rate in a positive feedback loop.
Kupper et al.\cite{kupper_end--life_2018} also developed a model for electrolyte depletion knees using percolation theory to capture the nonlinear knee behavior, a model we detail in the subsequent pathway section. 
Experimentally, Sieg et al.\cite{sieg_fast-charging_2022} attributed capacity knees during fast charging of large pouch cells to electrolyte dryout via careful coin cell diagnostic studies; while the electrode capacities remained healthy over life, lithium plating and decreased fast charging times could be tracked to decreased local electrolyte volumes.
While the principles of the electrolyte depletion knee pathway are clear, more work is needed to understand the mechanistic details. 

Additionally, robust work has linked the depletion of electrolyte additives to knees. Electrolyte additives have a disproportionate effect on lifetime relative to their amount in a cell; small quantities of electrolyte additives can often delay the occurrence of the knee by many cycles.\cite{ma_editors_2019, li_comparison_2017} Additive chemistry is complex; for instance, Burns et al.\cite{burns_predicting_2013} showed how electrolyte performance often improves with the number of additives used. Additives can certainly influence the onset of other knee pathways, including lithium plating knees via various mechanisms (e.g., electrolyte transport properties, SEI growth rate and thus porosity decrease rate, etc.) and resistance growth knees by controlling the rate of resistance growth\cite{ma_editors_2019}. However, the depletion of electrolyte additives is another demonstrated knee pathway. Here, we discuss perhaps the most widely studied additive depletion knee mechanism: fluoroethylene carbonate (FEC) depletion in silicon-containing cells.

\begin{figure}[ht]
\centering
\includegraphics[scale = 0.9]{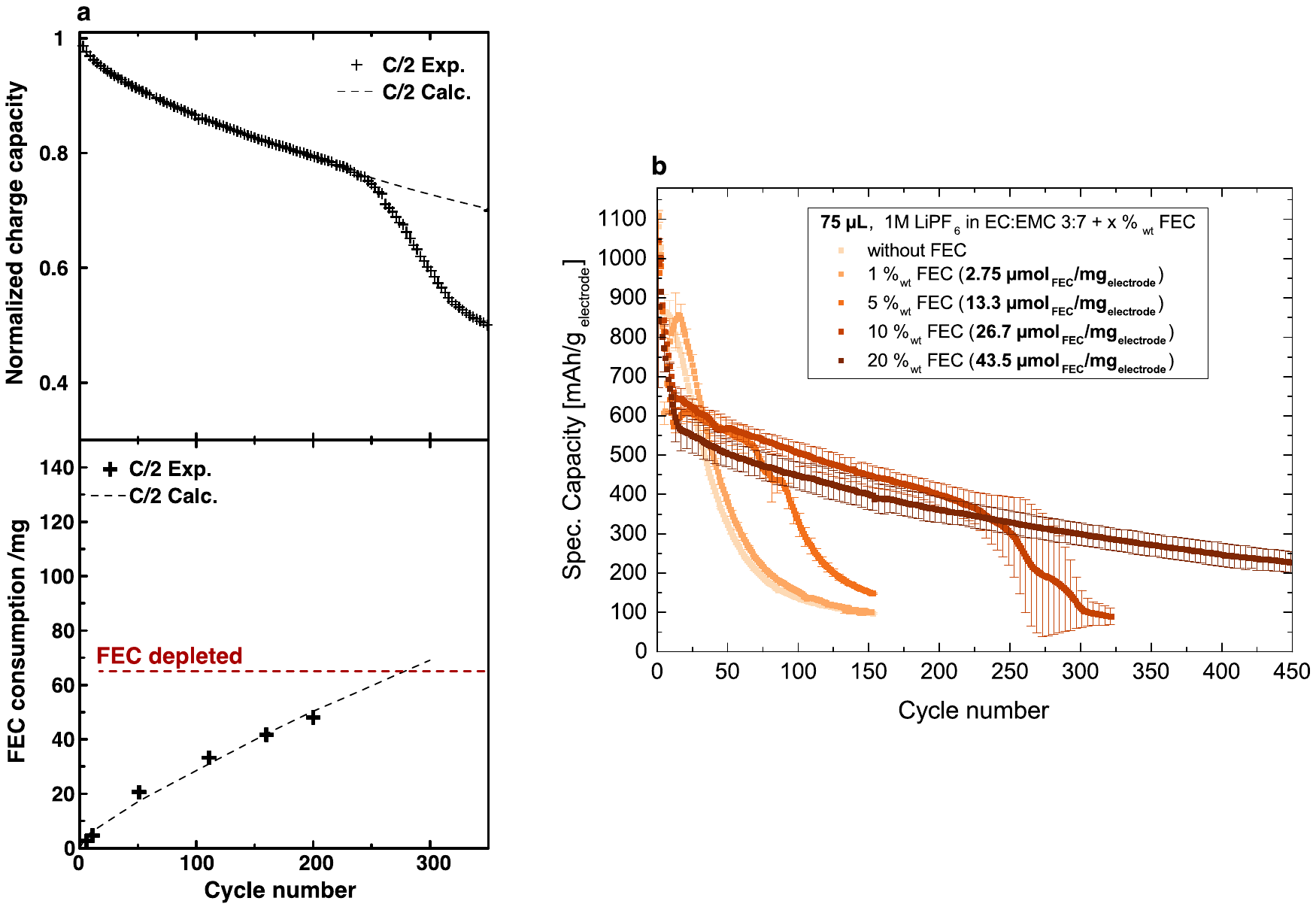}
\caption{Additive depletion knees.
(a) The capacity retention of a LCO/silicon-graphite pouch cell with 15\% Si in the negative electrode exhibits a knee at around 250 cycles, the same point at which the FEC is depleted from the electrolyte (from destructive gas chromatography-mass spectrometry measurements of FEC concentration). Adapted from Figure 8 of Petibon et al.\cite{petibon_studies_2016} CC-BY 4.0.
(b) In half cells with 100\% Si negative electrodes, the cycle number of the knee increases with the FEC concentration in the electrolyte.
Error bars reflect measurements from two nominally identical cells.
Adapted from Figure 1 of Jung et al.\cite{jung_consumption_2016} CC-BY 4.0.}  
\label{fig:fec_knee}
\end{figure}

FEC has been shown to substantially improve the capacity retention of silicon electrodes.\cite{choi_effect_2006, etacheri_effect_2012}
Among standard electrolyte components, FEC preferentially reacts at the surface of silicon particles; in fact, the rate of FEC consumption on silicon may be 10x that of graphite, in part due to its large volume expansion (around 300\%)\cite{wetjen_differentiating_2017}{}.
Petibon et al.\cite{petibon_studies_2016}{},
Jung et al.\cite{jung_consumption_2016}{},
and Wetjen et al.\cite{wetjen_differentiating_2017}
performed comprehensive studies of Si-containing full cells with FEC-containing electrolytes and commercially-representative volumes,
conclusively demonstrating that a knee occurs when FEC is depleted from the electrolyte.
Figure \ref{fig:fec_knee} displays key results from Petibon et al.\cite{petibon_studies_2016} and
Jung et al.\cite{jung_consumption_2016}{}, in which the dependence of the knee on FEC concentration was confirmed via destructive measurements of FEC concentration vs. cycle number\cite{petibon_studies_2016} and cycling cells with increasing initial FEC concentration\cite{jung_consumption_2016}.
Louli et al.\cite{louli_operando_2019} also corroborated these findings.
Earlier studies of the use of FEC in high-Si cells\cite{choi_effect_2006, etacheri_effect_2012} did not observe knees due to their use of high electrolyte volumes, which provided a large reservoir of FEC.
Other electrolyte components (namely, linear carbonates) are consumed only after the knee, since FEC can no longer be preferentially consumed\cite{petibon_studies_2016}; the cell polarization increases substantially after the knee\cite{petibon_studies_2016, jung_consumption_2016, wetjen_differentiating_2017}, possibly due to high reaction overpotential caused by reactions of silicon with these nonpreferred electrolyte components.
This knee pathway is exacerbated by high upper cutoff voltages \cite{petibon_studies_2016}{}, higher cycling rates (presumably due to more mechanical damage to the SEI layer) \cite{petibon_studies_2016}{}, and (presumably) high temperatures (due to higher SEI growth rates).

The electrolyte and additive depletion knee pathway, a clear threshold trajectory, has a number of interesting implications.
First, since laboratory-built cells are often filled with high electrolyte volumes, electrolyte-related knee mechanisms that are not present in lab testing may manifest in more commercially representative form factors.
As Wetjen et al.\cite{wetjen_differentiating_2017} emphasize,
maintaining representative electrolyte volumes in lab-scale cells is critical for accurately capturing this knee pathway in production-scale cells.
Second, nominally identical cells, cycled identically, but with different initial FEC concentrations exhibited minute electrochemical differences before the knee.\cite{jung_consumption_2016}
While the equivalent study has not been performed for electrolyte depletion, we expect a similar result.
Since only the electrolyte or additive \textit{consumed} in a given cycle manifests in the electrochemical signals from cycling (e.g., differential capacity or differential voltage analysis), the \textit{remaining} electrolyte or additive is not electrochemically detectable as it does not participate in reactions with the electrode.
However, the remaining electrolyte or additive amount is the key internal state variable for this pathway.
To estimate the remaining electrolyte or additive amount, the internal electrolyte/additive amount and the electrolyte/additive consumption over life must be known.
However, extracting electrolyte/additive consumption during cycling from electrochemical data is challenging since side reaction signals are often faint and occur concurrently with other electrochemical processes.
Furthermore, for commercial cells, the initial electrolyte/additive amount and the electrolyte/additive consumed during formation are unknown, although obtaining these values may be possible via electrolyte reverse engineering.
Thus, knee onset for this knee pathway is challenging to predict via standard electrochemical signals.
A proposal for future work is to evaluate electrochemical tests or other nondestructive probes that may be sensitive to remaining electrolyte/additive amounts.
Fortunately, ultrasonic probes appear well suited for detecting electrolyte loss in some cell form factors, namely pouch cells.\cite{knehr_understanding_2018, deng_ultrasonic_2020}

\subsection{Percolation-limited connectivity knees}

Percolation theory \cite{essam_percolation_1980, stauffer_introduction_1994} is commonly used to describe statistical properties of clusters of materials that are geometrically connected in porous media, including porous electrodes used in modern lithium-ion batteries\cite{ferguson_nonequilibrium_2012}. In a porous medium described by percolation theory, there exists a critical material parameter above which the probability of a spanning cluster, i.e., a cluster that spans the entire spatial extent of the porous medium, being formed tends towards one and below which this probability tends towards zero.\cite{ferguson_nonequilibrium_2012} In many percolating systems, this probability is highly sensitive to the value of the critical material parameter. For lithium-ion batteries, percolation theory can be used to describe both the ionic conductivity of the liquid electrolyte that fills the porous electrode and the electronic conductivity of the network of conductive additives. In battery modeling and experimentation, the electrode is often implicitly assumed to be sufficiently porous for the liquid electrolyte to completely percolate it. On the other hand, much effort has been spent on elucidating how the volume fraction of conductive additives may or may not give rise to a percolating electrically conducting network~\cite{chen_selection_2007, li_effects_2015, cerbelaud_understanding_2015, guzman_improved_2017}, which is especially important for ensuring electronic conduction is not rate limiting in electrically insulating active materials, such as lithium iron phosphate.\cite{li_effects_2015, guzman_improved_2017}

\begin{figure}[ht]
    \centering
    \includegraphics[scale=1.0]{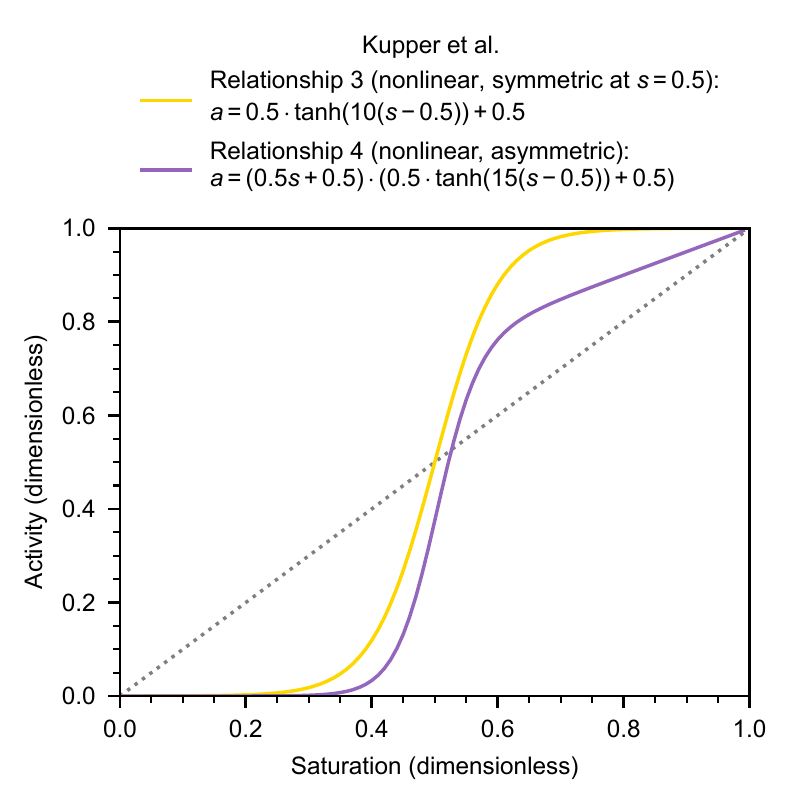}
    \caption{Two activity-saturation relationships describing percolation-limited electrolyte dry-out, adapted from Figure 5 of Kupper et al.\cite{kupper_end--life_2018} Relationships 3 and 4 model percolation of the liquid electrolyte where activity depends nonlinearly on saturation. The key feature of both relationships is the presence of a percolation threshold value ($s=0.5$), around which activity varies rapidly between $0$ and $1$.
    The sensitivity of activity to small changes in saturation is apparent.}
    \label{fig:percolation}
\end{figure}

Kupper et al.\cite{kupper_end--life_2018} proposed an electrolyte depletion knee mechanism that incorporates percolation theory. In this proposed electrolyte dry-out mechanism, they first define two new electrode descriptors: ``activity'', $a$, and ``saturation'', $s$, given by $a = \frac{\varepsilon_{\text{LiC}_6}}{\varepsilon_{\text{LiC}_6,\text{inactive}}+\varepsilon_{\text{LiC}_6}}$ and $s = \frac{\varepsilon_\text{elyt}}{\varepsilon_\text{elyt}+\varepsilon_\text{gas}}$, respectively. In these equations, $\varepsilon$ is the volume fraction of the active graphite ($\varepsilon_{\text{LiC}_6}$), inactive graphite ($\varepsilon_{\text{LiC}_6,\text{inactive}}$), electrolyte ($\varepsilon_{\text{elyte}}$) and gas ($\varepsilon_{\text{gas}}$, which is produced during SEI growth). Activity describes how much of the electrode material is active and available for reaction, while saturation describes the amount of pore space occupied by the liquid electrolyte. The loss of ionic contact of graphite caused by electrolyte dry-out is then described by a kinetic rate law that is proportional to the difference in activity and equilibrium activity, which is assumed to be a function of only saturation. To predict a knee in cell capacity, the equilibrium activity-saturation relationships were formulated to be nonlinear and contain a percolation threshold value, around which the equilibrium activity varies rapidly between $0$ and $1$. The functional forms of these relationships were assumed given the absence of theoretical or experimental guidance. Figure~\ref{fig:percolation} plots two such nonlinear relationships, named relationships $3$ and $4$, adapted from Figure 5 of Kupper et al.\cite{kupper_end--life_2018} The authors concluded that relationship 4 best fitted experimental aging data.

The knee caused by this electrolyte dry-out model is a threshold trajectory, where the threshold is the critical saturation value illustrated in Figure~\ref{fig:percolation}. Although Kupper et al.\cite{kupper_end--life_2018} did not provide convincing experimental validation to definitively prove that electrolyte dry-out resulted in sudden death of the cell and that relationship 4 was the most plausible activity-saturation relationship, the proposed electrolyte dry-out mechanism is plausible in principle and should be experimentally studied in more detail. Furthermore, a similar effect may apply for electronic conductivity networks; Guzmán et al. \cite{guzman_improved_2017} illustrated how the electronic conductivity of lithium iron phosphate electrodes exhibits a percolation threshold based on the conductive carbon content.
If this effect is experimentally validated, we expect that identifying both the activity-saturation relationship and nondestructive measurements of saturation during aging will be challenging; we note that ultrasonic probes have had success in detecting electrolyte loss\cite{knehr_understanding_2018, deng_ultrasonic_2020}{}.

\subsection{Mechanical deformation knees}

\begin{figure}[hp]
\centering
\includegraphics[scale = 1]{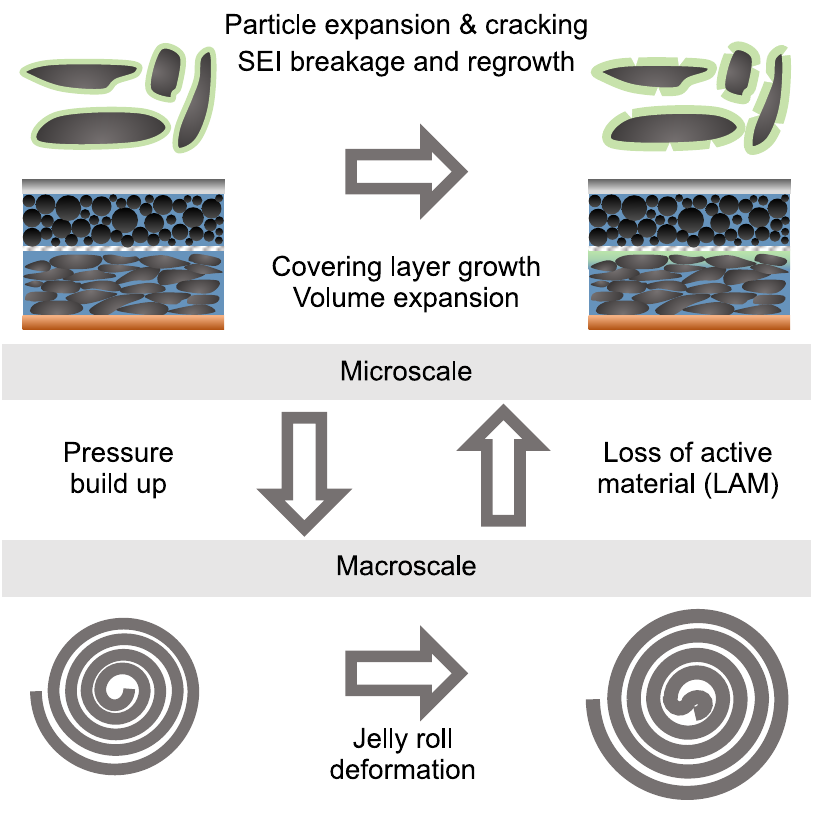}
\caption{Positive feedback mechanisms between mechanical effects that can lead to knees. During cycling, both particles and the SEI expand and crack, leading to the regrowth of additional SEI. In addition, covering layers may form on the surface of the negative electrode. Pressure due to volume expansion can subsequently lead to jelly roll deformation in cylindrical cells. This deformation can cause loss of active material, leading to more SEI/covering layer growth and potentially lithium plating.}
\label{fig:knee_mechanical}
\end{figure}

\begin{figure}[ht!]
\centering
\includegraphics[scale = 1.0]{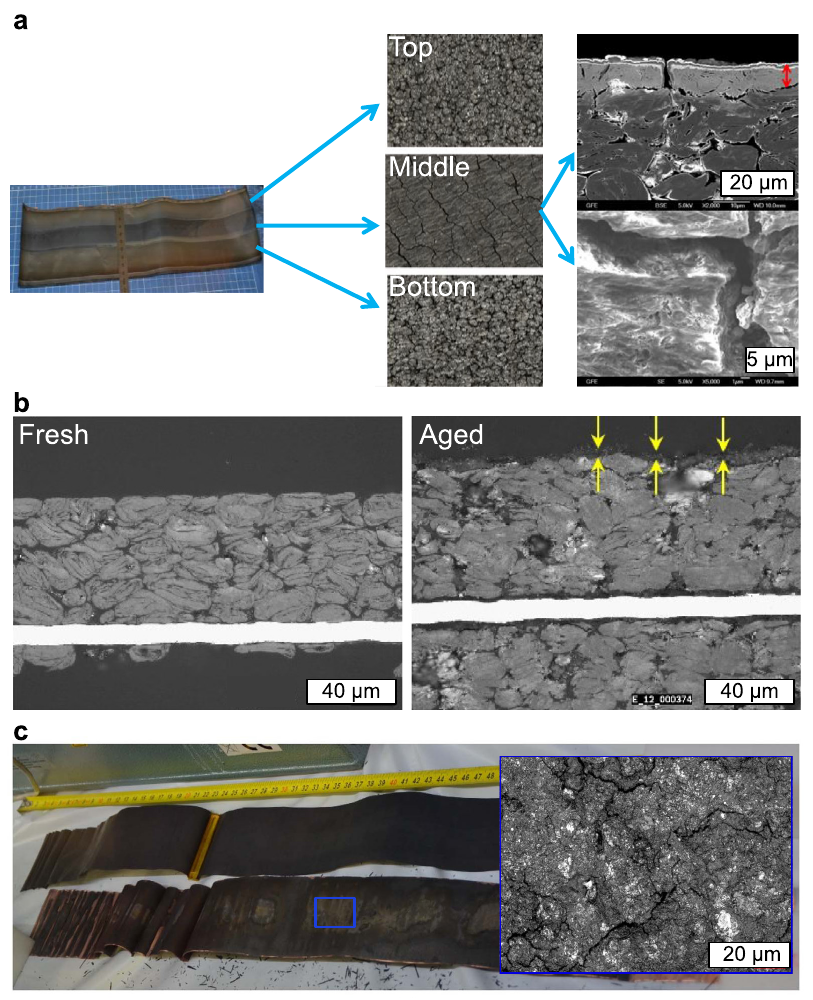}
\caption{``Covering layers'' as found via post-mortem analysis in cells with knees. Covering layers are commonly observed in cells with knees but are a poorly understood source of active material loss.
a) A covering layer in the middle of the negative electrode of an unwound cylindrical cell, along with laser microscope images of top, middle and bottom sections and scanning electron micrographs of the middle section with a covering layer. The red arrow corresponds to a thickness of 10 $\mathrm{\mu m}$. Reproduced with permission from Figure 5 of Lewerenz et al. \cite{lewerenz_post-mortem_2017} Copyright 2017, Elsevier. b) Optical microscopy cross-sections of fresh and aged electrodes with a visible covering layer. Reproduced with permission from Figure 12 of Stiaszny et al. \cite{stiaszny_electrochemical_2014} Copyright 2014, Elsevier. c) Covering layer in the middle of the negative electrode of an unwound cylindrical cell, as identified by post-mortem dissection and confocal laser microscopy. Reproduced from Figures 11 and 12 of Willenberg et al.\cite{willenberg_development_2020} CC-BY-4.0.} 
\label{fig:covering_layers}
\end{figure}

Both microscale mechanical effects occurring at the particle scale and macroscale mechanical effects occurring at the cell scale can be pathways for knees.
These mechanical degradation mechanisms often interact in positive feedback loops (Figure \ref{fig:knee_mechanical}).
Mechanical degradation mechanisms are closely tied to other knee pathways. For instance, Cannarella and Arnold~\cite{cannarella_stress_2014} showed how high external stack pressure can cause a lithium plating knee; Bach et al.\cite{bach_nonlinear_2016} demonstrated a link between heterogeneous pressure and localized lithium plating; and many loss of active material mechanisms are mechanical in nature (e.g., delamination, particle cracking).
Additionally, the growth of covering layers on the surface of the negative electrode (Figure \ref{fig:covering_layers}), often reported on cells with knees \cite{lewerenz_post-mortem_2017,willenberg_development_2020,stiaszny_electrochemical_2014, fang_capacity_2021}{}, may lead to additional mechanical stresses on the macroscale.
Here, we focus on knees more explicitly linked to mechanical effects.

At the microscale, repeated (de)intercalation can stress the electrode particles, which can then lead to both loss of active material through particle cracking and accelerated growth of side reaction products (e.g., SEI and its analogue on the positive electrode, often termed ``CEI'' for cathode-electrolyte interphase).
Reniers et al. \cite{reniers_review_2019} illustrated a positive feedback mechanism between the mechanical stress and loss of active material, leading to a snowballing knee. They combine a fatigue model for loss of active material due to stress from Laresgoiti et al. \cite{laresgoiti_modeling_2015} with a stress model from Dai et al. \cite{dai_simulation_2014}{}: higher stress causes higher loss of active material, which in turn increases the current density and hence causes higher stress.
Other authors have suggested that mechanical effects can accelerate SEI/CEI growth by causing SEI/CEI cracking and revealing new active surface area to grow \cite{pinson_theory_2013,kupper_end--life_2018,louli_operando_2019, jana_physical_2019}. Since growth of these interphasial layers is self-limiting and thus sublinear\cite{bloom_accelerated_2001, broussely_aging_2001, wright_calendar-_2002, smith_high_2011, attia_revisiting_2020}, this effect alone is not enough to lead to a knee, but it could accelerate the onset of knees in other pathways related to side reactions (i.e., lithium plating induced by pore clogging on the negative electrode \cite{lewerenz_post-mortem_2017} or resistance growth driven by side reactions on the positive electrode\cite{ma_editors_2019, jana_physical_2019}).
Overall, microscale mechanical deformation mechanisms are challenging to model given the difficulties of experimental validation and the complexity of their interactions.

At the mesoscale, an additional $\mathrm{LAM_{deNE}}$ mechanism is the growth of thick layers ($1 - 10\: \mathrm{\mu m}$) on the surface of the negative electrode at the separator-facing interface, sometimes termed ``covering layers''\cite{lewerenz_post-mortem_2017, lewerenz_systematic_2017, willenberg_development_2020}.
Covering layer growth-driven $\mathrm{LAM_{deNE}}$ may impede lithium-ion transport into the negative electrode during charging, effectively isolating portions of the electrode and resulting in an apparent loss of active material. This covering layer is commonly observed in cells with knees and is often attributed to manganese or iron dissolution from the positive electrode and/or electrolyte salt decomposition \cite{lewerenz_post-mortem_2017,lewerenz_systematic_2017,zhu_investigation_2021,stiaszny_electrochemical_2014,rahe_nanoscale_2019,keil_linear_2019,sarasketa-zabala_understanding_2015, li_degradation_2016, klett_non-uniform_2014, klett_uneven_2015, willenberg_high-precision_2020, wang_cycle-life_2011, fang_capacity_2021}{}, or possibly dead lithium agglomerates\cite{schindler_fast_2018}{}, but the root cause has not been definitively identified. Peculiarly, the size of these layers (microns) is much larger than typical reported SEI thicknesses (nanometers)\cite{peled_reviewsei_2017}. Furthermore, this phenomenon is almost exclusively observed in cylindrical cells.
Lewerenz et al.\cite{lewerenz_post-mortem_2017,lewerenz_systematic_2017} thoroughly documented covering layer growth, finding that increasing C rate and larger depth-of-discharge could lead to earlier onset of a knee. Earlier knee onsets were correlated with the presence of a thick covering layer on cells that contained knees; cells without knees also contained obvious covering layers, but with lower surface coverage and less thickness. These covering layers sometimes seem to lead directly to localized lithium plating due to the lack of active sites available for lithium insertion, with lithium observed at the covering layer/separator interface.\cite{zhu_investigation_2021}
Further investigation of these covering layers is needed to understand this seemingly ubiquitous mechanism for loss of active negative electrode material and its relationship to knees.

On the macroscale (cell level), mechanical degradation manifests differently depending on the form factor. For both pouch and prismatic cells, the external pressure can impact cell lifetime. Pouch and prismatic cells either without external pressure\cite{wunsch_investigation_2019} or with high external pressure\cite{cannarella_stress_2014} can show rapid knees, indicating an intermediate value of pressure is optimal to avoid knees and maximize lifetime. Cannarella et al.\cite{cannarella_stress_2014} also found that the surface layer is more pronounced with increasing external pressure. Heterogeneous pressure distributions from internal cell components (e.g., electrode tabs) or loading variation\cite{beck_inhomogeneities_2021} can also lead to knees. Knees due to mechanical heterogeneity can perhaps be modeled and predicted via a careful understanding of the porosity distributions within the cell. This pathway can be considered either a snowball or threshold trajectory.

For cylindrical cells, several studies\cite{waldmann_mechanical_2014, pfrang_long-term_2018, pfrang_geometrical_2019, carter_mechanical_2019, kok_virtual_2019} have identified jelly roll deformation, using X-ray computed tomography after cycle life testing, in which the jelly roll deforms inwards towards the core of the cell and can lead to a capacity knee. This deformation was attributed to an increase in internal pressure of the cell over life, which itself is driven by either side reaction growth (including covering layers)\cite{willenberg_development_2020}{}, lithium plating\cite{carter_mechanical_2019}{}, or high thermo-mechanical stress due to large radial temperature gradients within the cell\cite{waldmann_mechanical_2014}{}. Geometric heterogeneities (e.g., due to the internal tabs) can also exacerbate this failure mode.\cite{pfrang_long-term_2018, pfrang_geometrical_2019}
This pathway can be considered a threshold trajectory, where the jelly roll is eventually unable to accommodate the increase in internal volume and pressure. Modeling and predicting this mechanism, however, requires measuring and tracking the internal pressure over life, as well as knowing the internal pressure at which the jelly roll will deform. A center pin appears to help avoid this failure mode.\cite{waldmann_mechanical_2014, carter_mechanical_2019}

\subsection{Interactions, heterogeneity, and variation}

While the six knee pathways we have identified can occur independently, these pathways can clearly interact. For instance, loss of active material plays a central role in four of our six pathways (lithium plating, electrode saturation, percolation-driven connectivity, and mechanical deformation). This coupling between degradation mechanisms can create positive feedback mechanisms, a special case of snowball internal trajectories.
Reniers et al.\cite{reniers_review_2019} explored a number of interacting degradation mechanisms with a single-particle model, finding that many have positive feedback. These interactions can also occur across length scales (Figure \ref{fig:knee_mechanical}), as SEI growth on the nanometer level can drive lithium plating on the centimeter level.
Given the high sensitivity of snowball pathways to small changes in state, interacting knee pathways can create positive feedback mechanisms with high sensitivity to internal state.

As an extreme example of positive feedback coupling between knee mechanisms, consider a hypothetical “quadruple snowball” trajectory. Each individual component in this mechanism has been previously discussed. First, particle cracking leading to loss of active material can snowball since the local current density on the remaining active particles is continuously increasing, driving additional mechanical stress. Second, loss of active material itself can snowball with percolation-limited connectivity---i.e., if the active material fraction drops below the critical percolation threshold. Third, loss of active material from the negative electrode, upon saturation, can cause rate-dependent lithium plating to snowball; the local current density will keep rising on the remaining negative electrode active material, increasing the driving force for lithium plating over reversible interaction. Lastly, lithium plating can snowball due to its nucleation and growth kinetics. While this example is certainly contrived, feedback between multiple knee pathways is perhaps probable given the shared sensitivities of many of these mechanisms to the same levers. 

Heterogeneity within a cell may also exacerbate these knee pathways. Commercially-relevant form factors have electrochemical, thermal, and mechanical gradients due to intrinsic heterogeneity and inactive components; these gradients can drive localized degradation. For instance, heterogeneous particle distributions or porosity profiles on the electrode level can lead to localized lithium plating\cite{chung_particle_2014}. Furthermore, the presence and location of electrode tabs in cylindrical cells can create electrochemical, thermal, and mechanical gradients\cite{lee_three_2013, reimers_accurate_2014, senyshyn_homogeneity_2015, waldmann_influence_2015, waldmann_influence_2016, bach_nonlinear_2016, carter_detection_2019, yao_tab_2019, pfrang_geometrical_2019, li_optimal_2021}, in some cases also leading to localized lithium plating\cite{bach_nonlinear_2016, coron_impact_2020}{}. Heterogeneity can also arise from ambient factors, e.g., thermal gradients induced by environment or thermal management systems.\cite{werner_inhomogeneous_2020} Furthermore, local heterogeneity can lead to positive feedback for degradation; for instance, the temperature of a region that receives a higher local current density will rise, leading to even higher local current density. Given the sensitivity and positive feedback of many knee mechanisms, heterogeneity can certainly exacerbate the presence of knees.

Finally, we consider the impact of sample and testing variation. Nominally identical cells cycled identically often show differences in knee behavior. This sampling variability includes both intrinsic variability from manufacturing (component-level variation, cell assembly, etc.) and extrinsic variability from testing (cycler calibration, temperature control, etc.). These sources of variability require rigorous equipment calibration to distinguish.

\begin{figure}[ht!]
\centering
\includegraphics[scale=1]{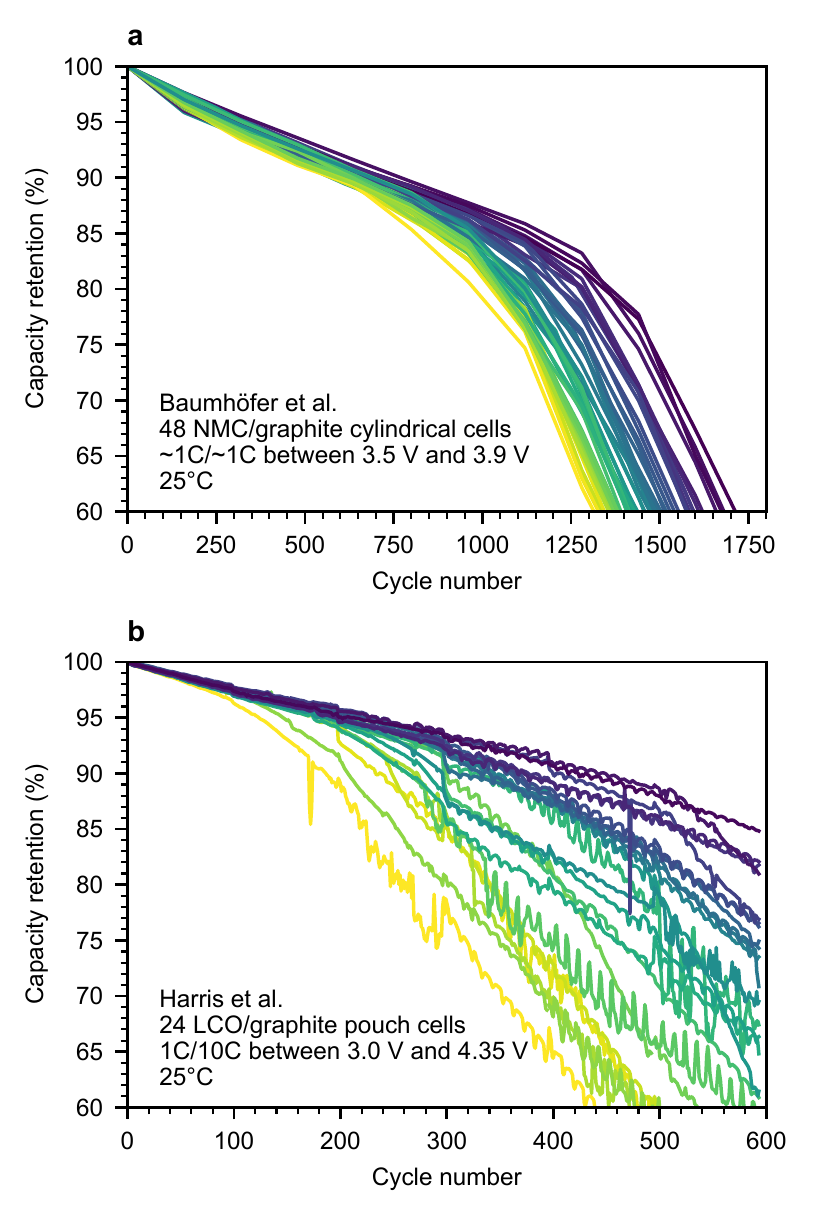}
\caption{Experimental studies of sample and testing variation with knees, illustrating the sensitivity of knees to these sources of variation.
(a) Retention vs. cycle number for 48 commercial NMC/graphite cylindrical cells. Adapted from Figure 6 of Baumhöfer et al. \cite{baumhofer_production_2014}
(b) Retention vs. cycle number for 24 commercial lithium cobalt oxide (LCO)/graphite pouch cells. Adapted from Figure 2a of Harris et al.\cite{harris_failure_2017}
}
\label{fig:var_exp}
\end{figure}

The magnitude of sampling variability is a function of the cell design, manufacturing variability, and testing conditions. Sampling variability may increase with more aggressive cell designs (e.g., higher silicon content), more manual cell assembly processes, and more aggressive testing conditions (particularly for test setups with no or poor temperature control). The magnitude of sampling variability can be estimated using studies with fairly large sample sizes (i.e., at least $\sim$10 cells)\cite{dechent_estimation_2021}{}; Beck et al.\cite{beck_inhomogeneities_2021} provide a detailed review of cell-to-cell variation. Baumhöfer et al.\cite{baumhofer_production_2014} and Harris et al.\cite{harris_failure_2017} studied this type of variation in 48 cells and 24 cells, respectively, finding widely varying knee locations across their datasets (Figure \ref{fig:var_exp}).
While the knee pathway and internal state trajectories are unknown for these datasets, the Harris et al.\cite{harris_failure_2017} dataset has much larger variability, perhaps due to its aggressive 10C discharge rate.
These studies did not identify a correlation between beginning-of-life capacity and end-of-life capacity, suggesting that differences in initial internal state trajectories did not manifest in the initial capacity measurements (possibly implying that hidden or threshold internal state trajectories caused these knees). In general, sampling and testing variation also poses challenges for accurate knee prediction; identifying the manufacturing and testing variation of the internal state variable of interest is needed to evaluate the accuracy of knee prediction methods in real-world settings.
Lifetime variability highlights the high sensitivity of knees to manufacturing and testing variability.

\begin{figure}[ht!]
\centering
\includegraphics[scale=0.7]{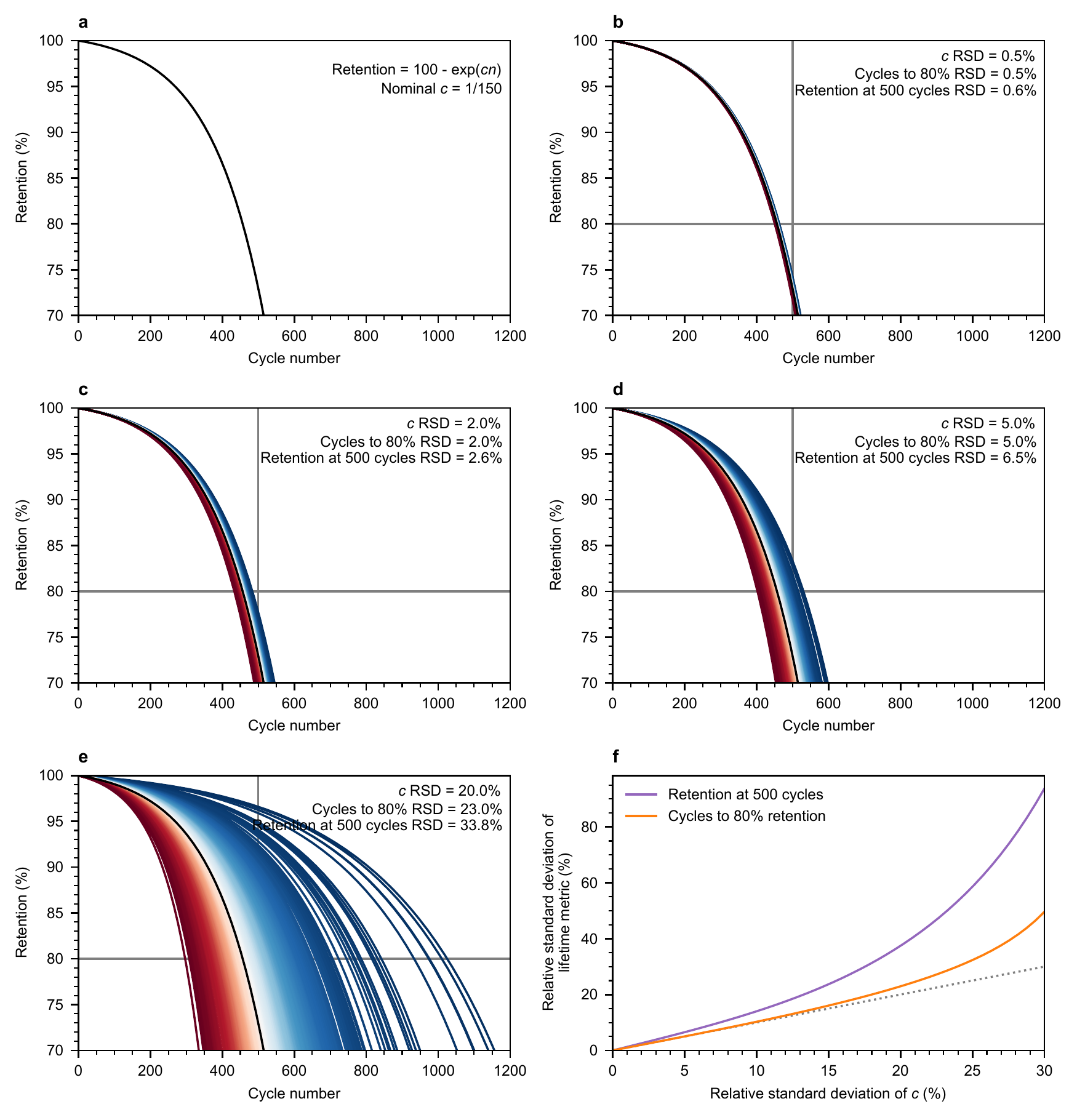}
\caption{Impact of cell-to-cell and/or testing variation on knees.
(a) A simple, one-parameter exponential model is defined to simulate a retention curve with a knee, $Q=100 - \exp(cn)$, where $n$ represents cycle number.
The adjustable parameter, $c$, is given an initial value of 1/150.
The retention model is simulated for 500 cells with a normal distribution of values for $c$, with relative standard deviations (RSDs) of (b) 0.5\%, (c) 2\%, (d) 5\%, and (e) 20\%. The RSDs of both the number of cycles to 80\% and the retention at 500 cycles is tracked; these RSDs are either equivalent or larger than the RSDs of $c$ (f).
Note that the distributions of both the number of cycles to 80\% and the retention at 500 cycles are non-Gaussian and skewed, despite Gaussian distributions of $c$.}
\label{fig:var_model}
\end{figure}

In Figure \ref{fig:var_model}, we develop a simple model to consider the sensitivity of knees on cell-to-cell/testing variation. We propose a simple one-parameter exponential functional form for a cell retention curve with a knee, $Q=100 - \exp(cn)$, where $n$ represents cycle number. In Figure \ref{fig:var_model}a, we plot this function for $c=1/150$. We then visualize the distribution of retention curves if $c$ is normally distributed with various relative standard deviations (RSDs), including 0.5\% (\ref{fig:var_model}b), 2\% (\ref{fig:var_model}c), 5\% (\ref{fig:var_model}d), and 20\% (\ref{fig:var_model}e). For each distribution of retention curves, we also track the RSDs of two lifetime metrics: the number of cycles until 80\% retention and the retention at 500 cycles. We find that the RSDs of the two output metrics always equal or exceed the RSD of $c$ (Figure \ref{fig:var_model}f). Moreover, despite Gaussian input variation, the distribution of the number of cycles until 80\% retention and the retention at 500 cycles are non-Gaussian and skewed (illustrated most clearly in Figure \ref{fig:var_model}e). While simplistic, this model illustrates how cell-to-cell variation can have an outsize effect on knee location given the nonlinear dependencies of lifetime. This exercise could be repeated for other internal state trajectories and all of their functional forms.

In summary, the impact of interactions, heterogeneity, and variation on knee pathways and internal state trajectories is complex, poorly understood, and an opportunity for future work.

\section{Factors influencing the knee}

With a foundation for the fundamentals of knee pathways and internal state trajectories in place, we surveyed the literature to identify empirical case studies in which the knee point can be controlled via changes to a single variable. Table \ref{tab:experimental_summary} classifies these case studies into three categories based on the nature of the variable: cell design, testing conditions, and sampling/testing variability (a special case of these two categories). Some cell designs and testing conditions have a consistent impact on the emergence of the knee; for example, higher charging rates and wider cycling voltage ranges accelerate the appearance of the knee. However, the impact of other variables (e.g., discharging rate and rest times) is less clear and may depend on the specific cell design and operating conditions.

\subsection{Cell design}

While the dependence of knees on cell usage conditions has been studied extensively, less attention has been focused on the dependence of knees on cell design---likely due to the challenges of representative lab-scale cell fabrication. Ma et al.\cite{ma_editors_2019}{}, one of the most comprehensive works on the impact of cell design on knees, studied the impact of various electrodes and electrolytes on the location of the knee. These knees were classified as resistance ``pseudo-knees'' due to increased electrolyte oxidation on the positive electrode, as evidenced by the strong dependence of the knee severity on discharge rate as well as positive electrode impedance measurements. For electrode design, Ma et al.\cite{ma_editors_2019} and Klein et al.\cite{klein_demonstrating_2021} found that positive electrode particle coatings and low positive electrode loadings delayed the knee. Ma et al. \cite{ma_editors_2019} and Glazier et al.\cite{glazier_analysis_2017} also found that the graphite type (i.e., natural or artificial) can substantially impact the knee location; while natural graphite has larger irreversible expansion and thus higher parasitic reaction rates\cite{glazier_analysis_2017}{}, the root cause of the knee in this case is unclear.

Electrode loadings (i.e., capacity per unit area) can also lead to knees via the lithium plating pathway.
Rate-independent plating can occur if the ratio of negative electrode loading to positive electrode loading is too low (i.e., $n$:$p< 1$) \cite{deichmann_investigating_2020}; however, rate-dependent lithium plating can occur at low loading ratios if the negative electrode is too thick or the porosity is too low.\cite{waldmann_mechanical_2014, yang_modeling_2017, klein_demonstrating_2021}

Additionally, small changes in the electrolyte can play an outsized role on the lifetime performance. Ma et al.\cite{ma_editors_2019} demonstrated the sensitivity of the knee location to the electrolyte additive mixture; specifically, high methyl acetate (MA) concentrations (MA is used to increase electrolyte transport capability) and low LiPF$_6$ concentrations consistently led to earlier knees. These knees were all attributed to increased electrolyte oxidation on the positive electrode via the resistance growth-induced pathway. Ma et al.\cite{ma_editors_2019} also identified other electrolyte systems with a strong knee sensitivity.
Note that while Ma et al.\cite{ma_editors_2019} found that lower salt concentrations accelerated the knee in NMC/graphite pouch cells,
Wang et al.\cite{wang_systematic_2014} found that higher salt concentrations accelerated the knee in lithium cobalt oxide (LCO)/graphite pouch cells over a similar range of salt concentrations; both studies attributed the observed trends to positive electrode impedance growth.
Additionally, as previously discussed, the additive depletion pathway can be a direct cause of knees for some cell designs (e.g., cells with high silicon content in the negative electrode and low FEC content in the electrolyte).\cite{petibon_studies_2016, jung_consumption_2016}

Furthermore, mechanical deformation knees are naturally highly sensitive to the cell form factor. For instance, deformation of the core \cite{pfrang_long-term_2018,carter_mechanical_2019,willenberg_development_2020} can only occur in wound cells, primarily cylindrical cells. The presence of a mandrel in the core may prevent this mechanical deformation.\cite{carter_mechanical_2019}

Lastly, the formation protocol can influence the location of the knee. For instance, Weng et al.\cite{weng_predicting_2021} found that NMC/graphite cells formed with a fast formation protocol that emphasizes time at high SOCs exhibited later knees than cells formed with a slower baseline formation protocol, which was attributed the creation of well-passivating SEI at high potentials to decrease the amount of side reaction product formed during use.\cite{attia_benefits_2021}
Klein et al.\cite{klein_demonstrating_2021} found that decreasing the upper cutoff voltage in formation from 4.5 V to 4.3 V delayed the knee in NMC/graphite cells, which was attributed to decreased transition metal dissolution to cause plating. In principle, the formation protocol can be optimized to avoid or delay knees for a given use case.

\subsection{Testing conditions}

The sensitivity of knees to testing conditions has been extensively explored in the literature.
We note that many studies use accelerated aging protocols, which may introduce failure modes that are unrepresentative of real-world usage.
The representativeness of an aging profile to the target application must be considered when evaluating the sensitivity of knees to test conditions.
Field data may enable design of more representative test conditions.\cite{sulzer_challenge_2021}

\subsubsection{Charging rate}
Many studies have found that increasing the charging rate accelerates the onset of the knee.\cite{lewerenz_systematic_2017,lewerenz_post-mortem_2017, petzl_lithium_2015, burns_-situ_2015, waldmann_optimization_2015, schuster_nonlinear_2015, severson_data-driven_2019, schindler_fast_2018, keil_linear_2019} However, the critical charging rate leading to knees varies substantially, with knees appearing at C rates as low as 0.5C--1C\cite{waldmann_optimization_2015, willenberg_development_2020} and as high as 8C \cite{lewerenz_systematic_2017}{}. This critical charging rate is likely a function of cell design, temperature, and temperature control (e.g., air- vs. liquid-cooled).

High charging rates most commonly accelerate knees via lithium plating and covering layer growth. Lewerenz et al.\cite{lewerenz_systematic_2017,lewerenz_post-mortem_2017} suggested increased charging rates accelerated both lithium plating and covering layer growth in cylindrical LFP/graphite cells, demonstrating that knees occur reliably across a set of test replicates at a charging rate of 8C and occur less reliably at charging rates as low as 2C. Petzl et al.\cite{petzl_lithium_2015} and Burns et al.\cite{burns_-situ_2015} also found evidence of increased charging rates driving lithium plating after knees were observed. Note that interpretation of post-mortem analysis may be convoluted by the rapid degradation that occurs after the knee; in other words, lithium plating observed in a cell after a knee may be a cause or an effect of the knee.

\subsubsection{Discharging rate}

Unlike charging rate, the effect of discharging rate on knee location is mixed (Figure \ref{fig:discharge-rest_cycle}a--b).
In some systems, an increased discharging rate
accelerates the knee onset.
Omar et al.\cite{omar_lithium_2014} found that a higher discharging rate (1C to 15C) accelerated the knee for cylindrical LFP/graphite cells (Figure \ref{fig:discharge-rest_cycle}a).
Diao et al.\cite{diao_accelerated_2019} showed no effect of discharge rate except at 60°C, where the cells discharged at 2C degraded almost twice as quickly as the cells discharged at 0.7C or 1C.
High discharging rates may lead to earlier knees if they lead to higher temperatures, which accelerate electrolyte reduction (i.e., SEI growth driving pore clogging) and electrolyte oxidation (i.e., positive electrode resistance growth driving resistance growth pseudo-knees).
High discharging rates may be associated with mechanical stress on electrode particles as well, accelerating side reaction rates.\cite{christensen_mathematical_2006, allen_quantifying_2021, dubarry_cell_2014, sun_accelerated_2018}
Additionally, high discharging rates can lead to resistance pseudo-knees when the resistance growth or lower cutoff voltage is high (Figure \ref{fig:dcr_knee}).\cite{ma_editors_2019, mandli_analysis_2019}
Note that SEI growth does not occur appreciably during discharge in carbonaceous negative electrodes\cite{attia_electrochemical_2019, das_electrochemical_2019}.

In other systems, an increased discharging rate can delay the onset of the knee.
Keil et al.\cite{keil_linear_2019} found that increasing discharging current from 1C to 2C led to the elimination of the knee in nickel manganese cobalt oxide (NMC)/graphite cylindrical cells (Figure \ref{fig:discharge-rest_cycle}b).
Similarly, Atalay et al.\cite{atalay_theory_2020} found that increasing the discharge rate from 1C to 4C decelerated the knee point for nickel cobalt aluminum oxide (NCA)/graphite cylindrical cells.
Lastly, Keil et al.\cite{keil_charging_2016} illustrated how discharging current had no effect on cylindrical cells with blended transition metal oxide positive electrodes and graphite negative electrodes, but a lower discharging current (3A, 2.7C) led to faster degradation than a higher discharging current (5A, 4.5C) for an LFP/graphite cylindrical cell when charged at 4.5C.
The authors did not identify a mechanism. 

While more work is needed to understand these results, one hypothesis for these observations is decreased calendar aging for cells with faster discharge rates.
In other words, cells with less time spent cycling simply have less calendar aging. In Figure S2b, we revisualized the Keil et al.\cite{keil_linear_2019} dataset shown in Figure \ref{fig:discharge-rest_cycle}b using estimated time on the $x$ axis; we found that the knee locations appeared closer together, suggesting calendar aging is at least partially responsible for the discharge rate sensitivity. This hypothesis further highlights the sensitivity of the apparent severity of the knee to the choice of $x$ axis (as illustrated in Figure 2).

\subsubsection{Voltage limits} 
\begin{figure}[ht!]
\centering
\includegraphics[scale=1.0]{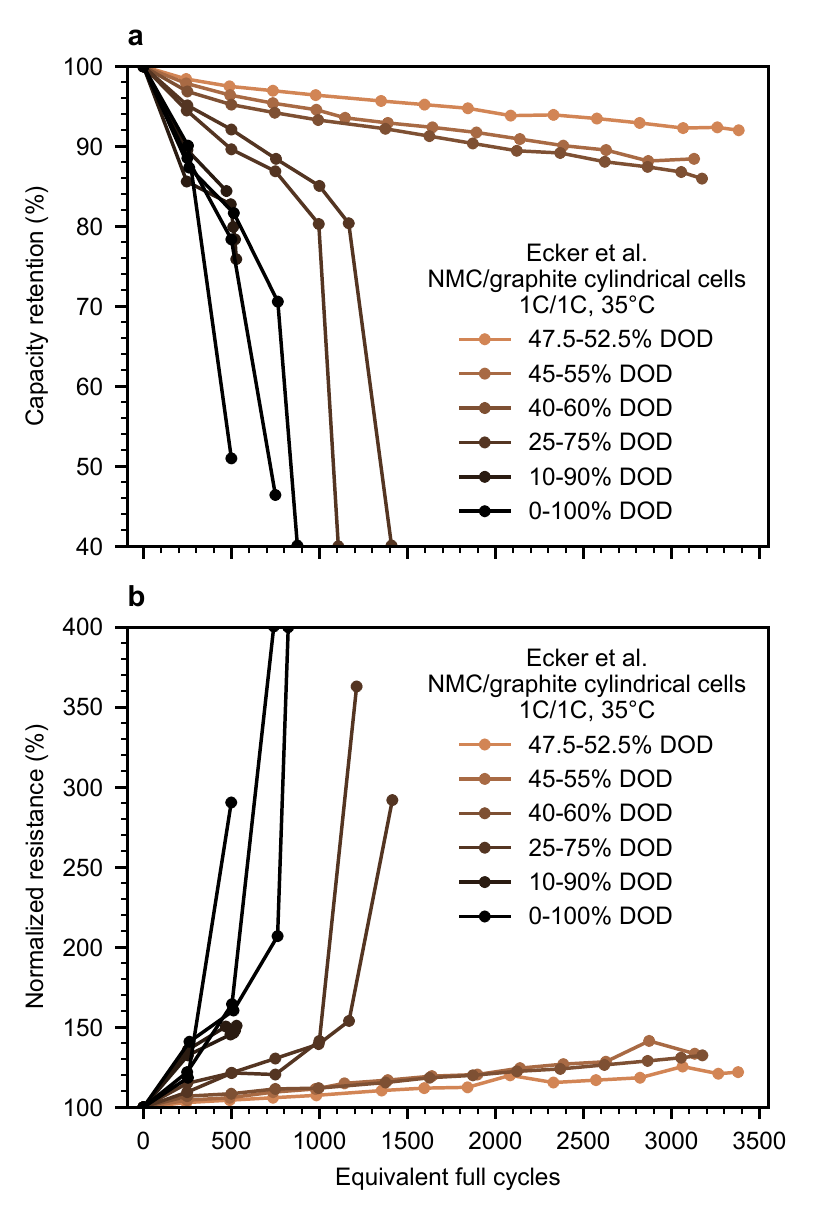}
\caption{Sensitivity of knees to voltage window/depth of discharge, and the correlation between capacity knees and ``resistance elbows''.
(a) Normalized capacity and (b) normalized resistance vs. equivalent full cycles for NMC/graphite cylindrical cells cycled at 1 C/1C and 35$^{\circ}$C.
Different depths of discharge around a mean SOC of 50\% are compared. Each trend displays a single cell. The number of equivalent full cycles increased for cells with smaller depths of discharge. Furthermore, the correlation between capacity knees and resistance elbows is evident.
Adapted from Figure 12 of Ecker et al.\cite{ecker_calendar_2014}
}
\label{fig:ecker_capacity_and_resistance}
\end{figure}

A wider voltage window generally accelerates the onset of the knee point \cite{ecker_calendar_2014, pfrang_long-term_2018, klett_non-uniform_2014, ma_novel_2019, petzl_lithium_2015, schuster_nonlinear_2015}. In one of the broadest studies, Ecker et al. \cite{ecker_calendar_2014} considered six DODs (100, 80, 50, 20, 10, and 0.5 \%) with up to six voltage windows per DOD. The authors found that the EFC systematically decreased with increased DOD (Figure \ref{fig:ecker_capacity_and_resistance}). By 1000 EFC, all cells with DODs greater than 25-75\% had a capacity below 80\% and exhibited a knee. When varying the voltage window with fixed DOD, the authors observed the highest degradation in cells cycled at low and high SOCs; the lowest degradation was observed for a midpoint SOC of 50\%. Other studies also observed accelerated degradation at extreme SOCs. \cite{aiken_accelerated_2020,ma_novel_2019, zhu_investigation_2021}

The impact of the voltage window on knee onset is typically attributed to resistance growth stemming from enhanced expansion and cracking of the positive electrode during intercalation, driving electrolyte oxidation \cite{broussely_main_2005, ma_editors_2019, aiken_accelerated_2020}. For some positive electrodes, transition metal migration (often manganese or iron) may also be exacerbated by high voltages. \cite{joshi_effects_2014, gilbert_transition_2017, ma_novel_2019}

\subsubsection{Rests}

\begin{figure}[ht!]
\centering
\includegraphics[scale = 1.0]{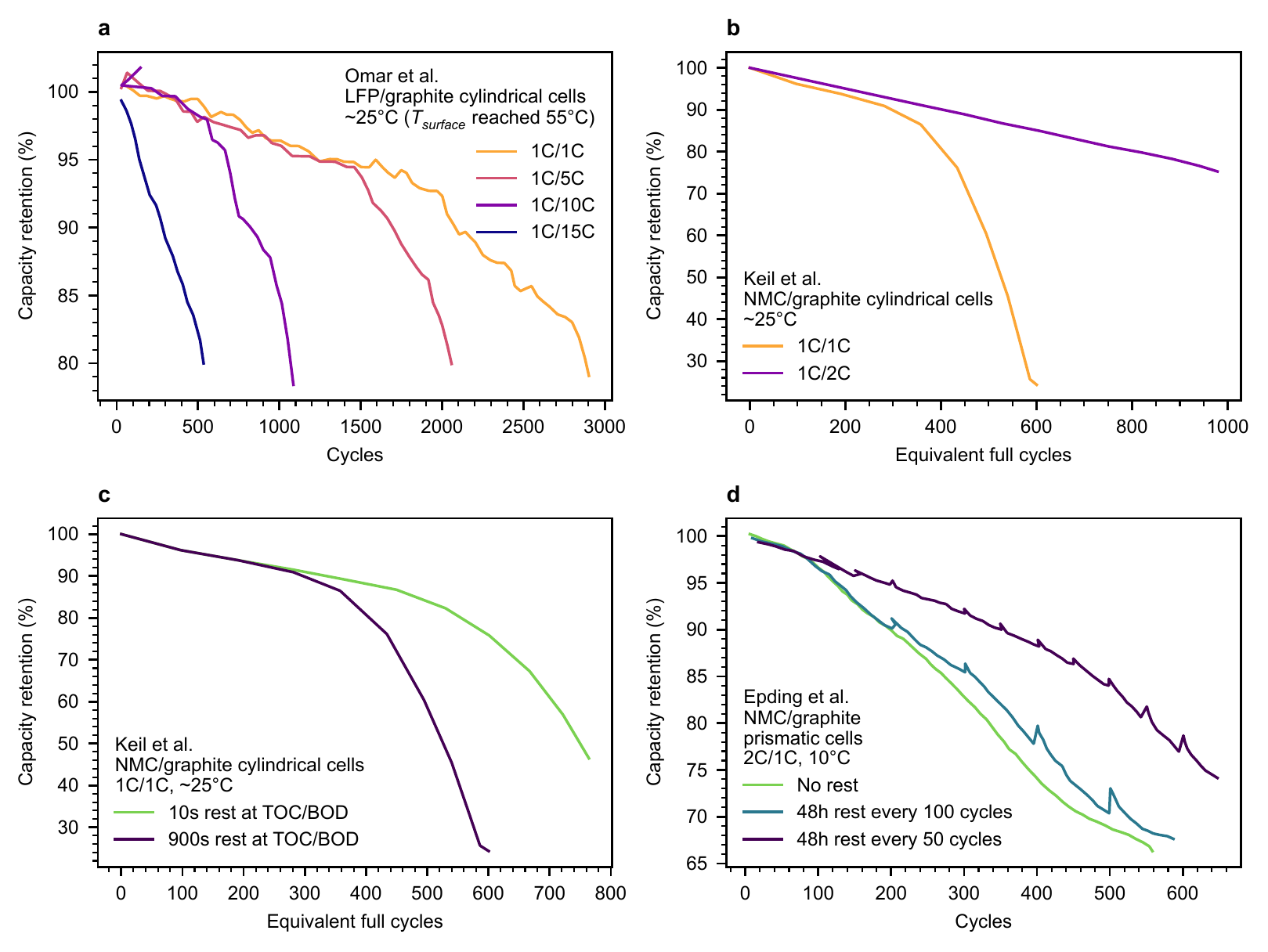}
\caption{Mixed effects of discharge rate and rest time on knee onset, depending on the testing conditions. (a) Higher discharge rate can accelerate knee onset. Adapted from Figure 8 of Omar et al.\cite{omar_lithium_2014}
Note that while the environmental test temperature was $\sim$25\degree C, the temperature at the surface reached as high as 55\degree C for the high rate discharge tests.
(b) Lower discharge rate can accelerate knee onset. Adapted from Figure 2a of Keil et al.\cite{keil_linear_2019} (c) Longer rest time can accelerate knee onset. Here, TOC and BOD refer to top-of-charge and bottom-of-discharge, respectively. Adapted from  Figure 2a of Keil et al.\cite{keil_linear_2019} (d) Shorter rest time can accelerate knee onset. Adapted from Figure 1a of Epding et al.\cite{epding_investigation_2019}. To isolate the influence of calendar aging, these data are replotted with an estimated time axis in Figure S2.}
\label{fig:discharge-rest_cycle}
\end{figure}

Like discharging rate, the effect of rests during cycling on the knee occurrence is mixed (Figure \ref{fig:discharge-rest_cycle}c--d).
Keil et al.\cite{keil_linear_2019} found that decreasing rest time from 900 seconds to 10 seconds after both charge and discharge delayed the knee in NMC/graphite cylindrical cells (Figure \ref{fig:discharge-rest_cycle}c).
Ma et al.\cite{ma_editors_2019} found a similar result: removing the 30-minute rests after both charge and discharge delayed the knee, but only with an upper cutoff voltage of 4.3 V. The rest time had no effect at 4.1 V.
These observations were rationalized by less time at high potential when plotted as a function of cycle number, which can induce knees driven by side reaction product buildup.
In fact, if the data in Figure \ref{fig:discharge-rest_cycle}c is replotted as a function of estimated time instead of cycle number, the curves become much more similar (Figure S2c)---suggesting that the degradation is primarily driven by time and not cycle number.
In contrast, Epding et al.\cite{epding_investigation_2019} found that longer rest times between cycles delayed the knee occurrence (Figure \ref{fig:discharge-rest_cycle}d). The authors proposed that these rests offered reversibly plated lithium time to reintercalate; another possibility is that the rests allowed for greater utilization of cycleable lithium\cite{rashid_effect_2015}.
Note that periodic rests interspersed throughout a cycling test can substantially improve lifetime in some cases\cite{raj_investigation_2020}.

Increased rest periods may delay the onset of knees driven by high overpotentials (e.g., rate-dependent lithium plating at high rates or low temperatures).
For knees driven by side reaction product buildup (e.g., porosity decrease, positive electrode resistance growth, etc.), increased rest may be beneficial or harmful for lifetime.
Longer rests at high temperature and high state of charge may exacerbate the onset of these knees; in fact, Ecker et al.\cite{ecker_calendar_2014} found that NMC/graphite cells stored at 50\degree C, 100\% SOC (4.162 V) exhibited a knee after 100 days of storage, while cells stored at 50\degree C and  85\% SOC (4.018 V) or below did not exhibit knees even after 400 days of storage.
However, increased LLI from high temperature, high SOC rests may delay electrode saturation knees by increasing the amount of LAM required for knee onset.\cite{mao_calendar_2017, uddin_viability_2018}
Increased rest after high rate cycling will also allow for the internal temperature of the cell to decrease, which will decrease side reaction rates and delay knee onset; fully accounting for the relationship between internal cell temperature, heating due to (dis)charging, and cooling during rests requires detailed cell thermal characterization and modeling.
More work is needed to understand the sensitivity of knees to rest at both low and high state of charge across different cell designs and usage conditions.

\subsubsection{Temperature}
The effect of temperature on knee onset depends on the test conditions. For example, some studies have found that knee onset is minimized at 25\degree C \cite{zhang_accelerated_2019, waldmann_temperature_2014, waldmann_optimization_2015} or 35\degree C \cite{schuster_nonlinear_2015}{}. At lower temperatures, knees are primarily attributed to lithium plating; at elevated temperatures, knees are attributed to side reaction mechanisms, such as lithium plating induced by SEI growth-induced porosity decrease, positive electrode resistance growth, electolyte depletion, and additive depletion (Figure \ref{fig:temperature_and_pressure}a).\cite{broussely_main_2005, zhang_accelerated_2019,schuster_nonlinear_2015,waldmann_temperature_2014,waldmann_optimization_2015,stevens_using_2014} In general, the temperature that minimizes degradation is reportedly lower for LFP cells than NMC cells \cite{preger_degradation_2020} and lower for power cells than energy cells\cite{yang_understanding_2018}.

\subsubsection{Pressure}
\begin{figure}[ht!]
\centering
\includegraphics[scale = 1.0]{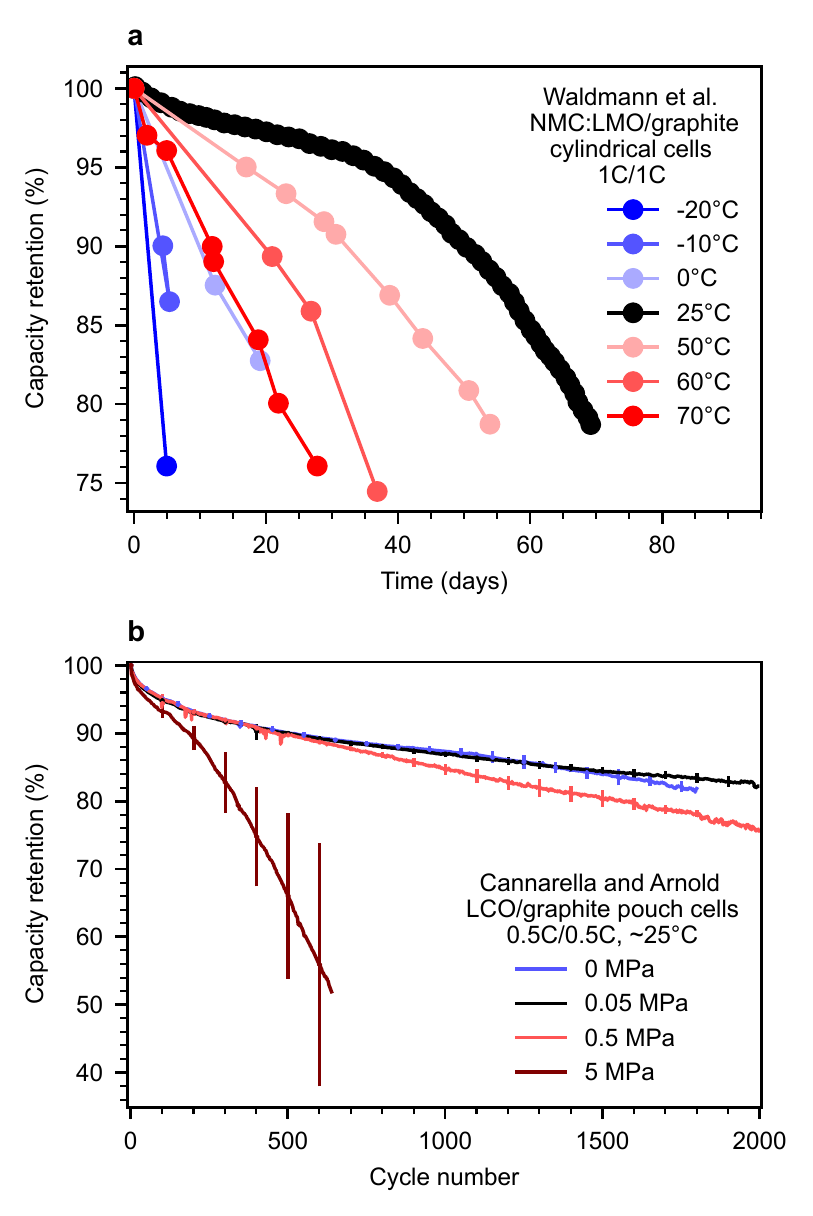}
\caption{Transition in degradation mechanism based on (a) environmental temperature and (b) applied pressure, illustrating an intermediate optimum for each variable that balances two competing degradation mechanisms.
In panel (a), only data measured at 25°C is shown (i.e., all points for the 25°C data and periodic measurements for data collected at other temperatures).
In panel (b), error bars represent one standard deviation of the average of three cells. 
Adapted from Figure 2 of Waldmann et al.\cite{waldmann_temperature_2014}{} and Figure 5 of Cannarella and Arnold\cite{cannarella_stress_2014}{}.}
\label{fig:temperature_and_pressure}
\end{figure}

Like temperature, studies of pressure dependence in pouch and prismatic cells have demonstrated that lifetime performance is optimized at an intermediate value. Cannarella and Arnold\cite{cannarella_stress_2014} demonstrated that the knee point can be accelerated by either an absence or an excess of pressure (Figure \ref{fig:temperature_and_pressure}b).
Additionally, Wünsch et al.\cite{wunsch_investigation_2019}. increased the cycle life of 37 Ah NMC/graphite pouch cells from 500 cycles (no bracing) to 3200 cycles (optimal spring compression) while investigating various methods of bracing.

For pouch and prismatic cells, some applied pressure is needed to enhance ionic and electronic conductivity and particle contact. However, when too much pressure is applied, the local porosity will decrease, which can drive rate-dependent lithium plating.
Furthermore, mechanical stress from high pressure can be unevenly distributed throughout the cell. This heterogeneity can drive delamination, surface film formation, and uneven lithium distribution and can cause heterogeneous lithium plating.
Heterogeneous compression in the test fixture can accelerate the knee point even in cylindrical cells \cite{bach_nonlinear_2016}.

\section{Modeling and predicting knees}

Finally, with both a fundamental and empirical understanding in hand, we turn to knee modeling and prediction efforts.
First, we examine the relationship between knees and the number of cycles to end-of-life, as well as the relationship between knees and resistance elbows.
We then offer an outlook on knee modeling and prediction work based on our findings in this work.


\subsubsection{Empirical relationship between knees, resistance elbows, and end-of-life}

\begin{figure}[ht]
\centering
\includegraphics[scale=1.0]{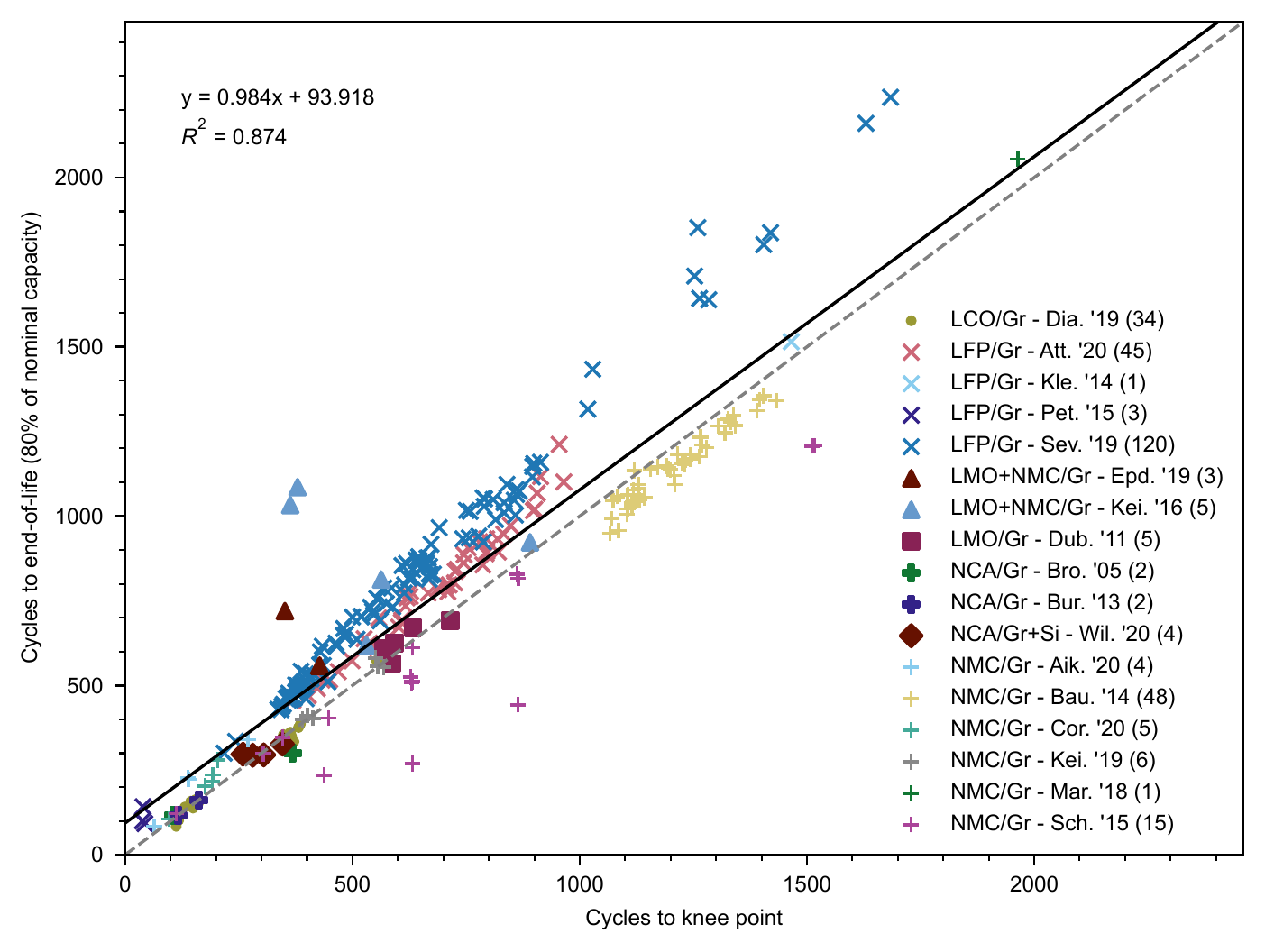}
  \label{fig:kneepoint2EOL}
\caption{Relationship between knee point and capacity end-of-life (defined as 80\% of nominal capacity) for a total of 303 cells across 17 datasets. The two metrics are linearly correlated ($R^2 = 0.874$). The Bacon-Watts algorithm was employed for knee point identification. See Table SI for details on the data.}
\label{fig:knees2EOL}
\end{figure}

While predicting knees is important, predicting end-of-life is more directly relevant for estimating product performance and warranty costs. In Figure \ref{fig:knees2EOL}, we display the relationship between knee locations and end-of-life (defined as 80\% of nominal capacity) across 17 datasets (303 cells) with knees and different knee pathways, cell types (chemistry, geometry, size, lab-made vs. commercial, etc.), and testing conditions. The data used to generate Figure \ref{fig:knees2EOL} is summarized in Table SI; data was obtained via direct access to the corresponding databases when possible\cite{baumhofer_production_2014,diao_accelerated_2019,severson_data-driven_2019,willenberg_high-precision_2020,attia_closed-loop_2020} or via \textit{WebPlotDigitizer}\cite{Rohatgi2021}{}.
We find a strong linear relationship from the knee point to the end-of-life ($R^2=0.874$). Apart from the data presented in Schuster et al.\cite{schuster_nonlinear_2015}{}, the knee points take place at or before the end-of-life.
Thus, predicting the knee location can often provide an estimate of end-of-life as well.
For the Severson et al.\cite{severson_data-driven_2019} method, this correlation holds independent of the knee point identification algorithm used (Figure S3).
However, we note that the reverse relationship (using the knee location to predict end-of-life) does not necessarily hold, as we only selected cells with knees in this study and many cells reach 80\% capacity without exhibiting a knee (i.e., exhibit linear or sublinear degradation).


We also explored the relationship between capacity knees and resistance ``elbows''.
Many aging studies have shown that capacity knee onset is nearly always correlated to the onset of rapid resistance growth, even across different cell chemistries, form factors, and test protocols. Table \ref{tab:dcr_growth_papers} reports the relative capacity and resistance at the knee onset point from several studies that reported both capacity fade and resistance growth; all studies measured resistance using a direct-current pulse except for Schuster et al. \cite{schuster_nonlinear_2015}{}, which used electrochemical impedance spectroscopy. Of all these aging studies, cells with capacity knees always displayed resistance elbows, and with the exception of the work by Martinez-Laserna et al. \cite{martinez-laserna_technical_2018}{}, the reverse is also true. For instance, the Ecker et al.\cite{ecker_calendar_2014} dataset previously discussed (Figure \ref{fig:ecker_capacity_and_resistance}) found a strong correlation between capacity knees and resistance elbows.
However, we cannot infer causality without more careful study; in many cases, the resistance elbow may occur with the capacity knee simply because the local current densities have increased with the loss of active capacity.
We expect that a resistance elbow will almost always follow a capacity knee, but a capacity knee will not necessarily follow a resistance elbow (i.e., during low rate cycling).
We note that our resistance growth-induced knee pathway (Figure \ref{fig:dcr_knee}) illustrated a capacity knee driven by linearly increasing resistance.
This correlation may provide opportunities to estimate either capacity or resistance from measurements of the other, e.g., in situations where one of these values is easy to measure and the other is challenging.
Lastly, this correlation highlights the value of periodic diagnostic tests at multiple rates to disentangle low-rate and high-rate effects.

\subsection{Modeling and prediction outlook}

As this work has demonstrated, knees in lithium-ion batteries are complex given the variety of knee pathways, internal state trajectories, and the combined effects of interactions, heterogeneity, and variation. While three of the knee pathways (lithium plating, electrode saturation, and resistance growth) are largely dependent on bulk internal states (i.e., LLI, LAM, and resistance) and are thus straightforward to detect and model via electrochemistry, three knee pathways (electrolyte and additive depletion, percolation-limited connectivity, and microscale mechanical deformation) involve subtle effects that are challenging to detect via electrochemical signals (e.g., porosity decrease, remaining additive amount). 
Additionally, the nature of the three internal state trajectories that cause knees also pose major challenges (i.e., extrapolation of an exponential function for snowball trajectories, simultaneously tracking two internal states for hidden trajectories, etc.).
Furthermore, interactions, heterogeneity, and variability add additional layers of difficulty: accurate modeling of these knee phenomena requires an understanding of simultaneous interactions between many degradation mechanisms over multiple length scales, heterogeneous electrochemical/thermal/mechanical gradients within cells, and cell-to-cell variation along multiple dimensions of cell design and manufacturing processes.
A final complication is translating lab testing results to a variety of field usage conditions.
Given this complexity, modeling and predicting knees is undoubtedly a formidable task.

Despite these challenges, we remain optimistic about avenues for improving knee modeling and prediction. The primary goal of this work is to lay the foundation for a fundamental understanding of the physics of knees; this understanding enables an accurate assessment of the limits of today’s models and opportunities for future work. Our findings suggest that many knee pathways, such as electrode saturation or resistance growth, can be readily predicted from physics-driven or ``degradation mode'' (e.g., 'Alawa \cite{dubarry_synthesize_2012, dubarry_big_2020}) modeling. For other knee pathways, such as plating due to porosity decrease and additive depletion, the modeling is straightforward, but the methods to estimate internal state are not. We hope this work inspires future research into a comprehensive knee prediction toolkit that generalizes over a wide range of cell designs and use cases. 

Lastly, we discuss the role of data-driven methods in this space.
Data-driven models may be well suited for knee pathways with bulk electrochemical signals; for instance, the high predictive performance of features sourced from the cell voltage responses during discharge in Severson et al.\cite{severson_data-driven_2019} was largely attributed to the detectability of the knee pathway via bulk electrochemical signals.
However, quantifying the values of bulk internal states from field data remains challenging\cite{aitio_predicting_2021, bian_state--health_2021, sulzer_challenge_2021}.
Furthermore, data-driven models trained on cycling data will naturally be poorly suited for knee pathways with signals that are challenging to measure via electrochemistry.
To this end, two promising research directions are to develop input perturbations that magnify the response of hard-to-detect signals and to introduce additional characterization techniques into cell cycling tests.
Our review has highlighted the benefits of periodic low-rate diagnostic cycles; perhaps more targeted electrochemical input signals could provide major benefits.
As a whole, datasets that span a variety of knee pathways for various cell designs and use cases are needed for training generalizable data-driven models. Generation of synthetic data sets using physics-based models may be well suited for training generalizable models reliant on electrochemical signals to identify knee pathways or predict knees.\cite{dubarry_big_2020, kim_rapid_2021}

\section{Conclusions and future work}

Knees are a major challenge to developing long-lasting lithium-ion batteries.
In this work, we review the topic of ``knees''---i.e., superlinear aging trajectories---in lithium-ion battery lifetime aging trajectories (e.g., capacity vs. cycle number). We first define knees, illustrate the sensitivity of knees to the $x$ and $y$ variables used, and compare various knee point estimation algorithms. We then categorize knees presented in the literature into one of six knee ``pathways'' (lithium plating, electrode saturation, resistance growth, electrolyte and additive depletion, percolation-limited connectivity, and mechanical deformation) and one of three ``internal state trajectories'' (snowball, hidden, and threshold). Each of these pathway-trajectory pairs has different implications for modeling and prediction; while some pairs have internal states that can be measured and modeled via standard electrochemical signals and models, others are dependent on internal states that are challenging to detect via typical electrochemical signals (e.g., remaining additive amounts, local porosity distributions, etc.). We also evaluate the role of interactions, heterogeneity, and variation on knees, which add additional layers of complexity. Next, we discuss key cell design and usage conditions levers on the location of knees. Finally, we consider the outlook for knee modeling and prediction. Overall, accurate knee modeling and prediction is quite challenging, but we hope this work provides a starting point for a comprehensive knee prediction framework.

Our findings suggest much future work is needed on this topic.
First, a better understanding of the fundamentals of knee pathways and internal state trajectories will enable more accurate modeling and data generation efforts. Both experimental and modeling efforts can improve our understanding of knee fundamentals and reveal other knee pathways and internal state trajectories not captured in this work.
In particular, the ``covering layer'' phenomenon observed concurrently with knees in many studies deserves further study.
Second, larger battery aging datasets over a variety of cell designs and use cases will aid fundamental, modeling, and prediction efforts to capture a variety of knee pathways and internal state trajectories. Synthetic, lab-generated, and field-generated datasets would all be useful for this purpose.
Third, new characterization probes (or new state estimation methods using existing probes) that can capture subtle changes in internal state, such as local electrode porosity or remaining additive amount---ideally nondestructively---would enable quantitative detection of internal state trajectories over the lifetime of a cell.
Lastly, a variety of modeling and prediction approaches, spanning the physics-driven to data-driven continuum, may unlock accurate lifetime estimation to capture the most challenging lithium-ion battery degradation modes.
Many of these research directions will require substantial effort; however, with focused work on developing our understanding of knees, lithium-ion batteries can be designed for target applications and deployed with confidence that knees will be avoided within their lifetimes.

\section{Data and code availability}

All data and code are publicly available on GitHub (https://github.com/tinosulzer/kneepoint-review).

\section{Acknowledgments}

We thank Dr. Stephen Harris for providing the data used in Figure \ref{fig:var_exp}a and Dr. John Cannarella for providing the data used in Figure \ref{fig:temperature_and_pressure}b.
F.B.P. was supported by the Faraday Institution [EP/S003053/1 grant numbers FIRG003 and FIRG025]
P.D was supported by Bundesministerium für Bildung und Forschung (BMBF 03XP0302C).
G.d.R. acknowledges support from the {Funda{\c c}$\tilde{\text{a}}$o para a Ci$\hat{e}$ncia e a Tecnologia} (Portuguese Foundation for Science and Technology) through the project UIDB/00297/2020 (Centro de Matem\'atica e Aplica\c c$\tilde{\text{o}}$es CMA/FCT/UNL).
P.G. is supported by National Renewable Energy Laboratory which is operated by Alliance for Sustainable Energy, LLC, for the U.S. Department of Energy (DOE) under Contract No. DE-AC36-08GO28308. 
E.K. acknowledges funding by Agency for Science, Technology and Research (A*STAR) under the Career Development Fund (C210112037).
Y.P. was supported by the US Department of Energy Office of Electricity, Energy Storage Program under the direction of Dr.~Imre Gyuk. Sandia National Laboratories is a multi-mission laboratory managed and operated by National Technology and Engineering Solutions of Sandia, LLC., a wholly owned subsidiary of Honeywell International, Inc., for the U.S.~Department of Energy’s National Nuclear Security Administration under contract DE-NA-0003525.
This paper describes objective technical results and analysis. Any subjective views or opinions that might be expressed in the paper do not necessarily represent the views of the U.S.~Department of Energy or the United States Government.
S.S. thanks support from the Carnegie Mellon University Presidential Fellowship.

\section{CRediT author contributions statement}


Conceptualization, P.M.A.;
Data curation, P.M.A., P.D., P.G., R.G., Y.P.;
Investigation, P.M.A., P.D., G.d.R., P.G., Y.P., S.S.;
Methodology, P.M.A., G.d.R., S.S., V.S.;
Software, P.M.A., R.G., S.G., V.S.;
Supervision, P.M.A., D.A.H.;
Visualization, P.M.A., P.D., P.G., S.G., A.S., S.S., V.S.; 
Writing – original draft, P.M.A., A.B., F.B.P., P.D., G.d.R., P.G., S.G., O.L., E.K., Y.P., A.S., S.S., V.S.; 
Writing – review \& editing, P.M.A., G.d.R., M.D., P.G., S.G., D.A.H., E.K., Y.P., A.S., S.S., A.G.S.

\newpage
\bibliography{refs_zotero}

\providecommand{\latin}[1]{#1}
\makeatletter
\providecommand{\doi}
  {\begingroup\let\do\@makeother\dospecials
  \catcode`\{=1 \catcode`\}=2 \doi@aux}
\providecommand{\doi@aux}[1]{\endgroup\texttt{#1}}
\makeatother
\providecommand*\mcitethebibliography{\thebibliography}
\csname @ifundefined\endcsname{endmcitethebibliography}
  {\let\endmcitethebibliography\endthebibliography}{}
\begin{mcitethebibliography}{173}
\providecommand*\natexlab[1]{#1}
\providecommand*\mciteSetBstSublistMode[1]{}
\providecommand*\mciteSetBstMaxWidthForm[2]{}
\providecommand*\mciteBstWouldAddEndPuncttrue
  {\def\EndOfBibitem{\unskip.}}
\providecommand*\mciteBstWouldAddEndPunctfalse
  {\let\EndOfBibitem\relax}
\providecommand*\mciteSetBstMidEndSepPunct[3]{}
\providecommand*\mciteSetBstSublistLabelBeginEnd[3]{}
\providecommand*\EndOfBibitem{}
\mciteSetBstSublistMode{f}
\mciteSetBstMaxWidthForm{subitem}{(\alph{mcitesubitemcount})}
\mciteSetBstSublistLabelBeginEnd
  {\mcitemaxwidthsubitemform\space}
  {\relax}
  {\relax}

\bibitem[Hesse \latin{et~al.}(2017)Hesse, Schimpe, Kucevic, and
  Jossen]{hesse_lithium-ion_2017}
Hesse,~H.~C.; Schimpe,~M.; Kucevic,~D.; Jossen,~A. Lithium-{Ion} {Battery}
  {Storage} for the {Grid}—{A} {Review} of {Stationary} {Battery} {Storage}
  {System} {Design} {Tailored} for {Applications} in {Modern} {Power} {Grids}.
  \emph{Energies} \textbf{2017}, \emph{10}, 2107, Number: 12 Publisher:
  Multidisciplinary Digital Publishing Institute\relax
\mciteBstWouldAddEndPuncttrue
\mciteSetBstMidEndSepPunct{\mcitedefaultmidpunct}
{\mcitedefaultendpunct}{\mcitedefaultseppunct}\relax
\EndOfBibitem
\bibitem[Bocca and Baek(2020)Bocca, and Baek]{bocca_optimal_2020}
Bocca,~A.; Baek,~D. Optimal {Life}-{Cycle} {Costs} of {Batteries} for
  {Different} {Electric} {Cars}. 2020 {AEIT} {International} {Conference} of
  {Electrical} and {Electronic} {Technologies} for {Automotive} ({AEIT}
  {AUTOMOTIVE}). 2020; pp 1--6\relax
\mciteBstWouldAddEndPuncttrue
\mciteSetBstMidEndSepPunct{\mcitedefaultmidpunct}
{\mcitedefaultendpunct}{\mcitedefaultseppunct}\relax
\EndOfBibitem
\bibitem[Beltran \latin{et~al.}(2020)Beltran, Ayuso, and
  Pérez]{beltran_lifetime_2020}
Beltran,~H.; Ayuso,~P.; Pérez,~E. Lifetime {Expectancy} of {Li}-{Ion}
  {Batteries} used for {Residential} {Solar} {Storage}. \emph{Energies}
  \textbf{2020}, \emph{13}, 568, Number: 3 Publisher: Multidisciplinary Digital
  Publishing Institute\relax
\mciteBstWouldAddEndPuncttrue
\mciteSetBstMidEndSepPunct{\mcitedefaultmidpunct}
{\mcitedefaultendpunct}{\mcitedefaultseppunct}\relax
\EndOfBibitem
\bibitem[Harlow \latin{et~al.}(2019)Harlow, Ma, Li, Logan, Liu, Zhang, Ma,
  Glazier, Cormier, Genovese, Buteau, Cameron, Stark, and
  Dahn]{harlow_wide_2019}
Harlow,~J.~E.; Ma,~X.; Li,~J.; Logan,~E.; Liu,~Y.; Zhang,~N.; Ma,~L.;
  Glazier,~S.~L.; Cormier,~M. M.~E.; Genovese,~M. \latin{et~al.}  A {Wide}
  {Range} of {Testing} {Results} on an {Excellent} {Lithium}-{Ion} {Cell}
  {Chemistry} to be used as {Benchmarks} for {New} {Battery} {Technologies}.
  \emph{Journal of The Electrochemical Society} \textbf{2019}, \emph{166},
  A3031, Publisher: IOP Publishing\relax
\mciteBstWouldAddEndPuncttrue
\mciteSetBstMidEndSepPunct{\mcitedefaultmidpunct}
{\mcitedefaultendpunct}{\mcitedefaultseppunct}\relax
\EndOfBibitem
\bibitem[Harper \latin{et~al.}(2019)Harper, Sommerville, Kendrick, Driscoll,
  Slater, Stolkin, Walton, Christensen, Heidrich, Lambert, Abbott, Ryder,
  Gaines, and Anderson]{harper_recycling_2019}
Harper,~G.; Sommerville,~R.; Kendrick,~E.; Driscoll,~L.; Slater,~P.;
  Stolkin,~R.; Walton,~A.; Christensen,~P.; Heidrich,~O.; Lambert,~S.
  \latin{et~al.}  Recycling lithium-ion batteries from electric vehicles.
  \emph{Nature} \textbf{2019}, \emph{575}, 75--86, Bandiera\_abtest: a
  Cg\_type: Nature Research Journals Number: 7781 Primary\_atype: Reviews
  Publisher: Nature Publishing Group Subject\_term: Batteries;Carbon and
  energy;Energy and society Subject\_term\_id:
  batteries;carbon-and-energy;energy-and-society\relax
\mciteBstWouldAddEndPuncttrue
\mciteSetBstMidEndSepPunct{\mcitedefaultmidpunct}
{\mcitedefaultendpunct}{\mcitedefaultseppunct}\relax
\EndOfBibitem
\bibitem[Ma \latin{et~al.}(2019)Ma, Harlow, Li, Ma, Hall, Buteau, Genovese,
  Cormier, and Dahn]{ma_editors_2019}
Ma,~X.; Harlow,~J.~E.; Li,~J.; Ma,~L.; Hall,~D.~S.; Buteau,~S.; Genovese,~M.;
  Cormier,~M.; Dahn,~J.~R. Editors' {Choice}—{Hindering} {Rollover} {Failure}
  of {Li}[{Ni} $_{\textrm{0.5}}$ {Mn} $_{\textrm{0.3}}$ {Co} $_{\textrm{0.2}}$
  ]{O} $_{\textrm{2}}$ /{Graphite} {Pouch} {Cells} during {Long}-{Term}
  {Cycling}. \emph{Journal of The Electrochemical Society} \textbf{2019},
  \emph{166}, A711--A724\relax
\mciteBstWouldAddEndPuncttrue
\mciteSetBstMidEndSepPunct{\mcitedefaultmidpunct}
{\mcitedefaultendpunct}{\mcitedefaultseppunct}\relax
\EndOfBibitem
\bibitem[Keil and Jossen(2020)Keil, and Jossen]{keil_electrochemical_2020}
Keil,~J.; Jossen,~A. Electrochemical {Modeling} of {Linear} and {Nonlinear}
  {Aging} of {Lithium}-{Ion} {Cells}. \emph{Journal of The Electrochemical
  Society} \textbf{2020}, \emph{167}, 110535, Publisher: IOP Publishing\relax
\mciteBstWouldAddEndPuncttrue
\mciteSetBstMidEndSepPunct{\mcitedefaultmidpunct}
{\mcitedefaultendpunct}{\mcitedefaultseppunct}\relax
\EndOfBibitem
\bibitem[Preger \latin{et~al.}(2020)Preger, Barkholtz, Fresquez, Campbell,
  Juba, Romàn-Kustas, Ferreira, and Chalamala]{preger_degradation_2020}
Preger,~Y.; Barkholtz,~H.~M.; Fresquez,~A.; Campbell,~D.~L.; Juba,~B.~W.;
  Romàn-Kustas,~J.; Ferreira,~S.~R.; Chalamala,~B. Degradation of {Commercial}
  {Lithium}-{Ion} {Cells} as a {Function} of {Chemistry} and {Cycling}
  {Conditions}. \emph{Journal of The Electrochemical Society} \textbf{2020},
  \emph{167}, 120532\relax
\mciteBstWouldAddEndPuncttrue
\mciteSetBstMidEndSepPunct{\mcitedefaultmidpunct}
{\mcitedefaultendpunct}{\mcitedefaultseppunct}\relax
\EndOfBibitem
\bibitem[Bloom \latin{et~al.}(2001)Bloom, Cole, Sohn, Jones, Polzin, Battaglia,
  Henriksen, Motloch, Richardson, Unkelhaeuser, Ingersoll, and
  Case]{bloom_accelerated_2001}
Bloom,~I.; Cole,~B.~W.; Sohn,~J.~J.; Jones,~S.~A.; Polzin,~E.~G.;
  Battaglia,~V.~S.; Henriksen,~G.~L.; Motloch,~C.; Richardson,~R.;
  Unkelhaeuser,~T. \latin{et~al.}  An accelerated calendar and cycle life study
  of {Li}-ion cells. \emph{Journal of Power Sources} \textbf{2001}, \emph{101},
  238--247\relax
\mciteBstWouldAddEndPuncttrue
\mciteSetBstMidEndSepPunct{\mcitedefaultmidpunct}
{\mcitedefaultendpunct}{\mcitedefaultseppunct}\relax
\EndOfBibitem
\bibitem[Broussely \latin{et~al.}(2001)Broussely, Herreyre, Biensan, Kasztejna,
  Nechev, and Staniewicz]{broussely_aging_2001}
Broussely,~M.; Herreyre,~S.; Biensan,~P.; Kasztejna,~P.; Nechev,~K.;
  Staniewicz,~R. Aging mechanism in {Li} ion cells and calendar life
  predictions. \emph{Journal of Power Sources} \textbf{2001}, \emph{97-98},
  13--21\relax
\mciteBstWouldAddEndPuncttrue
\mciteSetBstMidEndSepPunct{\mcitedefaultmidpunct}
{\mcitedefaultendpunct}{\mcitedefaultseppunct}\relax
\EndOfBibitem
\bibitem[Wright \latin{et~al.}(2002)Wright, Motloch, Belt, Christophersen, Ho,
  Richardson, Bloom, Jones, Battaglia, Henriksen, Unkelhaeuser, Ingersoll,
  Case, Rogers, and Sutula]{wright_calendar-_2002}
Wright,~R.~B.; Motloch,~C.~G.; Belt,~J.~R.; Christophersen,~J.~P.; Ho,~C.~D.;
  Richardson,~R.~A.; Bloom,~I.; Jones,~S.~A.; Battaglia,~V.~S.;
  Henriksen,~G.~L. \latin{et~al.}  Calendar- and cycle-life studies of advanced
  technology development program generation 1 lithium-ion batteries.
  \emph{Journal of Power Sources} \textbf{2002}, \emph{110}, 445--470\relax
\mciteBstWouldAddEndPuncttrue
\mciteSetBstMidEndSepPunct{\mcitedefaultmidpunct}
{\mcitedefaultendpunct}{\mcitedefaultseppunct}\relax
\EndOfBibitem
\bibitem[Smith \latin{et~al.}(2011)Smith, Burns, Zhao, Xiong, and
  Dahn]{smith_high_2011}
Smith,~A.~J.; Burns,~J.~C.; Zhao,~X.; Xiong,~D.; Dahn,~J.~R. A {High}
  {Precision} {Coulometry} {Study} of the {SEI} {Growth} in {Li}/{Graphite}
  {Cells}. \emph{Journal of The Electrochemical Society} \textbf{2011},
  \emph{158}, A447--A452\relax
\mciteBstWouldAddEndPuncttrue
\mciteSetBstMidEndSepPunct{\mcitedefaultmidpunct}
{\mcitedefaultendpunct}{\mcitedefaultseppunct}\relax
\EndOfBibitem
\bibitem[Attia \latin{et~al.}(2020)Attia, Chueh, and
  Harris]{attia_revisiting_2020}
Attia,~P.~M.; Chueh,~W.~C.; Harris,~S.~J. Revisiting the t 0.5 {Dependence} of
  {SEI} {Growth}. \emph{Journal of The Electrochemical Society} \textbf{2020},
  \emph{167}, 090535, Publisher: The Electrochemical Society\relax
\mciteBstWouldAddEndPuncttrue
\mciteSetBstMidEndSepPunct{\mcitedefaultmidpunct}
{\mcitedefaultendpunct}{\mcitedefaultseppunct}\relax
\EndOfBibitem
\bibitem[Broussely \latin{et~al.}(2005)Broussely, Biensan, Bonhomme, Blanchard,
  Herreyre, Nechev, and Staniewicz]{broussely_main_2005}
Broussely,~M.; Biensan,~P.; Bonhomme,~F.; Blanchard,~P.; Herreyre,~S.;
  Nechev,~K.; Staniewicz,~R. Main aging mechanisms in {Li} ion batteries.
  \emph{Journal of Power Sources} \textbf{2005}, \emph{146}, 90--96\relax
\mciteBstWouldAddEndPuncttrue
\mciteSetBstMidEndSepPunct{\mcitedefaultmidpunct}
{\mcitedefaultendpunct}{\mcitedefaultseppunct}\relax
\EndOfBibitem
\bibitem[Diao \latin{et~al.}(2019)Diao, Saxena, Han, and
  Pecht]{diao_algorithm_2019}
Diao,~W.; Saxena,~S.; Han,~B.; Pecht,~M. Algorithm to {Determine} the {Knee}
  {Point} on {Capacity} {Fade} {Curves} of {Lithium}-{Ion} {Cells}.
  \emph{Energies} \textbf{2019}, \emph{12}, 2910\relax
\mciteBstWouldAddEndPuncttrue
\mciteSetBstMidEndSepPunct{\mcitedefaultmidpunct}
{\mcitedefaultendpunct}{\mcitedefaultseppunct}\relax
\EndOfBibitem
\bibitem[Fermín-Cueto \latin{et~al.}(2020)Fermín-Cueto, McTurk, Allerhand,
  Medina-Lopez, Anjos, Sylvester, and dos
  Reis]{fermin-cueto_identification_2020}
Fermín-Cueto,~P.; McTurk,~E.; Allerhand,~M.; Medina-Lopez,~E.; Anjos,~M.~F.;
  Sylvester,~J.; dos Reis,~G. Identification and machine learning prediction of
  knee-point and knee-onset in capacity degradation curves of lithium-ion
  cells. \emph{Energy and AI} \textbf{2020}, \emph{1}, 100006\relax
\mciteBstWouldAddEndPuncttrue
\mciteSetBstMidEndSepPunct{\mcitedefaultmidpunct}
{\mcitedefaultendpunct}{\mcitedefaultseppunct}\relax
\EndOfBibitem
\bibitem[Schuster \latin{et~al.}(2015)Schuster, Bach, Fleder, Müller, Brand,
  Sextl, and Jossen]{schuster_nonlinear_2015}
Schuster,~S.~F.; Bach,~T.; Fleder,~E.; Müller,~J.; Brand,~M.; Sextl,~G.;
  Jossen,~A. Nonlinear aging characteristics of lithium-ion cells under
  different operational conditions. \emph{Journal of Energy Storage}
  \textbf{2015}, \emph{1}, 44--53\relax
\mciteBstWouldAddEndPuncttrue
\mciteSetBstMidEndSepPunct{\mcitedefaultmidpunct}
{\mcitedefaultendpunct}{\mcitedefaultseppunct}\relax
\EndOfBibitem
\bibitem[Bach \latin{et~al.}(2016)Bach, Schuster, Fleder, Müller, Brand,
  Lorrmann, Jossen, and Sextl]{bach_nonlinear_2016}
Bach,~T.~C.; Schuster,~S.~F.; Fleder,~E.; Müller,~J.; Brand,~M.~J.;
  Lorrmann,~H.; Jossen,~A.; Sextl,~G. Nonlinear aging of cylindrical
  lithium-ion cells linked to heterogeneous compression. \emph{Journal of
  Energy Storage} \textbf{2016}, \emph{5}, 212--223\relax
\mciteBstWouldAddEndPuncttrue
\mciteSetBstMidEndSepPunct{\mcitedefaultmidpunct}
{\mcitedefaultendpunct}{\mcitedefaultseppunct}\relax
\EndOfBibitem
\bibitem[Yang \latin{et~al.}(2017)Yang, Leng, Zhang, Ge, and
  Wang]{yang_modeling_2017}
Yang,~X.-G.; Leng,~Y.; Zhang,~G.; Ge,~S.; Wang,~C.-Y. Modeling of lithium
  plating induced aging of lithium-ion batteries: {Transition} from linear to
  nonlinear aging. \emph{Journal of Power Sources} \textbf{2017}, \emph{360},
  28--40\relax
\mciteBstWouldAddEndPuncttrue
\mciteSetBstMidEndSepPunct{\mcitedefaultmidpunct}
{\mcitedefaultendpunct}{\mcitedefaultseppunct}\relax
\EndOfBibitem
\bibitem[Mandli \latin{et~al.}(2019)Mandli, Kaushik, Patil, Naha, Hariharan,
  Kolake, Han, and Choi]{mandli_analysis_2019}
Mandli,~A.~R.; Kaushik,~A.; Patil,~R.; Naha,~A.; Hariharan,~K.; Kolake,~S.;
  Han,~S.; Choi,~W. Analysis of the effect of resistance increase on the
  capacity fade of lithium ion batteries. \emph{International Journal of Energy
  Research} \textbf{2019}, \emph{43}, 2044--2056\relax
\mciteBstWouldAddEndPuncttrue
\mciteSetBstMidEndSepPunct{\mcitedefaultmidpunct}
{\mcitedefaultendpunct}{\mcitedefaultseppunct}\relax
\EndOfBibitem
\bibitem[Keil \latin{et~al.}(2019)Keil, Paul, Baran, Keil, Gilles, and
  Jossen]{keil_linear_2019}
Keil,~J.; Paul,~N.; Baran,~V.; Keil,~P.; Gilles,~R.; Jossen,~A. Linear and
  {Nonlinear} {Aging} of {Lithium}-{Ion} {Cells} {Investigated} by
  {Electrochemical} {Analysis} and {In}-{Situ} {Neutron} {Diffraction}.
  \emph{Journal of The Electrochemical Society} \textbf{2019}, \emph{166},
  A3908--A3917\relax
\mciteBstWouldAddEndPuncttrue
\mciteSetBstMidEndSepPunct{\mcitedefaultmidpunct}
{\mcitedefaultendpunct}{\mcitedefaultseppunct}\relax
\EndOfBibitem
\bibitem[Atalay \latin{et~al.}(2020)Atalay, Sheikh, Mariani, Merla, Bower, and
  Widanage]{atalay_theory_2020}
Atalay,~S.; Sheikh,~M.; Mariani,~A.; Merla,~Y.; Bower,~E.; Widanage,~W.~D.
  Theory of battery ageing in a lithium-ion battery: {Capacity} fade, nonlinear
  ageing and lifetime prediction. \emph{Journal of Power Sources}
  \textbf{2020}, \emph{478}, 229026\relax
\mciteBstWouldAddEndPuncttrue
\mciteSetBstMidEndSepPunct{\mcitedefaultmidpunct}
{\mcitedefaultendpunct}{\mcitedefaultseppunct}\relax
\EndOfBibitem
\bibitem[Müller \latin{et~al.}(2019)Müller, Dufaux, and
  Birke]{muller_model-based_2019}
Müller,~D.; Dufaux,~T.; Birke,~K.~P. Model-{Based} {Investigation} of
  {Porosity} {Profiles} in {Graphite} {Anodes} {Regarding} {Sudden}-{Death} and
  {Second}-{Life} of {Lithium} {Ion} {Cells}. \emph{Batteries} \textbf{2019},
  \emph{5}, 49\relax
\mciteBstWouldAddEndPuncttrue
\mciteSetBstMidEndSepPunct{\mcitedefaultmidpunct}
{\mcitedefaultendpunct}{\mcitedefaultseppunct}\relax
\EndOfBibitem
\bibitem[Willenberg \latin{et~al.}(2020)Willenberg, Dechent, Fuchs, Teuber,
  Eckert, Graff, Kürten, Sauer, and Figgemeier]{willenberg_development_2020}
Willenberg,~L.~K.; Dechent,~P.; Fuchs,~G.; Teuber,~M.; Eckert,~M.; Graff,~M.;
  Kürten,~N.; Sauer,~D.~U.; Figgemeier,~E. The development of jelly roll
  deformation in 18650 lithium-ion batteries at low state of charge.
  \emph{Journal of the Electrochemical Society} \textbf{2020}, \relax
\mciteBstWouldAddEndPunctfalse
\mciteSetBstMidEndSepPunct{\mcitedefaultmidpunct}
{}{\mcitedefaultseppunct}\relax
\EndOfBibitem
\bibitem[Kupper \latin{et~al.}(2018)Kupper, Weißhar, Rißmann, and
  Bessler]{kupper_end--life_2018}
Kupper,~C.; Weißhar,~B.; Rißmann,~S.; Bessler,~W.~G. End-of-{Life}
  {Prediction} of a {Lithium}-{Ion} {Battery} {Cell} {Based} on {Mechanistic}
  {Aging} {Models} of the {Graphite} {Electrode}. \emph{Journal of The
  Electrochemical Society} \textbf{2018}, \emph{165}, A3468--A3480\relax
\mciteBstWouldAddEndPuncttrue
\mciteSetBstMidEndSepPunct{\mcitedefaultmidpunct}
{\mcitedefaultendpunct}{\mcitedefaultseppunct}\relax
\EndOfBibitem
\bibitem[Lin \latin{et~al.}(2013)Lin, Park, Liu, Lee, Sastry, and
  Lu]{lin_comprehensive_2013}
Lin,~X.; Park,~J.; Liu,~L.; Lee,~Y.; Sastry,~A.~M.; Lu,~W. A {Comprehensive}
  {Capacity} {Fade} {Model} and {Analysis} for {Li}-{Ion} {Batteries}.
  \emph{Journal of The Electrochemical Society} \textbf{2013}, \emph{160},
  A1701--A1710\relax
\mciteBstWouldAddEndPuncttrue
\mciteSetBstMidEndSepPunct{\mcitedefaultmidpunct}
{\mcitedefaultendpunct}{\mcitedefaultseppunct}\relax
\EndOfBibitem
\bibitem[Dubarry \latin{et~al.}(2020)Dubarry, Baure, and
  Anseán]{dubarry_perspective_2020}
Dubarry,~M.; Baure,~G.; Anseán,~D. Perspective on {State}-of-{Health}
  {Determination} in {Lithium}-{Ion} {Batteries}. \emph{Journal of
  Electrochemical Energy Conversion and Storage} \textbf{2020}, \emph{17},
  044701\relax
\mciteBstWouldAddEndPuncttrue
\mciteSetBstMidEndSepPunct{\mcitedefaultmidpunct}
{\mcitedefaultendpunct}{\mcitedefaultseppunct}\relax
\EndOfBibitem
\bibitem[Pugalenthi \latin{et~al.}(2020)Pugalenthi, Park, and
  Raghavan]{pugalenthi_piecewise_2020}
Pugalenthi,~K.; Park,~H.; Raghavan,~N. Piecewise {Model}-{Based} {Online}
  {Prognosis} of {Lithium}-{Ion} {Batteries} {Using} {Particle} {Filters}.
  \emph{IEEE Access} \textbf{2020}, \emph{8}, 153508--153516\relax
\mciteBstWouldAddEndPuncttrue
\mciteSetBstMidEndSepPunct{\mcitedefaultmidpunct}
{\mcitedefaultendpunct}{\mcitedefaultseppunct}\relax
\EndOfBibitem
\bibitem[Fang \latin{et~al.}(2021)Fang, Dong, Ge, Fu, Li, and
  Zhang]{fang_capacity_2021}
Fang,~R.; Dong,~P.; Ge,~H.; Fu,~J.; Li,~Z.; Zhang,~J. Capacity plunge of
  lithium-ion batteries induced by electrolyte drying-out: {Experimental} and
  {Modeling} {Study}. \emph{Journal of Energy Storage} \textbf{2021},
  \emph{42}, 103013\relax
\mciteBstWouldAddEndPuncttrue
\mciteSetBstMidEndSepPunct{\mcitedefaultmidpunct}
{\mcitedefaultendpunct}{\mcitedefaultseppunct}\relax
\EndOfBibitem
\bibitem[Braco \latin{et~al.}(2020)Braco, San~Martín, Berrueta, Sanchis, and
  Ursúa]{braco_experimental_2020}
Braco,~E.; San~Martín,~I.; Berrueta,~A.; Sanchis,~P.; Ursúa,~A. Experimental
  assessment of cycling ageing of lithium-ion second-life batteries from
  electric vehicles. \emph{Journal of Energy Storage} \textbf{2020}, \emph{32},
  101695\relax
\mciteBstWouldAddEndPuncttrue
\mciteSetBstMidEndSepPunct{\mcitedefaultmidpunct}
{\mcitedefaultendpunct}{\mcitedefaultseppunct}\relax
\EndOfBibitem
\bibitem[{IEEE Power and Energy
  Society}(2020)]{ieee_power_and_energy_society_ieee_2020}
{IEEE Power and Energy Society}, {IEEE} {Recommended} {Practice} for {Sizing}
  {Lead}-{Acid} {Batteries} for {Stationary} {Applications}. \emph{IEEE Std
  485-2020 (Revision of IEEE Std 485-2010)} \textbf{2020}, 1--69, Conference
  Name: IEEE Std 485-2020 (Revision of IEEE Std 485-2010)\relax
\mciteBstWouldAddEndPuncttrue
\mciteSetBstMidEndSepPunct{\mcitedefaultmidpunct}
{\mcitedefaultendpunct}{\mcitedefaultseppunct}\relax
\EndOfBibitem
\bibitem[Strange \latin{et~al.}(2021)Strange, Li, Gilchrist, and dos
  Reis]{strange_elbows_2021}
Strange,~C.; Li,~S.; Gilchrist,~R.; dos Reis,~G. Elbows of {Internal}
  {Resistance} {Rise} {Curves} in {Li}-{Ion} {Cells}. \emph{Energies}
  \textbf{2021}, \emph{14}, 1206\relax
\mciteBstWouldAddEndPuncttrue
\mciteSetBstMidEndSepPunct{\mcitedefaultmidpunct}
{\mcitedefaultendpunct}{\mcitedefaultseppunct}\relax
\EndOfBibitem
\bibitem[Wang \latin{et~al.}(2011)Wang, Liu, Hicks-Garner, Sherman, Soukiazian,
  Verbrugge, Tataria, Musser, and Finamore]{wang_cycle-life_2011}
Wang,~J.; Liu,~P.; Hicks-Garner,~J.; Sherman,~E.; Soukiazian,~S.;
  Verbrugge,~M.; Tataria,~H.; Musser,~J.; Finamore,~P. Cycle-life model for
  graphite-{LiFePO4} cells. \emph{Journal of Power Sources} \textbf{2011},
  \emph{196}, 3942--3948\relax
\mciteBstWouldAddEndPuncttrue
\mciteSetBstMidEndSepPunct{\mcitedefaultmidpunct}
{\mcitedefaultendpunct}{\mcitedefaultseppunct}\relax
\EndOfBibitem
\bibitem[Ahnert and Abel(2007)Ahnert, and Abel]{ahnert_numerical_2007}
Ahnert,~K.; Abel,~M. Numerical differentiation of experimental data: local
  versus global methods. \emph{Computer Physics Communications} \textbf{2007},
  \emph{177}, 764--774\relax
\mciteBstWouldAddEndPuncttrue
\mciteSetBstMidEndSepPunct{\mcitedefaultmidpunct}
{\mcitedefaultendpunct}{\mcitedefaultseppunct}\relax
\EndOfBibitem
\bibitem[Van~Breugel \latin{et~al.}(2020)Van~Breugel, Kutz, and
  Brunton]{van_breugel_numerical_2020}
Van~Breugel,~F.; Kutz,~J.~N.; Brunton,~B.~W. Numerical {Differentiation} of
  {Noisy} {Data}: {A} {Unifying} {Multi}-{Objective} {Optimization}
  {Framework}. \emph{IEEE Access} \textbf{2020}, \emph{8}, 196865--196877,
  Conference Name: IEEE Access\relax
\mciteBstWouldAddEndPuncttrue
\mciteSetBstMidEndSepPunct{\mcitedefaultmidpunct}
{\mcitedefaultendpunct}{\mcitedefaultseppunct}\relax
\EndOfBibitem
\bibitem[Severson \latin{et~al.}(2019)Severson, Attia, Jin, Perkins, Jiang,
  Yang, Chen, Aykol, Herring, Fraggedakis, Bazant, Harris, Chueh, and
  Braatz]{severson_data-driven_2019}
Severson,~K.~A.; Attia,~P.~M.; Jin,~N.; Perkins,~N.; Jiang,~B.; Yang,~Z.;
  Chen,~M.~H.; Aykol,~M.; Herring,~P.~K.; Fraggedakis,~D. \latin{et~al.}
  Data-driven prediction of battery cycle life before capacity degradation.
  \emph{Nature Energy} \textbf{2019}, \emph{4}, 383--391\relax
\mciteBstWouldAddEndPuncttrue
\mciteSetBstMidEndSepPunct{\mcitedefaultmidpunct}
{\mcitedefaultendpunct}{\mcitedefaultseppunct}\relax
\EndOfBibitem
\bibitem[Satopaa \latin{et~al.}(2011)Satopaa, Albrecht, Irwin, and
  Raghavan]{satopaa_finding_2011}
Satopaa,~V.; Albrecht,~J.; Irwin,~D.; Raghavan,~B. Finding a "{Kneedle}" in a
  {Haystack}: {Detecting} {Knee} {Points} in {System} {Behavior}. 2011 31st
  {International} {Conference} on {Distributed} {Computing} {Systems}
  {Workshops}. 2011; pp 166--171, ISSN: 2332-5666\relax
\mciteBstWouldAddEndPuncttrue
\mciteSetBstMidEndSepPunct{\mcitedefaultmidpunct}
{\mcitedefaultendpunct}{\mcitedefaultseppunct}\relax
\EndOfBibitem
\bibitem[Greenbank and Howey(2021)Greenbank, and
  Howey]{greenbank_automated_2021}
Greenbank,~S.; Howey,~D. Automated feature extraction and selection for
  data-driven models of rapid battery capacity fade and end of life. \emph{IEEE
  Transactions on Industrial Informatics} \textbf{2021}, \relax
\mciteBstWouldAddEndPunctfalse
\mciteSetBstMidEndSepPunct{\mcitedefaultmidpunct}
{}{\mcitedefaultseppunct}\relax
\EndOfBibitem
\bibitem[Zhang \latin{et~al.}(2019)Zhang, Wang, Gao, Wang, Mu, and
  Zhang]{zhang_accelerated_2019}
Zhang,~C.; Wang,~Y.; Gao,~Y.; Wang,~F.; Mu,~B.; Zhang,~W. Accelerated fading
  recognition for lithium-ion batteries with {Nickel}-{Cobalt}-{Manganese}
  cathode using quantile regression method. \emph{Applied Energy}
  \textbf{2019}, \emph{256}, 113841, Publisher: Elsevier\relax
\mciteBstWouldAddEndPuncttrue
\mciteSetBstMidEndSepPunct{\mcitedefaultmidpunct}
{\mcitedefaultendpunct}{\mcitedefaultseppunct}\relax
\EndOfBibitem
\bibitem[Aitio and Howey(2021)Aitio, and Howey]{aitio_predicting_2021}
Aitio,~A.; Howey,~D.~A. Predicting battery end of life from solar off-grid
  system field data using machine learning. \emph{Joule} \textbf{2021},
  \emph{5}, 3204--3220\relax
\mciteBstWouldAddEndPuncttrue
\mciteSetBstMidEndSepPunct{\mcitedefaultmidpunct}
{\mcitedefaultendpunct}{\mcitedefaultseppunct}\relax
\EndOfBibitem
\bibitem[Gao \latin{et~al.}(2021)Gao, Han, Fraggedakis, Das, Zhou, Yeh, Xu,
  Chueh, Li, and Bazant]{gao_interplay_2021}
Gao,~T.; Han,~Y.; Fraggedakis,~D.; Das,~S.; Zhou,~T.; Yeh,~C.-N.; Xu,~S.;
  Chueh,~W.~C.; Li,~J.; Bazant,~M.~Z. Interplay of {Lithium} {Intercalation}
  and {Plating} on a {Single} {Graphite} {Particle}. \emph{Joule}
  \textbf{2021}, \emph{5}, 393--414\relax
\mciteBstWouldAddEndPuncttrue
\mciteSetBstMidEndSepPunct{\mcitedefaultmidpunct}
{\mcitedefaultendpunct}{\mcitedefaultseppunct}\relax
\EndOfBibitem
\bibitem[Baure and Dubarry(2019)Baure, and Dubarry]{baure_synthetic_2019}
Baure,~G.; Dubarry,~M. Synthetic vs. {Real} {Driving} {Cycles}: {A}
  {Comparison} of {Electric} {Vehicle} {Battery} {Degradation}.
  \emph{Batteries} \textbf{2019}, \emph{5}, 42, Number: 2 Publisher:
  Multidisciplinary Digital Publishing Institute\relax
\mciteBstWouldAddEndPuncttrue
\mciteSetBstMidEndSepPunct{\mcitedefaultmidpunct}
{\mcitedefaultendpunct}{\mcitedefaultseppunct}\relax
\EndOfBibitem
\bibitem[Dubarry and Beck(2020)Dubarry, and Beck]{dubarry_big_2020}
Dubarry,~M.; Beck,~D. Big data training data for artificial intelligence-based
  {Li}-ion diagnosis and prognosis. \emph{Journal of Power Sources}
  \textbf{2020}, \emph{479}, 228806\relax
\mciteBstWouldAddEndPuncttrue
\mciteSetBstMidEndSepPunct{\mcitedefaultmidpunct}
{\mcitedefaultendpunct}{\mcitedefaultseppunct}\relax
\EndOfBibitem
\bibitem[Ely and García(2013)Ely, and García]{ely_heterogeneous_2013}
Ely,~D.; García,~E. Heterogeneous {Nucleation} and {Growth} of {Lithium}
  {Electrodeposits} on {Negative} {Electrodes}. \emph{Journal of the
  Electrochemical Society} \textbf{2013}, \emph{160}, A662--A668\relax
\mciteBstWouldAddEndPuncttrue
\mciteSetBstMidEndSepPunct{\mcitedefaultmidpunct}
{\mcitedefaultendpunct}{\mcitedefaultseppunct}\relax
\EndOfBibitem
\bibitem[Pei \latin{et~al.}(2017)Pei, Zheng, Shi, Li, and
  Cui]{pei_nanoscale_2017}
Pei,~A.; Zheng,~G.; Shi,~F.; Li,~Y.; Cui,~Y. Nanoscale {Nucleation} and
  {Growth} of {Electrodeposited} {Lithium} {Metal}. \emph{Nano Letters}
  \textbf{2017}, \emph{17}, 1132--1139\relax
\mciteBstWouldAddEndPuncttrue
\mciteSetBstMidEndSepPunct{\mcitedefaultmidpunct}
{\mcitedefaultendpunct}{\mcitedefaultseppunct}\relax
\EndOfBibitem
\bibitem[Waldmann \latin{et~al.}(2018)Waldmann, Hogg, and
  Wohlfahrt-Mehrens]{waldmann_li_2018}
Waldmann,~T.; Hogg,~B.-I.; Wohlfahrt-Mehrens,~M. Li plating as unwanted side
  reaction in commercial {Li}-ion cells – {A} review. \emph{Journal of Power
  Sources} \textbf{2018}, \emph{384}, 107--124\relax
\mciteBstWouldAddEndPuncttrue
\mciteSetBstMidEndSepPunct{\mcitedefaultmidpunct}
{\mcitedefaultendpunct}{\mcitedefaultseppunct}\relax
\EndOfBibitem
\bibitem[Deichmann \latin{et~al.}(2020)Deichmann, Torres-Castro, Lamb,
  Karulkar, Ivanov, Grosso, Gray, Langendorf, and
  Garzon]{deichmann_investigating_2020}
Deichmann,~E.; Torres-Castro,~L.; Lamb,~J.; Karulkar,~M.; Ivanov,~S.;
  Grosso,~C.; Gray,~L.; Langendorf,~J.; Garzon,~F. Investigating the {Effects}
  of {Lithium} {Deposition} on the {Abuse} {Responseof} {Lithium}-{Ion}
  {Batteries}. \emph{Journal of The Electrochemical Society} \textbf{2020},
  \emph{167}, 090552\relax
\mciteBstWouldAddEndPuncttrue
\mciteSetBstMidEndSepPunct{\mcitedefaultmidpunct}
{\mcitedefaultendpunct}{\mcitedefaultseppunct}\relax
\EndOfBibitem
\bibitem[Martin \latin{et~al.}(2020)Martin, Genovese, Louli, Weber, and
  Dahn]{martin_cycling_2020}
Martin,~C.; Genovese,~M.; Louli,~A.~J.; Weber,~R.; Dahn,~J.~R. Cycling
  {Lithium} {Metal} on {Graphite} to {FormHybrid} {Lithium}-{Ion}/{Lithium}
  {Metal} {Cells}. \emph{Joule} \textbf{2020}, \emph{4}, 1296--1310\relax
\mciteBstWouldAddEndPuncttrue
\mciteSetBstMidEndSepPunct{\mcitedefaultmidpunct}
{\mcitedefaultendpunct}{\mcitedefaultseppunct}\relax
\EndOfBibitem
\bibitem[Anseán \latin{et~al.}(2017)Anseán, Dubarry, Devie, Liaw, García,
  Viera, and González]{ansean_operando_2017}
Anseán,~D.; Dubarry,~M.; Devie,~A.; Liaw,~B.~Y.; García,~V.~M.; Viera,~J.~C.;
  González,~M. Operando lithium plating quantification and early detection of
  a commercial {LiFePO4} cell cycled under dynamic driving schedule.
  \emph{Journal of Power Sources} \textbf{2017}, \emph{356}, 36--46\relax
\mciteBstWouldAddEndPuncttrue
\mciteSetBstMidEndSepPunct{\mcitedefaultmidpunct}
{\mcitedefaultendpunct}{\mcitedefaultseppunct}\relax
\EndOfBibitem
\bibitem[Dubarry \latin{et~al.}(2018)Dubarry, Baure, and
  Devie]{dubarry_durability_2018}
Dubarry,~M.; Baure,~G.; Devie,~A. Durability and {Reliability} of {EV}
  {Batteries} under {Electric} {Utility} {Grid} {Operations}: {Path}
  {Dependence} of {Battery} {Degradation}. \emph{Journal of The Electrochemical
  Society} \textbf{2018}, \emph{165}, A773--A783\relax
\mciteBstWouldAddEndPuncttrue
\mciteSetBstMidEndSepPunct{\mcitedefaultmidpunct}
{\mcitedefaultendpunct}{\mcitedefaultseppunct}\relax
\EndOfBibitem
\bibitem[Liu \latin{et~al.}(2010)Liu, Wang, Hicks-Garner, Sherman, Soukiazian,
  Verbrugge, Tataria, Musser, and Finamore]{liu_aging_2010}
Liu,~P.; Wang,~J.; Hicks-Garner,~J.; Sherman,~E.; Soukiazian,~S.;
  Verbrugge,~M.; Tataria,~H.; Musser,~J.; Finamore,~P. Aging {Mechanisms} of
  {LiFePO4} {Batteries} {Deduced} by {Electrochemical} and {Structural}
  {Analyses}. \emph{Journal of The Electrochemical Society} \textbf{2010},
  \emph{157}, A499--A507\relax
\mciteBstWouldAddEndPuncttrue
\mciteSetBstMidEndSepPunct{\mcitedefaultmidpunct}
{\mcitedefaultendpunct}{\mcitedefaultseppunct}\relax
\EndOfBibitem
\bibitem[Cannarella and Arnold(2014)Cannarella, and
  Arnold]{cannarella_stress_2014}
Cannarella,~J.; Arnold,~C.~B. Stress evolution and capacity fade in constrained
  lithium-ion pouch cells. \emph{Journal of Power Sources} \textbf{2014},
  \emph{245}, 745--751\relax
\mciteBstWouldAddEndPuncttrue
\mciteSetBstMidEndSepPunct{\mcitedefaultmidpunct}
{\mcitedefaultendpunct}{\mcitedefaultseppunct}\relax
\EndOfBibitem
\bibitem[Somerville \latin{et~al.}(2016)Somerville, Bareño, Trask, Jennings,
  McGordon, Lyness, and Bloom]{somerville_effect_2016}
Somerville,~L.; Bareño,~J.; Trask,~S.; Jennings,~P.; McGordon,~A.; Lyness,~C.;
  Bloom,~I. The effect of charging rate on the graphite electrode of commercial
  lithium-ion cells: {A} post-mortem study. \emph{Journal of Power Sources}
  \textbf{2016}, \emph{335}, 189--196\relax
\mciteBstWouldAddEndPuncttrue
\mciteSetBstMidEndSepPunct{\mcitedefaultmidpunct}
{\mcitedefaultendpunct}{\mcitedefaultseppunct}\relax
\EndOfBibitem
\bibitem[Willenberg \latin{et~al.}(2020)Willenberg, Dechent, Fuchs, Sauer, and
  Figgemeier]{willenberg_high-precision_2020}
Willenberg,~L.~K.; Dechent,~P.; Fuchs,~G.; Sauer,~D.~U.; Figgemeier,~E.
  High-{Precision} {Monitoring} of {Volume} {Change} of {Commercial}
  {Lithium}-{Ion} {Batteries} by {Using} {Strain} {Gauges}.
  \emph{Sustainability} \textbf{2020}, \emph{12}, 557\relax
\mciteBstWouldAddEndPuncttrue
\mciteSetBstMidEndSepPunct{\mcitedefaultmidpunct}
{\mcitedefaultendpunct}{\mcitedefaultseppunct}\relax
\EndOfBibitem
\bibitem[Takahashi and Srinivasan(2015)Takahashi, and
  Srinivasan]{takahashi_examination_2015}
Takahashi,~K.; Srinivasan,~V. Examination of {Graphite} {Particle} {Cracking}
  as a {Failure} {Mode} in {Lithium}-{Ion} {Batteries}: {A}
  {Model}-{Experimental} {Study}. \emph{Journal of The Electrochemical Society}
  \textbf{2015}, \emph{162}, A635--A645\relax
\mciteBstWouldAddEndPuncttrue
\mciteSetBstMidEndSepPunct{\mcitedefaultmidpunct}
{\mcitedefaultendpunct}{\mcitedefaultseppunct}\relax
\EndOfBibitem
\bibitem[Mao \latin{et~al.}(2017)Mao, Farkhondeh, Pritzker, Fowler, and
  Chen]{mao_calendar_2017}
Mao,~Z.; Farkhondeh,~M.; Pritzker,~M.; Fowler,~M.; Chen,~Z. Calendar {Aging}
  and {Gas} {Generation} in {Commercial} {Graphite}/{NMC}-{LMO} {Lithium}-{Ion}
  {Pouch} {Cell}. \emph{Journal of The Electrochemical Society} \textbf{2017},
  \emph{164}, A3469\relax
\mciteBstWouldAddEndPuncttrue
\mciteSetBstMidEndSepPunct{\mcitedefaultmidpunct}
{\mcitedefaultendpunct}{\mcitedefaultseppunct}\relax
\EndOfBibitem
\bibitem[Waldmann \latin{et~al.}(2014)Waldmann, Wilka, Kasper, Fleischhammer,
  and Wohlfahrt-Mehrens]{waldmann_temperature_2014}
Waldmann,~T.; Wilka,~M.; Kasper,~M.; Fleischhammer,~M.; Wohlfahrt-Mehrens,~M.
  Temperature dependent ageing mechanisms in {Lithium}-ion batteries – {A}
  {Post}-{Mortem} study. \emph{Journal of Power Sources} \textbf{2014},
  \emph{262}, 129--135\relax
\mciteBstWouldAddEndPuncttrue
\mciteSetBstMidEndSepPunct{\mcitedefaultmidpunct}
{\mcitedefaultendpunct}{\mcitedefaultseppunct}\relax
\EndOfBibitem
\bibitem[Petzl \latin{et~al.}(2015)Petzl, Kasper, and
  Danzer]{petzl_lithium_2015}
Petzl,~M.; Kasper,~M.; Danzer,~M.~A. Lithium plating in a commercial
  lithium-ion battery – {A} low-temperature aging study. \emph{Journal of
  Power Sources} \textbf{2015}, \emph{275}, 799--807\relax
\mciteBstWouldAddEndPuncttrue
\mciteSetBstMidEndSepPunct{\mcitedefaultmidpunct}
{\mcitedefaultendpunct}{\mcitedefaultseppunct}\relax
\EndOfBibitem
\bibitem[Waldmann \latin{et~al.}(2015)Waldmann, Kasper, and
  Wohlfahrt-Mehrens]{waldmann_optimization_2015}
Waldmann,~T.; Kasper,~M.; Wohlfahrt-Mehrens,~M. Optimization of {Charging}
  {Strategy} by {Prevention} of {Lithium} {Deposition} on {Anodes} in
  high-energy {Lithium}-ion {Batteries} – {Electrochemical} {Experiments}.
  \emph{Electrochimica Acta} \textbf{2015}, \emph{178}, 525--532\relax
\mciteBstWouldAddEndPuncttrue
\mciteSetBstMidEndSepPunct{\mcitedefaultmidpunct}
{\mcitedefaultendpunct}{\mcitedefaultseppunct}\relax
\EndOfBibitem
\bibitem[Burns \latin{et~al.}(2015)Burns, Stevens, and Dahn]{burns_-situ_2015}
Burns,~J.~C.; Stevens,~D.~A.; Dahn,~J.~R. In-{Situ} {Detection} of {Lithium}
  {Plating} {Using} {High} {Precision} {Coulometry}. \emph{Journal of The
  Electrochemical Society} \textbf{2015}, \emph{162}, A959--A964\relax
\mciteBstWouldAddEndPuncttrue
\mciteSetBstMidEndSepPunct{\mcitedefaultmidpunct}
{\mcitedefaultendpunct}{\mcitedefaultseppunct}\relax
\EndOfBibitem
\bibitem[Yang and Wang(2018)Yang, and Wang]{yang_understanding_2018}
Yang,~X.-G.; Wang,~C.-Y. Understanding the trilemma of fast charging, energy
  density and cycle life of lithium-ion batteries. \emph{Journal of Power
  Sources} \textbf{2018}, \emph{402}, 489--498\relax
\mciteBstWouldAddEndPuncttrue
\mciteSetBstMidEndSepPunct{\mcitedefaultmidpunct}
{\mcitedefaultendpunct}{\mcitedefaultseppunct}\relax
\EndOfBibitem
\bibitem[Coron \latin{et~al.}(2020)Coron, Geniès, Cugnet, and
  Thivel]{coron_impact_2020}
Coron,~E.; Geniès,~S.; Cugnet,~M.; Thivel,~P.~X. Impact of {Lithium}-{Ion}
  {Cell} {Condition} on {Its} {Second} {Life} {Viability}. \emph{Journal of The
  Electrochemical Society} \textbf{2020}, \emph{167}, 110556\relax
\mciteBstWouldAddEndPuncttrue
\mciteSetBstMidEndSepPunct{\mcitedefaultmidpunct}
{\mcitedefaultendpunct}{\mcitedefaultseppunct}\relax
\EndOfBibitem
\bibitem[Lewerenz \latin{et~al.}(2017)Lewerenz, Münnix, Schmalstieg, Käbitz,
  Knips, and Sauer]{lewerenz_systematic_2017}
Lewerenz,~M.; Münnix,~J.; Schmalstieg,~J.; Käbitz,~S.; Knips,~M.;
  Sauer,~D.~U. Systematic aging of commercial {LiFePO} 4 {\textbar}{Graphite}
  cylindrical cells including a theory explaining rise of capacity during
  aging. \emph{Journal of Power Sources} \textbf{2017}, \emph{345},
  254--263\relax
\mciteBstWouldAddEndPuncttrue
\mciteSetBstMidEndSepPunct{\mcitedefaultmidpunct}
{\mcitedefaultendpunct}{\mcitedefaultseppunct}\relax
\EndOfBibitem
\bibitem[Schindler \latin{et~al.}(2018)Schindler, Bauer, Cheetamun, and
  Danzer]{schindler_fast_2018}
Schindler,~S.; Bauer,~M.; Cheetamun,~H.; Danzer,~M.~A. Fast charging of
  lithium-ion cells: {Identification} of aging-minimal current profiles using a
  design of experiment approach and a mechanistic degradation analysis.
  \emph{Journal of Energy Storage} \textbf{2018}, \emph{19}, 364--378\relax
\mciteBstWouldAddEndPuncttrue
\mciteSetBstMidEndSepPunct{\mcitedefaultmidpunct}
{\mcitedefaultendpunct}{\mcitedefaultseppunct}\relax
\EndOfBibitem
\bibitem[Attia \latin{et~al.}(2020)Attia, Grover, Jin, Severson, Markov, Liao,
  Chen, Cheong, Perkins, Yang, Herring, Aykol, Harris, Braatz, Ermon, and
  Chueh]{attia_closed-loop_2020}
Attia,~P.~M.; Grover,~A.; Jin,~N.; Severson,~K.~A.; Markov,~T.~M.; Liao,~Y.-H.;
  Chen,~M.~H.; Cheong,~B.; Perkins,~N.; Yang,~Z. \latin{et~al.}  Closed-loop
  optimization of fast-charging protocols for batteries with machine learning.
  \emph{Nature} \textbf{2020}, \emph{578}, 397--402\relax
\mciteBstWouldAddEndPuncttrue
\mciteSetBstMidEndSepPunct{\mcitedefaultmidpunct}
{\mcitedefaultendpunct}{\mcitedefaultseppunct}\relax
\EndOfBibitem
\bibitem[Nemani \latin{et~al.}(2015)Nemani, Harris, and
  Smith]{nemani_design_2015}
Nemani,~V.~P.; Harris,~S.~J.; Smith,~K.~C. Design of {Bi}-{Tortuous},
  {Anisotropic} {Graphite} {Anodes} for {Fast} {Ion}-{Transport} in {Li}-{Ion}
  {Batteries}. \emph{Journal of The Electrochemical Society} \textbf{2015},
  \emph{162}, A1415--A1423\relax
\mciteBstWouldAddEndPuncttrue
\mciteSetBstMidEndSepPunct{\mcitedefaultmidpunct}
{\mcitedefaultendpunct}{\mcitedefaultseppunct}\relax
\EndOfBibitem
\bibitem[Usseglio-Viretta \latin{et~al.}(2020)Usseglio-Viretta, Mai,
  Colclasure, Doeff, Yi, and Smith]{usseglio-viretta_enabling_2020}
Usseglio-Viretta,~F.; Mai,~W.; Colclasure,~A.; Doeff,~M.; Yi,~E.; Smith,~K.
  Enabling fast charging of lithium-ion batteries through secondary- /dual-
  pore network: {Part} {I} - {Analytical} diffusion model. \emph{Electrochimica
  Acta} \textbf{2020}, \emph{342}, 136034\relax
\mciteBstWouldAddEndPuncttrue
\mciteSetBstMidEndSepPunct{\mcitedefaultmidpunct}
{\mcitedefaultendpunct}{\mcitedefaultseppunct}\relax
\EndOfBibitem
\bibitem[Liu and Arnold(2020)Liu, and Arnold]{liu_effects_2020}
Liu,~X.~M.; Arnold,~C.~B. Effects of {Current} {Density} on {Defect}-{Induced}
  {Capacity} {Fade} through {Localized} {Plating} in {Lithium}-{Ion}
  {Batteries}. \emph{Journal of The Electrochemical Society} \textbf{2020},
  \emph{167}, 130519, Publisher: The Electrochemical Society\relax
\mciteBstWouldAddEndPuncttrue
\mciteSetBstMidEndSepPunct{\mcitedefaultmidpunct}
{\mcitedefaultendpunct}{\mcitedefaultseppunct}\relax
\EndOfBibitem
\bibitem[Liu \latin{et~al.}(2018)Liu, Fang, Haataja, and Arnold]{liu_size_2018}
Liu,~X.~M.; Fang,~A.; Haataja,~M.~P.; Arnold,~C.~B. Size {Dependence} of
  {Transport} {Non}-{Uniformities} on {Localized} {Plating} in {Lithium}-{Ion}
  {Batteries}. \emph{Journal of The Electrochemical Society} \textbf{2018},
  \emph{165}, A1147--A1155\relax
\mciteBstWouldAddEndPuncttrue
\mciteSetBstMidEndSepPunct{\mcitedefaultmidpunct}
{\mcitedefaultendpunct}{\mcitedefaultseppunct}\relax
\EndOfBibitem
\bibitem[Fuchs \latin{et~al.}(2019)Fuchs, Willenberg, Ringbeck, and
  Sauer]{fuchs_post-mortem_2019}
Fuchs,~G.; Willenberg,~L.; Ringbeck,~F.; Sauer,~D.~U. Post-{Mortem} {Analysis}
  of {Inhomogeneous} {Induced} {Pressure} on {Commercial} {Lithium}-{Ion}
  {Pouch} {Cells} and {Their} {Effects}. \emph{Sustainability} \textbf{2019},
  \emph{11}, 6738\relax
\mciteBstWouldAddEndPuncttrue
\mciteSetBstMidEndSepPunct{\mcitedefaultmidpunct}
{\mcitedefaultendpunct}{\mcitedefaultseppunct}\relax
\EndOfBibitem
\bibitem[Okasinski \latin{et~al.}(2020)Okasinski, Shkrob, Chuang, Rodrigues,
  Raj, Dees, and Abraham]{okasinski_situ_2020}
Okasinski,~J.~S.; Shkrob,~I.~A.; Chuang,~A.; Rodrigues,~M.-T.~F.; Raj,~A.;
  Dees,~D.~W.; Abraham,~D.~P. \textit{{In} situ} {X}-ray spatial profiling
  reveals uneven compression of electrode assemblies and steep lateral
  gradients in lithium-ion coin cells. \emph{Physical Chemistry Chemical
  Physics} \textbf{2020}, \emph{22}, 21977--21987\relax
\mciteBstWouldAddEndPuncttrue
\mciteSetBstMidEndSepPunct{\mcitedefaultmidpunct}
{\mcitedefaultendpunct}{\mcitedefaultseppunct}\relax
\EndOfBibitem
\bibitem[Sikha \latin{et~al.}(2004)Sikha, Popov, and White]{sikha_effect_2004}
Sikha,~G.; Popov,~B.~N.; White,~R.~E. Effect of {Porosity} on the {Capacity}
  {Fade} of a {Lithium}-{Ion} {Battery}. \emph{Journal of The Electrochemical
  Society} \textbf{2004}, \emph{151}, A1104\relax
\mciteBstWouldAddEndPuncttrue
\mciteSetBstMidEndSepPunct{\mcitedefaultmidpunct}
{\mcitedefaultendpunct}{\mcitedefaultseppunct}\relax
\EndOfBibitem
\bibitem[Frisco \latin{et~al.}(2016)Frisco, Kumar, Whitacre, and
  Litster]{frisco_understanding_2016}
Frisco,~S.; Kumar,~A.; Whitacre,~J.~F.; Litster,~S. Understanding {Li}-{Ion}
  {Battery} {Anode} {Degradation} and {Pore} {Morphological} {Changes} through
  {Nano}-{Resolution} {X}-ray {Computed} {Tomography}. \emph{Journal of The
  Electrochemical Society} \textbf{2016}, \emph{163}, A2636--A2640\relax
\mciteBstWouldAddEndPuncttrue
\mciteSetBstMidEndSepPunct{\mcitedefaultmidpunct}
{\mcitedefaultendpunct}{\mcitedefaultseppunct}\relax
\EndOfBibitem
\bibitem[Rahe \latin{et~al.}(2019)Rahe, Kelly, Rad, Sauer, Mayer, and
  Figgemeier]{rahe_nanoscale_2019}
Rahe,~C.; Kelly,~S.~T.; Rad,~M.~N.; Sauer,~D.~U.; Mayer,~J.; Figgemeier,~E.
  Nanoscale {X}-ray imaging of ageing in automotive lithium ion battery cells.
  \emph{Journal of Power Sources} \textbf{2019}, \emph{433}, 126631\relax
\mciteBstWouldAddEndPuncttrue
\mciteSetBstMidEndSepPunct{\mcitedefaultmidpunct}
{\mcitedefaultendpunct}{\mcitedefaultseppunct}\relax
\EndOfBibitem
\bibitem[Klett \latin{et~al.}(2015)Klett, Svens, Tengstedt, Seyeux,
  Światowska, Lindbergh, and Wreland~Lindström]{klett_uneven_2015}
Klett,~M.; Svens,~P.; Tengstedt,~C.; Seyeux,~A.; Światowska,~J.;
  Lindbergh,~G.; Wreland~Lindström,~R. Uneven {Film} {Formation} across
  {Depth} of {Porous} {Graphite} {Electrodes} in {Cycled} {Commercial}
  {Li}-{Ion} {Batteries}. \emph{The Journal of Physical Chemistry C}
  \textbf{2015}, \emph{119}, 90--100\relax
\mciteBstWouldAddEndPuncttrue
\mciteSetBstMidEndSepPunct{\mcitedefaultmidpunct}
{\mcitedefaultendpunct}{\mcitedefaultseppunct}\relax
\EndOfBibitem
\bibitem[Sarasketa-Zabala \latin{et~al.}(2015)Sarasketa-Zabala, Aguesse,
  Villarreal, Rodriguez-Martinez, López, and
  Kubiak]{sarasketa-zabala_understanding_2015}
Sarasketa-Zabala,~E.; Aguesse,~F.; Villarreal,~I.; Rodriguez-Martinez,~L.~M.;
  López,~C.~M.; Kubiak,~P. Understanding {Lithium} {Inventory} {Loss} and
  {Sudden} {Performance} {Fade} in {Cylindrical} {Cells} during {Cycling} with
  {Deep}-{Discharge} {Steps}. \emph{The Journal of Physical Chemistry C}
  \textbf{2015}, \emph{119}, 896--906\relax
\mciteBstWouldAddEndPuncttrue
\mciteSetBstMidEndSepPunct{\mcitedefaultmidpunct}
{\mcitedefaultendpunct}{\mcitedefaultseppunct}\relax
\EndOfBibitem
\bibitem[Sarasketa-Zabala \latin{et~al.}(2015)Sarasketa-Zabala, Gandiaga,
  Martinez-Laserna, Rodriguez-Martinez, and
  Villarreal]{sarasketa-zabala_cycle_2015}
Sarasketa-Zabala,~E.; Gandiaga,~I.; Martinez-Laserna,~E.;
  Rodriguez-Martinez,~L.~M.; Villarreal,~I. Cycle ageing analysis of a
  {LiFePO4}/graphite cell with dynamic model validations: {Towards} realistic
  lifetime predictions. \emph{Journal of Power Sources} \textbf{2015},
  \emph{275}, 573--587\relax
\mciteBstWouldAddEndPuncttrue
\mciteSetBstMidEndSepPunct{\mcitedefaultmidpunct}
{\mcitedefaultendpunct}{\mcitedefaultseppunct}\relax
\EndOfBibitem
\bibitem[Xu \latin{et~al.}(2010)Xu, von Cresce, and
  Lee]{xu_differentiating_2010}
Xu,~K.; von Cresce,~A.; Lee,~U. Differentiating {Contributions} to “{Ion}
  {Transfer}” {Barrier} from {Interphasial} {Resistance} and {Li}+
  {Desolvation} at {Electrolyte}/{Graphite} {Interface}. \emph{Langmuir}
  \textbf{2010}, \emph{26}, 11538--11543\relax
\mciteBstWouldAddEndPuncttrue
\mciteSetBstMidEndSepPunct{\mcitedefaultmidpunct}
{\mcitedefaultendpunct}{\mcitedefaultseppunct}\relax
\EndOfBibitem
\bibitem[Pinson and Bazant(2013)Pinson, and Bazant]{pinson_theory_2013}
Pinson,~M.~B.; Bazant,~M.~Z. Theory of {SEI} {Formation} in {Rechargeable}
  {Batteries}: {Capacity} {Fade}, {Accelerated} {Aging} and {Lifetime}
  {Prediction}. \emph{Journal of The Electrochemical Society} \textbf{2013},
  \emph{160}, A243--A250\relax
\mciteBstWouldAddEndPuncttrue
\mciteSetBstMidEndSepPunct{\mcitedefaultmidpunct}
{\mcitedefaultendpunct}{\mcitedefaultseppunct}\relax
\EndOfBibitem
\bibitem[Li \latin{et~al.}(2017)Li, Lu, Zheng, Jiao, Luo, Wang, Xu, Zhang, and
  Xu]{li_li-desolvation_2017}
Li,~Q.; Lu,~D.; Zheng,~J.; Jiao,~S.; Luo,~L.; Wang,~C.-M.; Xu,~K.;
  Zhang,~J.-G.; Xu,~W. Li+-{Desolvation} {Dictating} {Lithium}-{Ion}
  {Battery}’s {Low}-{Temperature} {Performances}. \emph{ACS Applied Materials
  \& Interfaces} \textbf{2017}, \emph{9}, 42761--42768\relax
\mciteBstWouldAddEndPuncttrue
\mciteSetBstMidEndSepPunct{\mcitedefaultmidpunct}
{\mcitedefaultendpunct}{\mcitedefaultseppunct}\relax
\EndOfBibitem
\bibitem[Lu and Harris(2011)Lu, and Harris]{lu_lithium_2011}
Lu,~P.; Harris,~S.~J. Lithium transport within the solid electrolyte
  interphase. \emph{Electrochemistry Communications} \textbf{2011}, \emph{13},
  1035--1037\relax
\mciteBstWouldAddEndPuncttrue
\mciteSetBstMidEndSepPunct{\mcitedefaultmidpunct}
{\mcitedefaultendpunct}{\mcitedefaultseppunct}\relax
\EndOfBibitem
\bibitem[Xu(2014)]{xu_electrolytes_2014}
Xu,~K. Electrolytes and {Interphases} in {Li}-{Ion} {Batteries} and {Beyond}.
  \emph{Chemical Reviews} \textbf{2014}, \emph{114}, 11503--11618\relax
\mciteBstWouldAddEndPuncttrue
\mciteSetBstMidEndSepPunct{\mcitedefaultmidpunct}
{\mcitedefaultendpunct}{\mcitedefaultseppunct}\relax
\EndOfBibitem
\bibitem[Klett \latin{et~al.}(2014)Klett, Eriksson, Groot, Svens,
  Ciosek~Högström, Lindström, Berg, Gustafson, Lindbergh, and
  Edström]{klett_non-uniform_2014}
Klett,~M.; Eriksson,~R.; Groot,~J.; Svens,~P.; Ciosek~Högström,~K.;
  Lindström,~R.~W.; Berg,~H.; Gustafson,~T.; Lindbergh,~G.; Edström,~K.
  Non-uniform aging of cycled commercial {LiFePO4}//graphite cylindrical cells
  revealed by post-mortem analysis. \emph{Journal of Power Sources}
  \textbf{2014}, \emph{257}, 126--137\relax
\mciteBstWouldAddEndPuncttrue
\mciteSetBstMidEndSepPunct{\mcitedefaultmidpunct}
{\mcitedefaultendpunct}{\mcitedefaultseppunct}\relax
\EndOfBibitem
\bibitem[Ecker \latin{et~al.}(2014)Ecker, Nieto, Käbitz, Schmalstieg, Blanke,
  Warnecke, and Sauer]{ecker_calendar_2014}
Ecker,~M.; Nieto,~N.; Käbitz,~S.; Schmalstieg,~J.; Blanke,~H.; Warnecke,~A.;
  Sauer,~D.~U. Calendar and cycle life study of {Li}({NiMnCo}){O2}-based 18650
  lithium-ion batteries. \emph{Journal of Power Sources} \textbf{2014},
  \emph{248}, 839--851\relax
\mciteBstWouldAddEndPuncttrue
\mciteSetBstMidEndSepPunct{\mcitedefaultmidpunct}
{\mcitedefaultendpunct}{\mcitedefaultseppunct}\relax
\EndOfBibitem
\bibitem[Dubarry \latin{et~al.}(2012)Dubarry, Truchot, and
  Liaw]{dubarry_synthesize_2012}
Dubarry,~M.; Truchot,~C.; Liaw,~B.~Y. Synthesize battery degradation modes via
  a diagnostic and prognostic model. \emph{Journal of Power Sources}
  \textbf{2012}, \emph{219}, 204--216\relax
\mciteBstWouldAddEndPuncttrue
\mciteSetBstMidEndSepPunct{\mcitedefaultmidpunct}
{\mcitedefaultendpunct}{\mcitedefaultseppunct}\relax
\EndOfBibitem
\bibitem[Smith \latin{et~al.}(2017)Smith, Saxon, Keyser, Lundstrom, {Ziwei
  Cao}, and Roc]{smith_life_2017}
Smith,~K.; Saxon,~A.; Keyser,~M.; Lundstrom,~B.; {Ziwei Cao},; Roc,~A. Life
  prediction model for grid-connected {Li}-ion battery energy storage system.
  2017 {American} {Control} {Conference} ({ACC}). 2017; pp 4062--4068, ISSN:
  2378-5861\relax
\mciteBstWouldAddEndPuncttrue
\mciteSetBstMidEndSepPunct{\mcitedefaultmidpunct}
{\mcitedefaultendpunct}{\mcitedefaultseppunct}\relax
\EndOfBibitem
\bibitem[Baure \latin{et~al.}(2019)Baure, Devie, and
  Dubarry]{baure_battery_2019}
Baure,~G.; Devie,~A.; Dubarry,~M. Battery {Durability} and {Reliability} under
  {Electric} {Utility} {Grid} {Operations}: {Path} {Dependence} of {Battery}
  {Degradation}. \emph{Journal of The Electrochemical Society} \textbf{2019},
  \emph{166}, A1991--A2001\relax
\mciteBstWouldAddEndPuncttrue
\mciteSetBstMidEndSepPunct{\mcitedefaultmidpunct}
{\mcitedefaultendpunct}{\mcitedefaultseppunct}\relax
\EndOfBibitem
\bibitem[Baure and Dubarry(2020)Baure, and Dubarry]{baure_battery_2020}
Baure,~G.; Dubarry,~M. Battery durability and reliability under electric
  utility grid operations: 20-year forecast under different grid applications.
  \emph{Journal of Energy Storage} \textbf{2020}, \emph{29}, 101391\relax
\mciteBstWouldAddEndPuncttrue
\mciteSetBstMidEndSepPunct{\mcitedefaultmidpunct}
{\mcitedefaultendpunct}{\mcitedefaultseppunct}\relax
\EndOfBibitem
\bibitem[Sulzer \latin{et~al.}(2021)Sulzer, Mohtat, Pannala, Siegel, and
  Stefanopoulou]{sulzer_accelerated_2021}
Sulzer,~V.; Mohtat,~P.; Pannala,~S.; Siegel,~J.~B.; Stefanopoulou,~A.~G.
  Accelerated battery lifetime simulations using adaptive inter-cycle
  extrapolation algorithm. \emph{ECS ArXiv} \textbf{2021}, 1--21\relax
\mciteBstWouldAddEndPuncttrue
\mciteSetBstMidEndSepPunct{\mcitedefaultmidpunct}
{\mcitedefaultendpunct}{\mcitedefaultseppunct}\relax
\EndOfBibitem
\bibitem[Kindermann \latin{et~al.}(2017)Kindermann, Keil, Frank, and
  Jossen]{kindermann_sei_2017}
Kindermann,~F.~M.; Keil,~J.; Frank,~A.; Jossen,~A. A {SEI} {Modeling}
  {Approach} {Distinguishing} between {Capacity} and {Power} {Fade}.
  \emph{Journal of The Electrochemical Society} \textbf{2017}, \emph{164},
  E287--E294\relax
\mciteBstWouldAddEndPuncttrue
\mciteSetBstMidEndSepPunct{\mcitedefaultmidpunct}
{\mcitedefaultendpunct}{\mcitedefaultseppunct}\relax
\EndOfBibitem
\bibitem[Reniers \latin{et~al.}(2019)Reniers, Mulder, and
  Howey]{reniers_review_2019}
Reniers,~J.~M.; Mulder,~G.; Howey,~D.~A. Review and {Performance} {Comparison}
  of {Mechanical}-{Chemical} {Degradation} {Models} for {Lithium}-{Ion}
  {Batteries}. \emph{Journal of The Electrochemical Society} \textbf{2019},
  \emph{166}, A3189--A3200\relax
\mciteBstWouldAddEndPuncttrue
\mciteSetBstMidEndSepPunct{\mcitedefaultmidpunct}
{\mcitedefaultendpunct}{\mcitedefaultseppunct}\relax
\EndOfBibitem
\bibitem[Laresgoiti \latin{et~al.}(2015)Laresgoiti, Käbitz, Ecker, and
  Sauer]{laresgoiti_modeling_2015}
Laresgoiti,~I.; Käbitz,~S.; Ecker,~M.; Sauer,~D.~U. Modeling mechanical
  degradation in lithium ion batteries during cycling: {Solid} electrolyte
  interphase fracture. \emph{Journal of Power Sources} \textbf{2015},
  \emph{300}, 112--122\relax
\mciteBstWouldAddEndPuncttrue
\mciteSetBstMidEndSepPunct{\mcitedefaultmidpunct}
{\mcitedefaultendpunct}{\mcitedefaultseppunct}\relax
\EndOfBibitem
\bibitem[Ai \latin{et~al.}(2020)Ai, Kraft, Sturm, Jossen, and
  Wu]{ai_electrochemical_2020}
Ai,~W.; Kraft,~L.; Sturm,~J.; Jossen,~A.; Wu,~B. Electrochemical
  {Thermal}-{Mechanical} {Modelling} of {Stress} {Inhomogeneity} in
  {Lithium}-{Ion} {Pouch} {Cells}. \emph{Journal of The Electrochemical
  Society} \textbf{2020}, \emph{167}, 013512\relax
\mciteBstWouldAddEndPuncttrue
\mciteSetBstMidEndSepPunct{\mcitedefaultmidpunct}
{\mcitedefaultendpunct}{\mcitedefaultseppunct}\relax
\EndOfBibitem
\bibitem[Mohtat \latin{et~al.}(2020)Mohtat, Lee, Sulzer, Siegel, and
  Stefanopoulou]{mohtat_differential_2020}
Mohtat,~P.; Lee,~S.; Sulzer,~V.; Siegel,~J.~B.; Stefanopoulou,~A.~G.
  Differential {Expansion} and {Voltage} {Model} for {Li}-ion {Batteries} at
  {Practical} {Charging} {Rates}. \emph{Journal of The Electrochemical Society}
  \textbf{2020}, \emph{167}, 110561\relax
\mciteBstWouldAddEndPuncttrue
\mciteSetBstMidEndSepPunct{\mcitedefaultmidpunct}
{\mcitedefaultendpunct}{\mcitedefaultseppunct}\relax
\EndOfBibitem
\bibitem[Fear \latin{et~al.}(2018)Fear, Juarez-Robles, Jeevarajan, and
  Mukherjee]{fear_elucidating_2018}
Fear,~C.; Juarez-Robles,~D.; Jeevarajan,~J.~A.; Mukherjee,~P.~P. Elucidating
  {Copper} {Dissolution} {Phenomenon} in {Li}-{Ion} {Cells} under
  {Overdischarge} {Extremes}. \emph{Journal of The Electrochemical Society}
  \textbf{2018}, \emph{165}, A1639--A1647\relax
\mciteBstWouldAddEndPuncttrue
\mciteSetBstMidEndSepPunct{\mcitedefaultmidpunct}
{\mcitedefaultendpunct}{\mcitedefaultseppunct}\relax
\EndOfBibitem
\bibitem[Carter \latin{et~al.}(2018)Carter, Huhman, Love, and
  Zenyuk]{carter_x-ray_2018}
Carter,~R.; Huhman,~B.; Love,~C.~T.; Zenyuk,~I.~V. X-ray computed tomography
  comparison of individual and parallel assembled commercial lithium iron
  phosphate batteries at end of life after high rate cycling. \emph{Journal of
  Power Sources} \textbf{2018}, \emph{381}, 46--55\relax
\mciteBstWouldAddEndPuncttrue
\mciteSetBstMidEndSepPunct{\mcitedefaultmidpunct}
{\mcitedefaultendpunct}{\mcitedefaultseppunct}\relax
\EndOfBibitem
\bibitem[Keil \latin{et~al.}(2016)Keil, Schuster, Wilhelm, Travi, Hauser, Karl,
  and Jossen]{keil_calendar_2016}
Keil,~P.; Schuster,~S.~F.; Wilhelm,~J.; Travi,~J.; Hauser,~A.; Karl,~R.~C.;
  Jossen,~A. Calendar {Aging} of {Lithium}-{Ion} {Batteries} {I}. {Impact} of
  the {Graphite} {Anode} on {Capacity} {Fade}. \emph{Journal of The
  Electrochemical Society} \textbf{2016}, \emph{163}, A1872--A1880\relax
\mciteBstWouldAddEndPuncttrue
\mciteSetBstMidEndSepPunct{\mcitedefaultmidpunct}
{\mcitedefaultendpunct}{\mcitedefaultseppunct}\relax
\EndOfBibitem
\bibitem[Safari and Delacourt(2011)Safari, and Delacourt]{safari_aging_2011}
Safari,~M.; Delacourt,~C. Aging of a {Commercial} {Graphite}/{LiFePO4} {Cell}.
  \emph{Journal of The Electrochemical Society} \textbf{2011}, \emph{158},
  A1123--A1135\relax
\mciteBstWouldAddEndPuncttrue
\mciteSetBstMidEndSepPunct{\mcitedefaultmidpunct}
{\mcitedefaultendpunct}{\mcitedefaultseppunct}\relax
\EndOfBibitem
\bibitem[Sieg \latin{et~al.}(2022)Sieg, Schmid, Rau, Gesterkamp, Storch, Spier,
  Birke, and Sauer]{sieg_fast-charging_2022}
Sieg,~J.; Schmid,~A.~U.; Rau,~L.; Gesterkamp,~A.; Storch,~M.; Spier,~B.;
  Birke,~K.~P.; Sauer,~D.~U. Fast-charging capability of lithium-ion cells:
  {Influence} of electrode aging and electrolyte consumption. \emph{Applied
  Energy} \textbf{2022}, \emph{305}, 117747\relax
\mciteBstWouldAddEndPuncttrue
\mciteSetBstMidEndSepPunct{\mcitedefaultmidpunct}
{\mcitedefaultendpunct}{\mcitedefaultseppunct}\relax
\EndOfBibitem
\bibitem[Stevens \latin{et~al.}(2014)Stevens, Ying, Fathi, Reimers, Harlow, and
  Dahn]{stevens_using_2014}
Stevens,~D.~A.; Ying,~R.~Y.; Fathi,~R.; Reimers,~J.~N.; Harlow,~J.~E.;
  Dahn,~J.~R. Using {High} {Precision} {Coulometry} {Measurements} to {Compare}
  the {Degradation} {Mechanisms} of {NMC}/{LMO} and {NMC}-{Only} {Automotive}
  {Scale} {Pouch} {Cells}. \emph{Journal of The Electrochemical Society}
  \textbf{2014}, \emph{161}, A1364, Publisher: IOP Publishing\relax
\mciteBstWouldAddEndPuncttrue
\mciteSetBstMidEndSepPunct{\mcitedefaultmidpunct}
{\mcitedefaultendpunct}{\mcitedefaultseppunct}\relax
\EndOfBibitem
\bibitem[Park \latin{et~al.}(2017)Park, Appiah, Byun, Jin, Ryou, and
  Lee]{park_semi-empirical_2017}
Park,~J.; Appiah,~W.~A.; Byun,~S.; Jin,~D.; Ryou,~M.-H.; Lee,~Y.~M.
  Semi-empirical long-term cycle life model coupled with an electrolyte
  depletion function for large-format graphite/{LiFePO} 4 lithium-ion
  batteries. \emph{Journal of Power Sources} \textbf{2017}, \emph{365},
  257--265\relax
\mciteBstWouldAddEndPuncttrue
\mciteSetBstMidEndSepPunct{\mcitedefaultmidpunct}
{\mcitedefaultendpunct}{\mcitedefaultseppunct}\relax
\EndOfBibitem
\bibitem[Li \latin{et~al.}(2017)Li, Cameron, Li, Glazier, Xiong, Chatzidakis,
  Allen, Botton, and Dahn]{li_comparison_2017}
Li,~J.; Cameron,~A.~R.; Li,~H.; Glazier,~S.; Xiong,~D.; Chatzidakis,~M.;
  Allen,~J.; Botton,~G.~A.; Dahn,~J.~R. Comparison of {Single} {Crystal} and
  {Polycrystalline} {LiNi0}.{5Mn0}.{3Co0}.{2O2} {Positive} {Electrode}
  {Materials} for {High} {Voltage} {Li}-{Ion} {Cells}. \emph{Journal of The
  Electrochemical Society} \textbf{2017}, \emph{164}, A1534, Publisher: IOP
  Publishing\relax
\mciteBstWouldAddEndPuncttrue
\mciteSetBstMidEndSepPunct{\mcitedefaultmidpunct}
{\mcitedefaultendpunct}{\mcitedefaultseppunct}\relax
\EndOfBibitem
\bibitem[Burns \latin{et~al.}(2013)Burns, Kassam, Sinha, Downie, Solnickova,
  Way, and Dahn]{burns_predicting_2013}
Burns,~J.~C.; Kassam,~A.; Sinha,~N.~N.; Downie,~L.~E.; Solnickova,~L.; Way,~B.;
  Dahn,~J.~R. Predicting and {Extending} the {Lifetime} of {Li}-{Ion}
  {Batteries}. \emph{Journal of The Electrochemical Society} \textbf{2013},
  \emph{160}, A1451--A1456\relax
\mciteBstWouldAddEndPuncttrue
\mciteSetBstMidEndSepPunct{\mcitedefaultmidpunct}
{\mcitedefaultendpunct}{\mcitedefaultseppunct}\relax
\EndOfBibitem
\bibitem[Choi \latin{et~al.}(2006)Choi, Yew, Lee, Sung, Kim, and
  Kim]{choi_effect_2006}
Choi,~N.-S.; Yew,~K.~H.; Lee,~K.~Y.; Sung,~M.; Kim,~H.; Kim,~S.-S. Effect of
  fluoroethylene carbonate additive on interfacial properties of silicon
  thin-film electrode. \emph{Journal of Power Sources} \textbf{2006},
  \emph{161}, 1254--1259\relax
\mciteBstWouldAddEndPuncttrue
\mciteSetBstMidEndSepPunct{\mcitedefaultmidpunct}
{\mcitedefaultendpunct}{\mcitedefaultseppunct}\relax
\EndOfBibitem
\bibitem[Etacheri \latin{et~al.}(2012)Etacheri, Haik, Goffer, Roberts, Stefan,
  Fasching, and Aurbach]{etacheri_effect_2012}
Etacheri,~V.; Haik,~O.; Goffer,~Y.; Roberts,~G.~A.; Stefan,~I.~C.;
  Fasching,~R.; Aurbach,~D. Effect of {Fluoroethylene} {Carbonate} ({FEC}) on
  the {Performance} and {Surface} {Chemistry} of {Si}-{Nanowire} {Li}-{Ion}
  {Battery} {Anodes}. \emph{Langmuir} \textbf{2012}, \emph{28}, 965--976,
  Publisher: American Chemical Society\relax
\mciteBstWouldAddEndPuncttrue
\mciteSetBstMidEndSepPunct{\mcitedefaultmidpunct}
{\mcitedefaultendpunct}{\mcitedefaultseppunct}\relax
\EndOfBibitem
\bibitem[Wetjen \latin{et~al.}(2017)Wetjen, Pritzl, Jung, Solchenbach, Ghadimi,
  and Gasteiger]{wetjen_differentiating_2017}
Wetjen,~M.; Pritzl,~D.; Jung,~R.; Solchenbach,~S.; Ghadimi,~R.;
  Gasteiger,~H.~A. Differentiating the {Degradation} {Phenomena} in
  {Silicon}-{Graphite} {Electrodes} for {Lithium}-{Ion} {Batteries}.
  \emph{Journal of The Electrochemical Society} \textbf{2017}, \emph{164},
  A2840, Publisher: IOP Publishing\relax
\mciteBstWouldAddEndPuncttrue
\mciteSetBstMidEndSepPunct{\mcitedefaultmidpunct}
{\mcitedefaultendpunct}{\mcitedefaultseppunct}\relax
\EndOfBibitem
\bibitem[Petibon \latin{et~al.}(2016)Petibon, Chevrier, Aiken, Hall, Hyatt,
  Shunmugasundaram, and Dahn]{petibon_studies_2016}
Petibon,~R.; Chevrier,~V.~L.; Aiken,~C.~P.; Hall,~D.~S.; Hyatt,~S.~R.;
  Shunmugasundaram,~R.; Dahn,~J.~R. Studies of the {Capacity} {Fade}
  {Mechanisms} of {LiCoO2}/{Si}-{Alloy}: {Graphite} {Cells}. \emph{Journal of
  The Electrochemical Society} \textbf{2016}, \emph{163}, A1146, Publisher: IOP
  Publishing\relax
\mciteBstWouldAddEndPuncttrue
\mciteSetBstMidEndSepPunct{\mcitedefaultmidpunct}
{\mcitedefaultendpunct}{\mcitedefaultseppunct}\relax
\EndOfBibitem
\bibitem[Jung \latin{et~al.}(2016)Jung, Metzger, Haering, Solchenbach, Marino,
  Tsiouvaras, Stinner, and Gasteiger]{jung_consumption_2016}
Jung,~R.; Metzger,~M.; Haering,~D.; Solchenbach,~S.; Marino,~C.;
  Tsiouvaras,~N.; Stinner,~C.; Gasteiger,~H.~A. Consumption of {Fluoroethylene}
  {Carbonate} ({FEC}) on {Si}-{C} {Composite} {Electrodes} for {Li}-{Ion}
  {Batteries}. \emph{Journal of The Electrochemical Society} \textbf{2016},
  \emph{163}, A1705, Publisher: IOP Publishing\relax
\mciteBstWouldAddEndPuncttrue
\mciteSetBstMidEndSepPunct{\mcitedefaultmidpunct}
{\mcitedefaultendpunct}{\mcitedefaultseppunct}\relax
\EndOfBibitem
\bibitem[Louli \latin{et~al.}(2019)Louli, Ellis, and Dahn]{louli_operando_2019}
Louli,~A.~J.; Ellis,~L.~D.; Dahn,~J.~R. Operando {Pressure} {Measurements}
  {Reveal} {Solid} {Electrolyte} {Interphase} {Growth} to {Rank} {Li}-{Ion}
  {Cell} {Performance}. \emph{Joule} \textbf{2019}, \emph{3}, 745--761\relax
\mciteBstWouldAddEndPuncttrue
\mciteSetBstMidEndSepPunct{\mcitedefaultmidpunct}
{\mcitedefaultendpunct}{\mcitedefaultseppunct}\relax
\EndOfBibitem
\bibitem[Knehr \latin{et~al.}(2018)Knehr, Hodson, Bommier, Davies, Kim, and
  Steingart]{knehr_understanding_2018}
Knehr,~K.~W.; Hodson,~T.; Bommier,~C.; Davies,~G.; Kim,~A.; Steingart,~D.~A.
  Understanding {Full}-{Cell} {Evolution} and {Non}-chemical {Electrode}
  {Crosstalk} of {Li}-{Ion} {Batteries}. \emph{Joule} \textbf{2018}, \emph{2},
  1146--1159\relax
\mciteBstWouldAddEndPuncttrue
\mciteSetBstMidEndSepPunct{\mcitedefaultmidpunct}
{\mcitedefaultendpunct}{\mcitedefaultseppunct}\relax
\EndOfBibitem
\bibitem[Deng \latin{et~al.}(2020)Deng, Huang, Shen, Huang, Ding, Luscombe,
  Johnson, Harlow, Gauthier, and Dahn]{deng_ultrasonic_2020}
Deng,~Z.; Huang,~Z.; Shen,~Y.; Huang,~Y.; Ding,~H.; Luscombe,~A.; Johnson,~M.;
  Harlow,~J.~E.; Gauthier,~R.; Dahn,~J.~R. Ultrasonic {Scanning} to {Observe}
  {Wetting} and “{Unwetting}” in {Li}-{Ion} {Pouch} {Cells}. \emph{Joule}
  \textbf{2020}, \emph{4}, 2017--2029\relax
\mciteBstWouldAddEndPuncttrue
\mciteSetBstMidEndSepPunct{\mcitedefaultmidpunct}
{\mcitedefaultendpunct}{\mcitedefaultseppunct}\relax
\EndOfBibitem
\bibitem[Essam(1980)]{essam_percolation_1980}
Essam,~J.~W. Percolation theory. \emph{Reports on Progress in Physics}
  \textbf{1980}, \emph{43}, 833--912, Publisher: IOP Publishing\relax
\mciteBstWouldAddEndPuncttrue
\mciteSetBstMidEndSepPunct{\mcitedefaultmidpunct}
{\mcitedefaultendpunct}{\mcitedefaultseppunct}\relax
\EndOfBibitem
\bibitem[Stauffer and Aharony(1994)Stauffer, and
  Aharony]{stauffer_introduction_1994}
Stauffer,~D.; Aharony,~A. \emph{Introduction {To} {Percolation} {Theory}:
  {Second} {Edition}}; 1994\relax
\mciteBstWouldAddEndPuncttrue
\mciteSetBstMidEndSepPunct{\mcitedefaultmidpunct}
{\mcitedefaultendpunct}{\mcitedefaultseppunct}\relax
\EndOfBibitem
\bibitem[Ferguson and Bazant(2012)Ferguson, and
  Bazant]{ferguson_nonequilibrium_2012}
Ferguson,~T.~R.; Bazant,~M.~Z. Nonequilibrium {Thermodynamics} of {Porous}
  {Electrodes}. \emph{Journal of The Electrochemical Society} \textbf{2012},
  \emph{159}, A1967--A1985\relax
\mciteBstWouldAddEndPuncttrue
\mciteSetBstMidEndSepPunct{\mcitedefaultmidpunct}
{\mcitedefaultendpunct}{\mcitedefaultseppunct}\relax
\EndOfBibitem
\bibitem[Chen \latin{et~al.}(2007)Chen, Wang, Liu, Song, Battaglia, and
  Sastry]{chen_selection_2007}
Chen,~Y.-H.; Wang,~C.-W.; Liu,~G.; Song,~X.-Y.; Battaglia,~V.~S.; Sastry,~A.~M.
  Selection of {Conductive} {Additives} in {Li}-{Ion} {Battery} {Cathodes}: {A}
  {Numerical} {Study}. \emph{Journal of The Electrochemical Society}
  \textbf{2007}, \emph{154}, A978\relax
\mciteBstWouldAddEndPuncttrue
\mciteSetBstMidEndSepPunct{\mcitedefaultmidpunct}
{\mcitedefaultendpunct}{\mcitedefaultseppunct}\relax
\EndOfBibitem
\bibitem[Li \latin{et~al.}(2015)Li, Meyer, Lim, Lee, Gent, Marchesini,
  Krishnan, Tyliszczak, Shapiro, Kilcoyne, and Chueh]{li_effects_2015}
Li,~Y.; Meyer,~S.; Lim,~J.; Lee,~S.~C.; Gent,~W.~E.; Marchesini,~S.;
  Krishnan,~H.; Tyliszczak,~T.; Shapiro,~D.; Kilcoyne,~A. L.~D. \latin{et~al.}
  Effects of {Particle} {Size}, {Electronic} {Connectivity}, and {Incoherent}
  {Nanoscale} {Domains} on the {Sequence} of {Lithiation} in {LiFePO4} {Porous}
  {Electrodes}. \emph{Advanced Materials} \textbf{2015}, \emph{27}, 6591--6597,
  \_eprint:
  https://onlinelibrary.wiley.com/doi/pdf/10.1002/adma.201502276\relax
\mciteBstWouldAddEndPuncttrue
\mciteSetBstMidEndSepPunct{\mcitedefaultmidpunct}
{\mcitedefaultendpunct}{\mcitedefaultseppunct}\relax
\EndOfBibitem
\bibitem[Cerbelaud \latin{et~al.}(2015)Cerbelaud, Lestriez, Videcoq, Ferrando,
  and Guyomard]{cerbelaud_understanding_2015}
Cerbelaud,~M.; Lestriez,~B.; Videcoq,~A.; Ferrando,~R.; Guyomard,~D.
  Understanding the {Structure} of {Electrodes} in {Li}-{Ion} {Batteries}: {A}
  {Numerical} {Study}. \emph{Journal of The Electrochemical Society}
  \textbf{2015}, \emph{162}, A1485--A1492\relax
\mciteBstWouldAddEndPuncttrue
\mciteSetBstMidEndSepPunct{\mcitedefaultmidpunct}
{\mcitedefaultendpunct}{\mcitedefaultseppunct}\relax
\EndOfBibitem
\bibitem[Guzmán \latin{et~al.}(2017)Guzmán, Vazquez-Arenas, Ramos-Sánchez,
  Bautista-Ramírez, and González]{guzman_improved_2017}
Guzmán,~G.; Vazquez-Arenas,~J.; Ramos-Sánchez,~G.; Bautista-Ramírez,~M.;
  González,~I. Improved performance of {LiFePO4} cathode for {Li}-ion
  batteries through percolation studies. \emph{Electrochimica Acta}
  \textbf{2017}, \emph{247}, 451--459\relax
\mciteBstWouldAddEndPuncttrue
\mciteSetBstMidEndSepPunct{\mcitedefaultmidpunct}
{\mcitedefaultendpunct}{\mcitedefaultseppunct}\relax
\EndOfBibitem
\bibitem[Lewerenz \latin{et~al.}(2017)Lewerenz, Warnecke, and
  Sauer]{lewerenz_post-mortem_2017}
Lewerenz,~M.; Warnecke,~A.; Sauer,~D.~U. Post-mortem analysis on
  {LiFePO4}{\textbar}{Graphite} cells describing the evolution \& composition
  of covering layer on anode and their impact on cell performance.
  \emph{Journal of Power Sources} \textbf{2017}, \emph{369}, 122--132\relax
\mciteBstWouldAddEndPuncttrue
\mciteSetBstMidEndSepPunct{\mcitedefaultmidpunct}
{\mcitedefaultendpunct}{\mcitedefaultseppunct}\relax
\EndOfBibitem
\bibitem[Stiaszny \latin{et~al.}(2014)Stiaszny, Ziegler, Krauß, Schmidt, and
  Ivers-Tiffée]{stiaszny_electrochemical_2014}
Stiaszny,~B.; Ziegler,~J.~C.; Krauß,~E.~E.; Schmidt,~J.~P.; Ivers-Tiffée,~E.
  Electrochemical characterization and post-mortem analysis of aged
  {LiMn2O4}–{Li}({Ni0}.{5Mn0}.{3Co0}.2){O2}/graphite lithium ion batteries.
  {Part} {I}: {Cycle} aging. \emph{Journal of Power Sources} \textbf{2014},
  \emph{251}, 439--450\relax
\mciteBstWouldAddEndPuncttrue
\mciteSetBstMidEndSepPunct{\mcitedefaultmidpunct}
{\mcitedefaultendpunct}{\mcitedefaultseppunct}\relax
\EndOfBibitem
\bibitem[Dai \latin{et~al.}(2014)Dai, Cai, and White]{dai_simulation_2014}
Dai,~Y.; Cai,~L.; White,~R.~E. Simulation and analysis of stress in a {Li}-ion
  battery with a blended {LiMn} 2 {O} 4 and {LiNi} 0.8 {Co} 0.15 {Al} 0.05 {O}
  2 cathode. \emph{Journal of Power Sources} \textbf{2014}, \emph{247},
  365--376\relax
\mciteBstWouldAddEndPuncttrue
\mciteSetBstMidEndSepPunct{\mcitedefaultmidpunct}
{\mcitedefaultendpunct}{\mcitedefaultseppunct}\relax
\EndOfBibitem
\bibitem[Jana \latin{et~al.}(2019)Jana, Shaver, and
  García]{jana_physical_2019}
Jana,~A.; Shaver,~G.~M.; García,~R.~E. Physical, on the fly, capacity
  degradation prediction of {LiNiMnCoO2}-graphite cells. \emph{Journal of Power
  Sources} \textbf{2019}, \emph{422}, 185--195\relax
\mciteBstWouldAddEndPuncttrue
\mciteSetBstMidEndSepPunct{\mcitedefaultmidpunct}
{\mcitedefaultendpunct}{\mcitedefaultseppunct}\relax
\EndOfBibitem
\bibitem[Zhu \latin{et~al.}(2021)Zhu, Knapp, Sorensen, Darma, Muller, Mereacre,
  Dai, Senyshyn, Wei, and Ehrenberg]{zhu_investigation_2021}
Zhu,~J.; Knapp,~M.; Sorensen,~D.~R.; Darma,~M. S.~D.; Muller,~M.; Mereacre,~L.;
  Dai,~H.; Senyshyn,~A.; Wei,~X.; Ehrenberg,~H. Investigation of capacity fade
  for 18650-type lithium-ion batteries cycled in different state of charge
  ({SoC}) ranges. \emph{Journal of Power Sources} \textbf{2021}, \emph{489},
  229422\relax
\mciteBstWouldAddEndPuncttrue
\mciteSetBstMidEndSepPunct{\mcitedefaultmidpunct}
{\mcitedefaultendpunct}{\mcitedefaultseppunct}\relax
\EndOfBibitem
\bibitem[Li \latin{et~al.}(2016)Li, L.~Danilov, Gao, Yang, and
  Notten]{li_degradation_2016}
Li,~D.; L.~Danilov,~D.; Gao,~L.; Yang,~Y.; Notten,~P.~H. Degradation
  {Mechanisms} of {C6}/{LiFePO4} {Batteries}: {Experimental} {Analyses} of
  {Cycling}-induced {Aging}. \emph{Electrochimica Acta} \textbf{2016},
  \emph{210}, 445--455\relax
\mciteBstWouldAddEndPuncttrue
\mciteSetBstMidEndSepPunct{\mcitedefaultmidpunct}
{\mcitedefaultendpunct}{\mcitedefaultseppunct}\relax
\EndOfBibitem
\bibitem[Peled and Menkin(2017)Peled, and Menkin]{peled_reviewsei_2017}
Peled,~E.; Menkin,~S. Review—{SEI}: {Past}, {Present} and {Future}.
  \emph{Journal of The Electrochemical Society} \textbf{2017}, \emph{164},
  A1703--A1719\relax
\mciteBstWouldAddEndPuncttrue
\mciteSetBstMidEndSepPunct{\mcitedefaultmidpunct}
{\mcitedefaultendpunct}{\mcitedefaultseppunct}\relax
\EndOfBibitem
\bibitem[Wünsch \latin{et~al.}(2019)Wünsch, Kaufman, and
  Sauer]{wunsch_investigation_2019}
Wünsch,~M.; Kaufman,~J.; Sauer,~D.~U. Investigation of the influence of
  different bracing of automotive pouch cells on cyclic liefetime and impedance
  spectra. \emph{Journal of Energy Storage} \textbf{2019}, \emph{21},
  149--155\relax
\mciteBstWouldAddEndPuncttrue
\mciteSetBstMidEndSepPunct{\mcitedefaultmidpunct}
{\mcitedefaultendpunct}{\mcitedefaultseppunct}\relax
\EndOfBibitem
\bibitem[Beck \latin{et~al.}(2021)Beck, Dechent, Junker, Sauer, and
  Dubarry]{beck_inhomogeneities_2021}
Beck,~D.; Dechent,~P.; Junker,~M.; Sauer,~D.~U.; Dubarry,~M. Inhomogeneities
  and {Cell}-to-{Cell} {Variations} in {Lithium}-{Ion} {Batteries}, a {Review}.
  \emph{Energies} \textbf{2021}, \emph{14}, 3276, Number: 11 Publisher:
  Multidisciplinary Digital Publishing Institute\relax
\mciteBstWouldAddEndPuncttrue
\mciteSetBstMidEndSepPunct{\mcitedefaultmidpunct}
{\mcitedefaultendpunct}{\mcitedefaultseppunct}\relax
\EndOfBibitem
\bibitem[Waldmann \latin{et~al.}(2014)Waldmann, Gorse, Samtleben, Schneider,
  Knoblauch, and Wohlfahrt-Mehrens]{waldmann_mechanical_2014}
Waldmann,~T.; Gorse,~S.; Samtleben,~T.; Schneider,~G.; Knoblauch,~V.;
  Wohlfahrt-Mehrens,~M. A {Mechanical} {Aging} {Mechanism} in {Lithium}-{Ion}
  {Batteries}. \emph{Journal of The Electrochemical Society} \textbf{2014},
  \emph{161}, A1742--A1747\relax
\mciteBstWouldAddEndPuncttrue
\mciteSetBstMidEndSepPunct{\mcitedefaultmidpunct}
{\mcitedefaultendpunct}{\mcitedefaultseppunct}\relax
\EndOfBibitem
\bibitem[Pfrang \latin{et~al.}(2018)Pfrang, Kersys, Kriston, Sauer, Rahe,
  Käbitz, and Figgemeier]{pfrang_long-term_2018}
Pfrang,~A.; Kersys,~A.; Kriston,~A.; Sauer,~D.~U.; Rahe,~C.; Käbitz,~S.;
  Figgemeier,~E. Long-term cycling induced jelly roll deformation in commercial
  18650 cells. \emph{Journal of Power Sources} \textbf{2018}, \emph{392},
  168--175\relax
\mciteBstWouldAddEndPuncttrue
\mciteSetBstMidEndSepPunct{\mcitedefaultmidpunct}
{\mcitedefaultendpunct}{\mcitedefaultseppunct}\relax
\EndOfBibitem
\bibitem[Pfrang \latin{et~al.}(2019)Pfrang, Kersys, Kriston, Sauer, Rahe,
  Käbitz, and Figgemeier]{pfrang_geometrical_2019}
Pfrang,~A.; Kersys,~A.; Kriston,~A.; Sauer,~D.~U.; Rahe,~C.; Käbitz,~S.;
  Figgemeier,~E. Geometrical {Inhomogeneities} as {Cause} of {Mechanical}
  {Failure} in {Commercial} 18650 {Lithium} {Ion} {Cells}. \emph{Journal of The
  Electrochemical Society} \textbf{2019}, \emph{166}, A3745, Publisher: IOP
  Publishing\relax
\mciteBstWouldAddEndPuncttrue
\mciteSetBstMidEndSepPunct{\mcitedefaultmidpunct}
{\mcitedefaultendpunct}{\mcitedefaultseppunct}\relax
\EndOfBibitem
\bibitem[Carter \latin{et~al.}(2019)Carter, Klein, Atkinson, and
  Love]{carter_mechanical_2019}
Carter,~R.; Klein,~E.~J.; Atkinson,~R.~W.; Love,~C.~T. Mechanical collapse as
  primary degradation mode in mandrel-free 18650 {Li}-ion cells operated at 0
  °{C}. \emph{Journal of Power Sources} \textbf{2019}, \emph{437},
  226820\relax
\mciteBstWouldAddEndPuncttrue
\mciteSetBstMidEndSepPunct{\mcitedefaultmidpunct}
{\mcitedefaultendpunct}{\mcitedefaultseppunct}\relax
\EndOfBibitem
\bibitem[Kok \latin{et~al.}(2019)Kok, Robinson, Weaving, Jnawali, Pham,
  Iacoviello, Brett, and Shearing]{kok_virtual_2019}
Kok,~M. D.~R.; Robinson,~J.~B.; Weaving,~J.~S.; Jnawali,~A.; Pham,~M.;
  Iacoviello,~F.; Brett,~D. J.~L.; Shearing,~P.~R. Virtual unrolling of
  spirally-wound lithium-ion cells for correlative degradation studies and
  predictive fault detection. \emph{Sustainable Energy \& Fuels} \textbf{2019},
  \emph{3}, 2972--2976\relax
\mciteBstWouldAddEndPuncttrue
\mciteSetBstMidEndSepPunct{\mcitedefaultmidpunct}
{\mcitedefaultendpunct}{\mcitedefaultseppunct}\relax
\EndOfBibitem
\bibitem[Chung \latin{et~al.}(2014)Chung, Shearing, Brandon, Harris, and
  García]{chung_particle_2014}
Chung,~D.-W.; Shearing,~P.~R.; Brandon,~N.~P.; Harris,~S.~J.; García,~R.~E.
  Particle {Size} {Polydispersity} in {Li}-{Ion} {Batteries}. \emph{Journal of
  The Electrochemical Society} \textbf{2014}, \emph{161}, A422, Publisher: IOP
  Publishing\relax
\mciteBstWouldAddEndPuncttrue
\mciteSetBstMidEndSepPunct{\mcitedefaultmidpunct}
{\mcitedefaultendpunct}{\mcitedefaultseppunct}\relax
\EndOfBibitem
\bibitem[Lee \latin{et~al.}(2013)Lee, Smith, Pesaran, and Kim]{lee_three_2013}
Lee,~K.-J.; Smith,~K.; Pesaran,~A.; Kim,~G.-H. Three dimensional thermal-,
  electrical-, and electrochemical-coupled model for cylindrical wound large
  format lithium-ion batteries. \emph{Journal of Power Sources} \textbf{2013},
  \emph{241}, 20--32\relax
\mciteBstWouldAddEndPuncttrue
\mciteSetBstMidEndSepPunct{\mcitedefaultmidpunct}
{\mcitedefaultendpunct}{\mcitedefaultseppunct}\relax
\EndOfBibitem
\bibitem[Reimers(2014)]{reimers_accurate_2014}
Reimers,~J.~N. Accurate and {Efficient} {Treatment} of {Foil} {Currents} in a
  {Spiral} {Wound} {Li}-{Ion} {Cell}. \emph{Journal of The Electrochemical
  Society} \textbf{2014}, \emph{161}, A118--A127\relax
\mciteBstWouldAddEndPuncttrue
\mciteSetBstMidEndSepPunct{\mcitedefaultmidpunct}
{\mcitedefaultendpunct}{\mcitedefaultseppunct}\relax
\EndOfBibitem
\bibitem[Senyshyn \latin{et~al.}(2015)Senyshyn, Mühlbauer, Dolotko, Hofmann,
  and Ehrenberg]{senyshyn_homogeneity_2015}
Senyshyn,~A.; Mühlbauer,~M.~J.; Dolotko,~O.; Hofmann,~M.; Ehrenberg,~H.
  Homogeneity of lithium distribution in cylinder-type {Li}-ion batteries.
  \emph{Scientific Reports} \textbf{2015}, \emph{5}, 18380\relax
\mciteBstWouldAddEndPuncttrue
\mciteSetBstMidEndSepPunct{\mcitedefaultmidpunct}
{\mcitedefaultendpunct}{\mcitedefaultseppunct}\relax
\EndOfBibitem
\bibitem[Waldmann \latin{et~al.}(2015)Waldmann, Bisle, Hogg, Stumpp, Danzer,
  Kasper, Axmann, and Wohlfahrt-Mehrens]{waldmann_influence_2015}
Waldmann,~T.; Bisle,~G.; Hogg,~B.-I.; Stumpp,~S.; Danzer,~M.~A.; Kasper,~M.;
  Axmann,~P.; Wohlfahrt-Mehrens,~M. Influence of {Cell} {Design} on
  {Temperatures} and {Temperature} {Gradients} in {Lithium}-{Ion} {Cells}: {An}
  {In} {Operando} {Study}. \emph{Journal of The Electrochemical Society}
  \textbf{2015}, \emph{162}, A921--A927\relax
\mciteBstWouldAddEndPuncttrue
\mciteSetBstMidEndSepPunct{\mcitedefaultmidpunct}
{\mcitedefaultendpunct}{\mcitedefaultseppunct}\relax
\EndOfBibitem
\bibitem[Waldmann \latin{et~al.}(2016)Waldmann, Geramifard, and
  Wohlfahrt-Mehrens]{waldmann_influence_2016}
Waldmann,~T.; Geramifard,~G.; Wohlfahrt-Mehrens,~M. Influence of current
  collecting tab design on thermal and electrochemical performance of
  cylindrical {Lithium}-ion cells during high current discharge. \emph{Journal
  of Energy Storage} \textbf{2016}, \emph{5}, 163--168\relax
\mciteBstWouldAddEndPuncttrue
\mciteSetBstMidEndSepPunct{\mcitedefaultmidpunct}
{\mcitedefaultendpunct}{\mcitedefaultseppunct}\relax
\EndOfBibitem
\bibitem[Carter \latin{et~al.}(2019)Carter, Klein, Kingston, and
  Love]{carter_detection_2019}
Carter,~R.; Klein,~E.~J.; Kingston,~T.~A.; Love,~C.~T. Detection of {Lithium}
  {Plating} {During} {Thermally} {Transient} {Charging} of {Li}-{Ion}
  {Batteries}. \emph{Frontiers in Energy Research} \textbf{2019}, \emph{7},
  144\relax
\mciteBstWouldAddEndPuncttrue
\mciteSetBstMidEndSepPunct{\mcitedefaultmidpunct}
{\mcitedefaultendpunct}{\mcitedefaultseppunct}\relax
\EndOfBibitem
\bibitem[Yao and Pecht(2019)Yao, and Pecht]{yao_tab_2019}
Yao,~X.; Pecht,~M.~G. Tab {Design} and {Failures} in {Cylindrical} {Li}-ion
  {Batteries}. \emph{IEEE Access} \textbf{2019}, \emph{7}, 24082--24095\relax
\mciteBstWouldAddEndPuncttrue
\mciteSetBstMidEndSepPunct{\mcitedefaultmidpunct}
{\mcitedefaultendpunct}{\mcitedefaultseppunct}\relax
\EndOfBibitem
\bibitem[Li \latin{et~al.}(2021)Li, Kirkaldy, Zhang, Gopalakrishnan,
  Amietszajew, Diaz, Barreras, Shams, Hua, Patel, Offer, and
  Marinescu]{li_optimal_2021}
Li,~S.; Kirkaldy,~N.; Zhang,~C.; Gopalakrishnan,~K.; Amietszajew,~T.;
  Diaz,~L.~B.; Barreras,~J.~V.; Shams,~M.; Hua,~X.; Patel,~Y. \latin{et~al.}
  Optimal cell tab design and cooling strategy for cylindrical lithium-ion
  batteries. \emph{Journal of Power Sources} \textbf{2021}, \emph{492},
  229594\relax
\mciteBstWouldAddEndPuncttrue
\mciteSetBstMidEndSepPunct{\mcitedefaultmidpunct}
{\mcitedefaultendpunct}{\mcitedefaultseppunct}\relax
\EndOfBibitem
\bibitem[Werner \latin{et~al.}(2020)Werner, Paarmann, Wiebelt, and
  Wetzel]{werner_inhomogeneous_2020}
Werner,~D.; Paarmann,~S.; Wiebelt,~A.; Wetzel,~T. Inhomogeneous {Temperature}
  {Distribution} {Affecting} the {Cyclic} {Aging} of {Li}-{Ion} {Cells}. {Part}
  {II}: {Analysis} and {Correlation}. \emph{Batteries} \textbf{2020}, \emph{6},
  12, Number: 1 Publisher: Multidisciplinary Digital Publishing Institute\relax
\mciteBstWouldAddEndPuncttrue
\mciteSetBstMidEndSepPunct{\mcitedefaultmidpunct}
{\mcitedefaultendpunct}{\mcitedefaultseppunct}\relax
\EndOfBibitem
\bibitem[Dechent \latin{et~al.}(2021)Dechent, Greenbank, Hildenbrand, Jbabdi,
  Sauer, and Howey]{dechent_estimation_2021}
Dechent,~P.; Greenbank,~S.; Hildenbrand,~F.; Jbabdi,~S.; Sauer,~D.~U.;
  Howey,~D.~A. Estimation of {Li}-{Ion} {Degradation} {Test} {Sample} {Sizes}
  {Required} to {Understand} {Cell}-to-{Cell} {Variability}. \emph{Batteries \&
  Supercaps} \textbf{2021}, \emph{4}, 1821--1829, \_eprint:
  https://onlinelibrary.wiley.com/doi/pdf/10.1002/batt.202100148\relax
\mciteBstWouldAddEndPuncttrue
\mciteSetBstMidEndSepPunct{\mcitedefaultmidpunct}
{\mcitedefaultendpunct}{\mcitedefaultseppunct}\relax
\EndOfBibitem
\bibitem[Baumhöfer \latin{et~al.}(2014)Baumhöfer, Brühl, Rothgang, and
  Sauer]{baumhofer_production_2014}
Baumhöfer,~T.; Brühl,~M.; Rothgang,~S.; Sauer,~D.~U. Production caused
  variation in capacity aging trend and correlation to initial cell
  performance. \emph{Journal of Power Sources} \textbf{2014}, \emph{247},
  332--338\relax
\mciteBstWouldAddEndPuncttrue
\mciteSetBstMidEndSepPunct{\mcitedefaultmidpunct}
{\mcitedefaultendpunct}{\mcitedefaultseppunct}\relax
\EndOfBibitem
\bibitem[Harris \latin{et~al.}(2017)Harris, Harris, and
  Li]{harris_failure_2017}
Harris,~S.~J.; Harris,~D.~J.; Li,~C. Failure statistics for commercial lithium
  ion batteries: {A} study of 24 pouch cells. \emph{Journal of Power Sources}
  \textbf{2017}, \emph{342}, 589--597\relax
\mciteBstWouldAddEndPuncttrue
\mciteSetBstMidEndSepPunct{\mcitedefaultmidpunct}
{\mcitedefaultendpunct}{\mcitedefaultseppunct}\relax
\EndOfBibitem
\bibitem[Klein \latin{et~al.}(2021)Klein, Bärmann, Stolz, Borzutzki,
  Schmiegel, Börner, Winter, Placke, and
  Kasnatscheew]{klein_demonstrating_2021}
Klein,~S.; Bärmann,~P.; Stolz,~L.; Borzutzki,~K.; Schmiegel,~J.-P.;
  Börner,~M.; Winter,~M.; Placke,~T.; Kasnatscheew,~J. Demonstrating
  {Apparently} {Inconspicuous} but {Sensitive} {Impacts} on the {Rollover}
  {Failure} of {Lithium}-{Ion} {Batteries} at a {High} {Voltage}. \emph{ACS
  Applied Materials \& Interfaces} \textbf{2021}, acsami.1c17408\relax
\mciteBstWouldAddEndPuncttrue
\mciteSetBstMidEndSepPunct{\mcitedefaultmidpunct}
{\mcitedefaultendpunct}{\mcitedefaultseppunct}\relax
\EndOfBibitem
\bibitem[Glazier \latin{et~al.}(2017)Glazier, Li, Louli, Allen, and
  Dahn]{glazier_analysis_2017}
Glazier,~S.~L.; Li,~J.; Louli,~A.~J.; Allen,~J.~P.; Dahn,~J.~R. An {Analysis}
  of {Artificial} and {Natural} {Graphite} in {Lithium} {Ion} {Pouch} {Cells}
  {Using} {Ultra}-{High} {Precision} {Coulometry}, {Isothermal}
  {Microcalorimetry}, {Gas} {Evolution}, {Long} {Term} {Cycling} and {Pressure}
  {Measurements}. \emph{Journal of The Electrochemical Society} \textbf{2017},
  \emph{164}, A3545, Publisher: IOP Publishing\relax
\mciteBstWouldAddEndPuncttrue
\mciteSetBstMidEndSepPunct{\mcitedefaultmidpunct}
{\mcitedefaultendpunct}{\mcitedefaultseppunct}\relax
\EndOfBibitem
\bibitem[Wang \latin{et~al.}(2014)Wang, Burns, and Dahn]{wang_systematic_2014}
Wang,~D.~Y.; Burns,~J.~C.; Dahn,~J.~R. A {Systematic} {Study} of the
  {Concentration} of {Lithium} {Hexafluorophosphate} ({LiPF6}) as a {Salt} for
  {LiCoO2}/{Graphite} {Pouch} {Cells}. \emph{Journal of The Electrochemical
  Society} \textbf{2014}, \emph{161}, A1278, Publisher: IOP Publishing\relax
\mciteBstWouldAddEndPuncttrue
\mciteSetBstMidEndSepPunct{\mcitedefaultmidpunct}
{\mcitedefaultendpunct}{\mcitedefaultseppunct}\relax
\EndOfBibitem
\bibitem[Weng \latin{et~al.}(2021)Weng, Mohtat, Attia, Sulzer, Lee, Less, and
  Stefanopoulou]{weng_predicting_2021}
Weng,~A.; Mohtat,~P.; Attia,~P.~M.; Sulzer,~V.; Lee,~S.; Less,~G.;
  Stefanopoulou,~A. Predicting the impact of formation protocols on battery
  lifetime immediately after manufacturing. \emph{Joule} \textbf{2021},
  \emph{5}, 2971--2992\relax
\mciteBstWouldAddEndPuncttrue
\mciteSetBstMidEndSepPunct{\mcitedefaultmidpunct}
{\mcitedefaultendpunct}{\mcitedefaultseppunct}\relax
\EndOfBibitem
\bibitem[Attia \latin{et~al.}(2021)Attia, Harris, and
  Chueh]{attia_benefits_2021}
Attia,~P.~M.; Harris,~S.~J.; Chueh,~W.~C. Benefits of {Fast} {Battery}
  {Formation} in a {Model} {System}. \emph{Journal of The Electrochemical
  Society} \textbf{2021}, \emph{168}, 050543, Publisher: The Electrochemical
  Society\relax
\mciteBstWouldAddEndPuncttrue
\mciteSetBstMidEndSepPunct{\mcitedefaultmidpunct}
{\mcitedefaultendpunct}{\mcitedefaultseppunct}\relax
\EndOfBibitem
\bibitem[Sulzer \latin{et~al.}(2021)Sulzer, Mohtat, Aitio, Lee, Yeh,
  Steinbacher, Khan, Lee, Siegel, Stefanopoulou, and
  Howey]{sulzer_challenge_2021}
Sulzer,~V.; Mohtat,~P.; Aitio,~A.; Lee,~S.; Yeh,~Y.~T.; Steinbacher,~F.;
  Khan,~M.~U.; Lee,~J.~W.; Siegel,~J.~B.; Stefanopoulou,~A.~G. \latin{et~al.}
  The challenge and opportunity of battery lifetime prediction from field data.
  \emph{Joule} \textbf{2021}, \relax
\mciteBstWouldAddEndPunctfalse
\mciteSetBstMidEndSepPunct{\mcitedefaultmidpunct}
{}{\mcitedefaultseppunct}\relax
\EndOfBibitem
\bibitem[Omar \latin{et~al.}(2014)Omar, Monem, Firouz, Salminen, Smekens,
  Hegazy, Gaulous, Mulder, Van~den Bossche, Coosemans, and
  Van~Mierlo]{omar_lithium_2014}
Omar,~N.; Monem,~M.~A.; Firouz,~Y.; Salminen,~J.; Smekens,~J.; Hegazy,~O.;
  Gaulous,~H.; Mulder,~G.; Van~den Bossche,~P.; Coosemans,~T. \latin{et~al.}
  Lithium iron phosphate based battery – {Assessment} of the aging parameters
  and development of cycle life model. \emph{Applied Energy} \textbf{2014},
  \emph{113}, 1575--1585\relax
\mciteBstWouldAddEndPuncttrue
\mciteSetBstMidEndSepPunct{\mcitedefaultmidpunct}
{\mcitedefaultendpunct}{\mcitedefaultseppunct}\relax
\EndOfBibitem
\bibitem[Diao \latin{et~al.}(2019)Diao, Saxena, and
  Pecht]{diao_accelerated_2019}
Diao,~W.; Saxena,~S.; Pecht,~M. Accelerated cycle life testing and capacity
  degradation modeling of {LiCoO2}-graphite cells. \emph{Journal of Power
  Sources} \textbf{2019}, \emph{435}, 226830\relax
\mciteBstWouldAddEndPuncttrue
\mciteSetBstMidEndSepPunct{\mcitedefaultmidpunct}
{\mcitedefaultendpunct}{\mcitedefaultseppunct}\relax
\EndOfBibitem
\bibitem[Christensen and Newman(2006)Christensen, and
  Newman]{christensen_mathematical_2006}
Christensen,~J.; Newman,~J. A {Mathematical} {Model} of {Stress} {Generation}
  and {Fracture} in {Lithium} {Manganese} {Oxide}. \emph{Journal of The
  Electrochemical Society} \textbf{2006}, \emph{153}, A1019, Publisher: IOP
  Publishing\relax
\mciteBstWouldAddEndPuncttrue
\mciteSetBstMidEndSepPunct{\mcitedefaultmidpunct}
{\mcitedefaultendpunct}{\mcitedefaultseppunct}\relax
\EndOfBibitem
\bibitem[Allen \latin{et~al.}(2021)Allen, Weddle, Verma, Mallarapu,
  Usseglio-Viretta, Finegan, Colclasure, Mai, Schmidt, Furat, Diercks, Tanim,
  and Smith]{allen_quantifying_2021}
Allen,~J.~M.; Weddle,~P.~J.; Verma,~A.; Mallarapu,~A.; Usseglio-Viretta,~F.;
  Finegan,~D.~P.; Colclasure,~A.~M.; Mai,~W.; Schmidt,~V.; Furat,~O.
  \latin{et~al.}  Quantifying the influence of charge rate and cathode-particle
  architectures on degradation of {Li}-ion cells through {3D} continuum-level
  damage models. \emph{Journal of Power Sources} \textbf{2021}, \emph{512},
  230415\relax
\mciteBstWouldAddEndPuncttrue
\mciteSetBstMidEndSepPunct{\mcitedefaultmidpunct}
{\mcitedefaultendpunct}{\mcitedefaultseppunct}\relax
\EndOfBibitem
\bibitem[Dubarry \latin{et~al.}(2014)Dubarry, Truchot, and
  Liaw]{dubarry_cell_2014}
Dubarry,~M.; Truchot,~C.; Liaw,~B.~Y. Cell degradation in commercial {LiFePO4}
  cells with high-power and high-energy designs. \emph{Journal of Power
  Sources} \textbf{2014}, \emph{258}, 408--419\relax
\mciteBstWouldAddEndPuncttrue
\mciteSetBstMidEndSepPunct{\mcitedefaultmidpunct}
{\mcitedefaultendpunct}{\mcitedefaultseppunct}\relax
\EndOfBibitem
\bibitem[Sun \latin{et~al.}(2018)Sun, Guan, Cheng, Zuo, Gao, Du, and
  Yin]{sun_accelerated_2018}
Sun,~S.; Guan,~T.; Cheng,~X.; Zuo,~P.; Gao,~Y.; Du,~C.; Yin,~G. Accelerated
  aging and degradation mechanism of {LiFePO4}/graphite batteries cycled at
  high discharge rates. \emph{RSC Advances} \textbf{2018}, \emph{8},
  25695--25703, Publisher: The Royal Society of Chemistry\relax
\mciteBstWouldAddEndPuncttrue
\mciteSetBstMidEndSepPunct{\mcitedefaultmidpunct}
{\mcitedefaultendpunct}{\mcitedefaultseppunct}\relax
\EndOfBibitem
\bibitem[Attia \latin{et~al.}(2019)Attia, Das, Harris, Bazant, and
  Chueh]{attia_electrochemical_2019}
Attia,~P.~M.; Das,~S.; Harris,~S.~J.; Bazant,~M.~Z.; Chueh,~W.~C.
  Electrochemical {Kinetics} of {SEI} {Growth} on {Carbon} {Black}: {Part} {I}.
  {Experiments}. \emph{Journal of The Electrochemical Society} \textbf{2019},
  \emph{166}, E97--E106\relax
\mciteBstWouldAddEndPuncttrue
\mciteSetBstMidEndSepPunct{\mcitedefaultmidpunct}
{\mcitedefaultendpunct}{\mcitedefaultseppunct}\relax
\EndOfBibitem
\bibitem[Das \latin{et~al.}(2019)Das, Attia, Chueh, and
  Bazant]{das_electrochemical_2019}
Das,~S.; Attia,~P.~M.; Chueh,~W.~C.; Bazant,~M.~Z. Electrochemical {Kinetics}
  of {SEI} {Growth} on {Carbon} {Black}: {Part} {II}. {Modeling}. \emph{Journal
  of The Electrochemical Society} \textbf{2019}, \emph{166}, E107--E118\relax
\mciteBstWouldAddEndPuncttrue
\mciteSetBstMidEndSepPunct{\mcitedefaultmidpunct}
{\mcitedefaultendpunct}{\mcitedefaultseppunct}\relax
\EndOfBibitem
\bibitem[Keil and Jossen(2016)Keil, and Jossen]{keil_charging_2016}
Keil,~P.; Jossen,~A. Charging protocols for lithium-ion batteries and their
  impact on cycle life—{An} experimental study with different 18650
  high-power cells. \emph{Journal of Energy Storage} \textbf{2016}, \emph{6},
  125--141\relax
\mciteBstWouldAddEndPuncttrue
\mciteSetBstMidEndSepPunct{\mcitedefaultmidpunct}
{\mcitedefaultendpunct}{\mcitedefaultseppunct}\relax
\EndOfBibitem
\bibitem[Ma \latin{et~al.}(2019)Ma, Yang, and Wang]{ma_novel_2019}
Ma,~Z.; Yang,~R.; Wang,~Z. A novel data-model fusion state-of-health estimation
  approach for lithium-ion batteries. \emph{Applied Energy} \textbf{2019},
  \emph{237}, 836--847\relax
\mciteBstWouldAddEndPuncttrue
\mciteSetBstMidEndSepPunct{\mcitedefaultmidpunct}
{\mcitedefaultendpunct}{\mcitedefaultseppunct}\relax
\EndOfBibitem
\bibitem[Aiken \latin{et~al.}(2020)Aiken, Harlow, Tingley, Hynes, Logan,
  Glazier, Keefe, and Dahn]{aiken_accelerated_2020}
Aiken,~C.~P.; Harlow,~J.~E.; Tingley,~R.; Hynes,~T.; Logan,~E.~R.;
  Glazier,~S.~L.; Keefe,~A.~S.; Dahn,~J.~R. Accelerated {Failure} in
  {Li}[{Ni0}.{5Mn0}.{3Co0}.2]{O2}/{Graphite} {Pouch} {Cells} {Due} to {Low}
  {LiPF6} {Concentration} and {Extended} {Time} at {High} {Voltage}.
  \emph{Journal of The Electrochemical Society} \textbf{2020}, \emph{167},
  130541, Publisher: The Electrochemical Society\relax
\mciteBstWouldAddEndPuncttrue
\mciteSetBstMidEndSepPunct{\mcitedefaultmidpunct}
{\mcitedefaultendpunct}{\mcitedefaultseppunct}\relax
\EndOfBibitem
\bibitem[Joshi \latin{et~al.}(2014)Joshi, Eom, Yushin, and
  Fuller]{joshi_effects_2014}
Joshi,~T.; Eom,~K.; Yushin,~G.; Fuller,~T.~F. Effects of {Dissolved}
  {Transition} {Metals} on the {Electrochemical} {Performance} and {SEI}
  {Growth} in {Lithium}-{Ion} {Batteries}. \emph{Journal of The Electrochemical
  Society} \textbf{2014}, \emph{161}, A1915--A1921\relax
\mciteBstWouldAddEndPuncttrue
\mciteSetBstMidEndSepPunct{\mcitedefaultmidpunct}
{\mcitedefaultendpunct}{\mcitedefaultseppunct}\relax
\EndOfBibitem
\bibitem[Gilbert \latin{et~al.}(2017)Gilbert, Shkrob, and
  Abraham]{gilbert_transition_2017}
Gilbert,~J.~A.; Shkrob,~I.~A.; Abraham,~D.~P. Transition {Metal} {Dissolution},
  {Ion} {Migration}, {Electrocatalytic} {Reduction} and {Capacity} {Loss} in
  {Lithium}-{Ion} {Full} {Cells}. \emph{Journal of The Electrochemical Society}
  \textbf{2017}, \emph{164}, A389--A399\relax
\mciteBstWouldAddEndPuncttrue
\mciteSetBstMidEndSepPunct{\mcitedefaultmidpunct}
{\mcitedefaultendpunct}{\mcitedefaultseppunct}\relax
\EndOfBibitem
\bibitem[Epding \latin{et~al.}(2019)Epding, Rumberg, Jahnke, Stradtmann, and
  Kwade]{epding_investigation_2019}
Epding,~B.; Rumberg,~B.; Jahnke,~H.; Stradtmann,~I.; Kwade,~A. Investigation of
  significant capacity recovery effects due to long rest periods during high
  current cyclic aging tests in automotive lithium ion cells and their
  influence on lifetime. \emph{Journal of Energy Storage} \textbf{2019},
  \emph{22}, 249--256\relax
\mciteBstWouldAddEndPuncttrue
\mciteSetBstMidEndSepPunct{\mcitedefaultmidpunct}
{\mcitedefaultendpunct}{\mcitedefaultseppunct}\relax
\EndOfBibitem
\bibitem[Rashid and Gupta(2015)Rashid, and Gupta]{rashid_effect_2015}
Rashid,~M.; Gupta,~A. Effect of {Relaxation} {Periods} over {Cycling}
  {Performance} of a {Li}-{Ion} {Battery}. \emph{Journal of The Electrochemical
  Society} \textbf{2015}, \emph{162}, A3145--A3153\relax
\mciteBstWouldAddEndPuncttrue
\mciteSetBstMidEndSepPunct{\mcitedefaultmidpunct}
{\mcitedefaultendpunct}{\mcitedefaultseppunct}\relax
\EndOfBibitem
\bibitem[Raj \latin{et~al.}(2020)Raj, Wang, Monroe, and
  Howey]{raj_investigation_2020}
Raj,~T.; Wang,~A.~A.; Monroe,~C.~W.; Howey,~D.~A. Investigation of
  {Path}-{Dependent} {Degradation} in {Lithium}-{Ion} {Batteries}.
  \emph{Batteries \& Supercaps} \textbf{2020}, \emph{3}, 1377--1385, \_eprint:
  https://onlinelibrary.wiley.com/doi/pdf/10.1002/batt.202000160\relax
\mciteBstWouldAddEndPuncttrue
\mciteSetBstMidEndSepPunct{\mcitedefaultmidpunct}
{\mcitedefaultendpunct}{\mcitedefaultseppunct}\relax
\EndOfBibitem
\bibitem[Uddin \latin{et~al.}(2018)Uddin, Dubarry, and
  Glick]{uddin_viability_2018}
Uddin,~K.; Dubarry,~M.; Glick,~M.~B. The viability of vehicle-to-grid
  operations from a battery technology and policy perspective. \emph{Energy
  Policy} \textbf{2018}, \emph{113}, 342--347\relax
\mciteBstWouldAddEndPuncttrue
\mciteSetBstMidEndSepPunct{\mcitedefaultmidpunct}
{\mcitedefaultendpunct}{\mcitedefaultseppunct}\relax
\EndOfBibitem
\bibitem[Rohatgi(2021)]{Rohatgi2021}
Rohatgi,~A. Webplotdigitizer: Version 4.5. 2021;
  \url{https://automeris.io/WebPlotDigitizer}\relax
\mciteBstWouldAddEndPuncttrue
\mciteSetBstMidEndSepPunct{\mcitedefaultmidpunct}
{\mcitedefaultendpunct}{\mcitedefaultseppunct}\relax
\EndOfBibitem
\bibitem[Martinez-Laserna \latin{et~al.}(2018)Martinez-Laserna,
  Sarasketa-Zabala, Villarreal~Sarria, Stroe, Swierczynski, Warnecke,
  Timmermans, Goutam, Omar, and Rodriguez]{martinez-laserna_technical_2018}
Martinez-Laserna,~E.; Sarasketa-Zabala,~E.; Villarreal~Sarria,~I.;
  Stroe,~D.-I.; Swierczynski,~M.; Warnecke,~A.; Timmermans,~J.-M.; Goutam,~S.;
  Omar,~N.; Rodriguez,~P. Technical {Viability} of {Battery} {Second} {Life}:
  {A} {Study} {From} the {Ageing} {Perspective}. \emph{IEEE Transactions on
  Industry Applications} \textbf{2018}, \emph{54}, 2703--2713\relax
\mciteBstWouldAddEndPuncttrue
\mciteSetBstMidEndSepPunct{\mcitedefaultmidpunct}
{\mcitedefaultendpunct}{\mcitedefaultseppunct}\relax
\EndOfBibitem
\bibitem[Bian \latin{et~al.}(2021)Bian, Wei, Li, Pou, Sauer, and
  Liu]{bian_state--health_2021}
Bian,~X.; Wei,~Z.~G.; Li,~W.; Pou,~J.; Sauer,~D.~U.; Liu,~L. State-of-{Health}
  {Estimation} of {Lithium}-ion {Batteries} by {Fusing} an
  {Open}-{Circuit}-{Voltage} {Model} and {Incremental} {Capacity} {Analysis}.
  \emph{IEEE Transactions on Power Electronics} \textbf{2021}, 1--1\relax
\mciteBstWouldAddEndPuncttrue
\mciteSetBstMidEndSepPunct{\mcitedefaultmidpunct}
{\mcitedefaultendpunct}{\mcitedefaultseppunct}\relax
\EndOfBibitem
\bibitem[Kim \latin{et~al.}(2021)Kim, Yi, Chen, Tanim, and
  Dufek]{kim_rapid_2021}
Kim,~S.; Yi,~Z.; Chen,~B.-R.; Tanim,~T.~R.; Dufek,~E.~J. Rapid {Failure} {Mode}
  {Classification} and {Quantification} in {Batteries}: {A} {Deep} {Learning}
  {Modeling} {Framework}. \emph{Energy Storage Materials} \textbf{2021},
  S2405829721003275\relax
\mciteBstWouldAddEndPuncttrue
\mciteSetBstMidEndSepPunct{\mcitedefaultmidpunct}
{\mcitedefaultendpunct}{\mcitedefaultseppunct}\relax
\EndOfBibitem
\end{mcitethebibliography}

\appendix

\newpage
\section{Appendix}



\newgeometry{margin=2.5cm}
\begin{landscape}
\begin{table}[p]
\caption{Summary of previous work studying the influence of experimental parameters (cell design, testing conditions, or cell-to-cell variation) on knee onset.}
\scalebox{0.5}{
\begin{tabular}{|c|c|l|l|l|l|l|}
\hline
\multicolumn{1}{|l|}{} & \multicolumn{1}{l|}{\textbf{Variable}} & \textbf{Reference} & \textbf{Cell Description} & \textbf{Range of Variable} & \textbf{Knee Acceleration} & \textbf{Proposed Mechanism(s)} \\ \hline
\multirow{9}{*}{
\textbf{Cell design}} & \textbf{Electrode loading} & Ma et al. 2019 \cite{ma_editors_2019} & Lab-made pouch NMC/Gr & 14.4--21.2 mg/cm2 & Higher positive electrode loading & Li plating \\ \cline{2-7} 
 & 
 \begin{tabular}[c]{@{}l@{}} \textbf{Positive electrode}\\ \textbf{coating}\end{tabular}
 & Ma et al. 2019 \cite{ma_editors_2019} & Lab-made pouch NMC/Gr & Ti-based coating & Uncoated positive electrode & Positive electrode impedance growth \\ \cline{2-7} 
 & \textbf{Graphite type} & Ma et al. 2019 \cite{ma_editors_2019} & Lab-made pouch NMC/Gr & Artificial (Kaijin AML-400), natural (BTR-918) & Natural graphite & N/A \\ 
 \cline{2-7} 
 & \multirow{3}{*}{
 \begin{tabular}[c]{@{}l@{}} \textbf{Additive package}\\ \textbf{and concentration}\end{tabular}
  } & Petibon et al. 2016 \cite{petibon_studies_2016} & Lab-made pouch LCO/Gr-Si & N/A & FEC consumed & SEI growth \\ 
 \cline{3-7} 
 &  & Jung et al. 2016 \cite{jung_consumption_2016} & Lab-made coin LFP/Gr-Si & 0--20 wt.\% FEC & FEC consumed & SEI growth \\ 
 \cline{3-7} 
 &  & Ma et al. 2019 \cite{ma_editors_2019} & Lab-made pouch NMC/Gr & 0--20\% methyl acetate additive & Higher methyl acetate concentration & Positive electrode impedance growth \\ 
 \cline{2-7} 
 & \multirow{3}{*}{\textbf{Salt concentration}} & Aiken et al. 2020 \cite{aiken_accelerated_2020} & Lab-made pouch NMC/Gr & 0.2--1.2M LiPF6 & Higher salt concentration & Electrolyte oxidation \\ 
 \cline{3-7} 
 &  & Ma et al. 2019 \cite{ma_editors_2019} & Lab-made pouch NMC/Gr & 1.2--1.5M LiPF6 & Lower salt concentration & Positive electrode impedance growth \\ 
 \cline{3-7} 
 &  & Wang et al. 2014 \cite{wang_systematic_2014} & Lab-made pouch LCO/Gr & 0.5--2M LiPF6 & Higher salt concentration & Positive electrode impedance growth \\ 
 \hline
\multirow{34}{*}{
\begin{tabular}[c]{@{}l@{}} \textbf{Testing}\\ \textbf{conditions}\end{tabular}
} & \multirow{8}{*}{\textbf{Charging rate}} & Lewerenz et al. 2017 \cite{lewerenz_systematic_2017}, \cite{lewerenz_post-mortem_2017} & OMT OMLIFE-8AH-HP LFP/Gr & 1-8C & Higher charging rate & Li plating, SEI growth \\ \cline{3-7} 
 &  & Petzl et al. 2015 \cite{petzl_lithium_2015} & Commercial 26650 LFP/Gr & 0.5--1C & Higher charging rate & Li plating \\ 
 \cline{3-7} 
 &  & Burns et al. 2015 \cite{burns_-situ_2015} & Panasonic 18650 NCA/Gr & 0.1--1C & Higher charging rate & Li plating \\ 
 \cline{3-7} 
 &  & Waldmann et al. 2015 \cite{waldmann_optimization_2015} & Cylindrical NCA/Gr & 0.25--1C, single vs. multi-step CC, optional CV & Higher charging rate, CV & Li plating \\ 
 \cline{3-7} 
 &  & Schuster et al. 2015 \cite{schuster_nonlinear_2015} & E-One Moli Energy IHR18650A NMC/Gr & 0.2--1C & Higher charging rate & Li plating, SEI growth \\ 
 \cline{3-7} 
 &  & Severson et al. 2019 \cite{severson_data-driven_2019} & A123 APR18650M1A LFP/Gr & 3.6--8C & Higher charging rate & LAM-induced Li plating, SEI growth \\ 
 \cline{3-7} 
 &  & Schindler et al. 2018 \cite{schindler_fast_2018} &  Samsung ICR18560-26F NMC/Gr & 
  \begin{tabular}[c]{@{}l@{}}0.25--2C with AC pulse, \\ current derating, current interrupt\end{tabular}
 & 
 \begin{tabular}[c]{@{}l@{}}Higher charging rate, \\ no AC pulse or current interrupt\end{tabular}
  & Li plating \\ 
 \cline{3-7} 
 &  & Keil et al. 2019 \cite{keil_linear_2019} & Cylindrical NMC/Gr & 0.7--1C & Higher charging rate & Li plating, SEI growth \\ 
 \cline{2-7} 
 & \multirow{5}{*}{\textbf{Discharging rate}} & Keil et al. 2016 \cite{keil_charging_2016} & \begin{tabular}[c]{@{}l@{}}a) Sanyo UR18650SA LMO+NMC/Gr\\ b) Sony US18650VT1 LMO+LCO/Gr\\ c) A123 APR18650M1A LFP/Gr\end{tabular} & \begin{tabular}[c]{@{}l@{}}a) 2.4--4C\\ b) 2.7--4.5C\\ c) 2.7--4.5C\end{tabular} & \begin{tabular}[c]{@{}l@{}}a) No difference\\ b) No difference\\ c) Lower discharging rate\end{tabular} & Li plating \\ 
 \cline{3-7} 
 &  & Keil et al. 2019 \cite{keil_linear_2019} & Cylindrical NMC/Gr & 1--2C & Lower discharging rate & Li plating, SEI growth \\ 
 \cline{3-7} 
 &  & Atalay et al. 2020 \cite{atalay_theory_2020} & Commercial 18650 NCA/Gr & 1--4C & Lower discharging rate & Li plating, SEI growth \\ 
 \cline{3-7} 
 &  & Omar et al. 2014 \cite{omar_lithium_2014} & Commercial cylindrical LFP/Gr & 1--15C & Higher discharging rate & SEI growth \\ 
 \cline{3-7} 
 &  & Diao et al. 2019 \cite{diao_accelerated_2019} & Pouch LCO/Gr & 0.7--2C & No difference at 10--45\degree C &  \\
 \cline{2-7} 
 & \multirow{8}{*}{\textbf{Voltage limits}} & Broussely et al. 2005 \cite{broussely_main_2005} & Saft VLE NCA/Gr & 50\%-100\% storage SOC & Higher SOC & Electrolyte oxidation \\ 
 \cline{3-7} 
 &  & Aiken et al. 2020 \cite{aiken_accelerated_2020} & Lab-made pouch NMC/Gr & 4.3--4.4V charge cutoff voltage & Higher voltage & Electrolyte oxidation \\ 
 \cline{3-7} 
 &  & 
 \begin{tabular}[c]{@{}l@{}} Ecker et al. 2014 \cite{ecker_calendar_2014},\\ Pfrang et al. 2018 \cite{pfrang_long-term_2018}\end{tabular}
 & Sanyo UR18650E NMC/Gr & \begin{tabular}[c]{@{}l@{}}1) 0.5\%--100\% DOD, 50\% SOC midpoint\\ 2) 10\% DOD and midpoint SOC of 10\%--95\%\end{tabular} & \begin{tabular}[c]{@{}l@{}}1) Higher DOD\\ 2) Extreme midpoints\end{tabular} & Mechanical deformation \\ 
 \cline{3-7} 
 &  & Klett et al. 2014 \cite{klett_non-uniform_2014} & Commercial 26650 LFP/Gr & 30--50\% vs. 5--95\% SOC & Higher DOD & SEI growth \\ 
 \cline{3-7} 
 &  & Schuster et al. 2015 \cite{schuster_nonlinear_2015} & E-One Moli Energy IHR18650A NMC/Gr & 0.56--1.2V DOD, 3.6V midpoint & Higher DOD & Li plating \\ 
 \cline{3-7} 
 &  & Ma et al. 2019 \cite{ma_novel_2019} & Commercial prismatic NMC+LMO/Gr & 0--20\%, 20--60\%, 60--100\%, 0--100\% SOC & \begin{tabular}[c]{@{}l@{}}1) Higher DOD\\ 2) Higher midpoint SOC\end{tabular} & Li plating \\ 
 \cline{3-7} 
 &  & Petzl et al. 2015 \cite{petzl_lithium_2015} & Commercial 26650 LFP/Gr & 0--80\% vs. 0--100\% SOC & Higher DOD & Li plating \\ 
 \cline{3-7} 
 &  & Zhu et al. 2021 \cite{zhu_investigation_2021} & Samsung INR 18650 25R NMC+NCA/Gr & 20--60\% DOD, 15--85\% SOC midpoint & Lower SOC & SEI growth \\ 
 \cline{2-7} 
 & \multirow{3}{*}{\textbf{Rests}} & Keil et al. 2019 \cite{keil_linear_2019} & Cylindrical NMC/Gr & 10--900s at TOC and BOD & Longer rest time & Li plating, SEI growth \\ 
 \cline{3-7} 
 &  & Ma et al. 2019 \cite{ma_editors_2019} & Lab-made pouch NMC/Gr & 0--30min at TOC and BOD & Longer rest time & Positive electrode impedance growth \\ 
 \cline{3-7} 
 &  & Epding et al. 2019 \cite{epding_investigation_2019} & Commercial prismatic NMC/Gr & 0--every 100 cycles & Shorter rest time & Li plating \\ 
 \cline{2-7} 
 & \multirow{7}{*}{\textbf{Temperature}} & Zhang et al. 2019 \cite{zhang_accelerated_2019} & NMC/Gr & 25--45\degree C & Temperature above and below 25\degree C & Li plating \\ 
 \cline{3-7} 
 &  & Broussely et al. 2005 \cite{broussely_main_2005} & Saft VLE NCA/Gr & 20--60\degree C & Higher temperature & Electrolyte oxidation \\ \cline{3-7} 
 &  & Schuster et al. 2015 \cite{schuster_nonlinear_2015} & E-One Moli Energy IHR18650A NMC/Gr & 25--50\degree C & Temperature above and below 35\degree C & Li plating, SEI growth \\ 
 \cline{3-7} 
 &  & Safari et al. 2011 \cite{safari_aging_2011} & Commercial 26650 LFP/Gr & 25--45\degree C & Higher temperature & LAM (graphite) \\ 
 \cline{3-7} 
 &  & Waldmann et al. 2014 \cite{waldmann_temperature_2014} & Commercial 18650 NMC+LMO/Gr & -20--70\degree C & Temperature above and below 25\degree C & Li plating, SEI growth \\ 
 \cline{3-7} 
 &  & Coron et al. 2020 \cite{coron_impact_2020} & \begin{tabular}[c]{@{}l@{}}Commercial 18650 NMC+LMO/Gr\\  Commercial 18650 NMC/Gr\end{tabular} & 0--25\degree C & Lower temperature & SEI growth, LAM \\ 
 \cline{3-7} 
 &  & Waldmann et al. 2015 \cite{waldmann_optimization_2015} & Cylindrical NCA/Gr & 0--60\degree C & Temperature below 25\degree C & Li plating, SEI growth \\ 
 \cline{2-7} 
 & \multirow{3}{*}{\textbf{Pressure}} & Wunsch et al. 2019 \cite{wunsch_investigation_2019} & Commercial pouch NMC/Gr & 4 bracing approaches & More rigid bracing or zero bracing & N/A \\ 
 \cline{3-7} 
 &  & Cannarella et al. 2014 \cite{cannarella_stress_2014} & Pouch LCO/Gr & 0--5 MPa & Higher stack pressure or zero pressure & LAM (graphite) or Li plating \\ \cline{3-7} 
 &  & Bach et al. 2016 \cite{bach_nonlinear_2016} & E-One Moli Energy IHR18650A NMC/Gr & With and without hose clamp & Heterogeneous compression & Li plating \\ 
 \hline
\multirow{4}{*}{
\begin{tabular}[c]{@{}l@{}} \textbf{Cell-to-cell}\\ \textbf{variation}\end{tabular}
} & \multicolumn{1}{l|}{\multirow{4}{*}{\textbf{}}} & Harris et al. 2017 \cite{harris_failure_2017} & Commercial pouch LCO/Gr & 24 cells & N/A & N/A \\ 
\cline{3-7} 
 & \multicolumn{1}{l|}{} & Baumhofer et al. 2014 \cite{baumhofer_production_2014} & Sanyo UR18650E NMC/Gr & 48 cells & N/A & N/A \\ 
 \cline{3-7} 
 & \multicolumn{1}{l|}{} & Willenberg et al. 2020 \cite{willenberg_high-precision_2020} & Samsung INR18650 35E NCA/Gr+Si & 4 cells & N/A & Mechanical deformation \\ 
 \cline{3-7} 
 & \multicolumn{1}{l|}{} & Stiaszny et al. 2014 \cite{stiaszny_electrochemical_2014} & Commercial 18650 NMC+LMO/Gr & 6 cells & N/A & N/A \\ 
 \hline
\end{tabular}
}
\label{tab:experimental_summary}
\end{table}

\end{landscape}
 \restoregeometry

\begin{table}[!ht]
    \centering
    \begin{tabular}{|c||c|c|c|c|}
        \hline
\multirow{2}{*}{Reference}
& Multiple test & Test        & Rel.~capacity & Rel.~resistance \\
& conditions?   & replicates? & at knee onset & at elbow onset \\
        \hline
        Wünsch et al.\cite{wunsch_investigation_2019} & Yes & No & 95\% & 100\% \\
        Willenberg et al.\cite{willenberg_development_2020} & No & Yes & 90\% & 120\% \\
        Rahe et al.\cite{rahe_nanoscale_2019} & No & No & 90\% & 160\% \\
        Martinez-Laserna et al.\cite{martinez-laserna_technical_2018} & Yes & No &  85\% & 170\% \\
        Klett et al.\cite{klett_non-uniform_2014} & No & No & 80\% & 110\% \\
        Lewerenz et al.\cite{lewerenz_systematic_2017, lewerenz_post-mortem_2017} & Yes & Yes & 80\% & 110\% \\
        Ecker et al.\cite{ecker_calendar_2014} & Yes & No & 80\% & 150\% \\
        Frisco et al.\cite{frisco_understanding_2016} & No & No & 80\% & 200\% \\
        Schuster et al.\cite{schuster_nonlinear_2015} & Yes & No & 80\% & 300\% \\
        Pfrang et al.\cite{pfrang_long-term_2018} & No & Yes & 75\% & 130\% \\
        Braco et al.\cite{braco_experimental_2020} & No & Yes & 70\% & 200\% \\
        Broussely et al.\cite{broussely_main_2005} & No & Yes & 70\% & 200\% \\
        \hline
    \end{tabular}
    \caption{Summary of studies reporting both capacity and resistance over cell lifetime with capacity knees and/or resistance elbows. All studies measure resistance using a direct-current pulse except for Schuster et al. \cite{schuster_nonlinear_2015}{}, which uses electrochemical impedance spectroscopy. Relative capacity and resistance values at capacity knee/resistance elbow onset are estimated either from a single measurement or roughly averaged from several measurements.}
    \label{tab:dcr_growth_papers}
\end{table}


\end{document}